\documentclass[preprint2]{aastex}

\bibpunct{(}{)}{;}{a}{}{,}
\usepackage{xspace}
\usepackage{amssymb}
\usepackage{amsmath}
\usepackage{verbatim}
\usepackage{enumerate}

\usepackage{natbib}
\usepackage{chngpage}
\usepackage{array}
\newcommand {\apgt} {\ {\raise-.5ex\hbox{$\buildrel>\over\sim$}}\ }
\newcommand {\aplt} {\ {\raise-.5ex\hbox{$\buildrel<\over\sim$}}\ }

\newcommand{\ro}[1]{\ensuremath{\textrm{#1}}}
\newcommand{\kmsec}{\ensuremath{\textrm{km}~ \ro{s}^{-1}}\xspace}
\newcommand{\ergs}{\ensuremath{~\textrm{erg s}^{-1}}}
\newcommand{\cm}{\ensuremath{~\textrm{cm}^{-2}}}
\newcommand{\ergsca}{\ensuremath{~\textrm{erg s}^{-1} \textrm{cm}^{-2} \textrm{arcsec}^{-2}}}
\newcommand{\Msol}{\ensuremath{M_{\odot}}\xspace}
\newcommand{\lya}{\ensuremath{\textrm{Ly}\alpha}\xspace}
\newcommand{\hi}{\ensuremath{\textrm{H\textsc{i}}}\xspace}

\newcommand{\dhi}{\ensuremath{n_{\ro{H\textsc{i}}}}\xspace}
\newcommand{\nhi}{\ensuremath{N_{\ro{H\textsc{i}}}}\xspace}
\newcommand{\df}{\ensuremath{~ \ro{d}}}
\newcommand{\dd}{\ensuremath{\ro{d}}}
\newcommand{\cN}{\ensuremath{\mathcal{N}}\xspace}
\newcommand{\ps}{Press-Schechter }

\renewcommand\arraystretch{1.4}

\newcommand{\MgIIdblt}{{\rm Mg}\kern 0.1em{\sc ii}~$\lambda\lambda 2796, 2803$}
\newcommand{\HI}{\hbox{{\rm H}\kern 0.1em{\sc i}}}
\newcommand{\Lya}{\hbox{{\rm Ly}\kern 0.1em$\alpha$}}
\newcommand{\Lyb}{\hbox{{\rm Ly}\kern 0.1em$\beta$}}
\newcommand{\MgII}{\hbox{{\rm Mg}\kern 0.1em{\sc ii}}}
\newcommand{\cmsq}{\hbox{cm$^{-2}$}}

\renewcommand {\lesssim} {\,\raisebox{-0.6ex}{$\stackrel{\textstyle<}{\textstyle\sim}$}\,}
\renewcommand {\gtrsim} {\,\raisebox{-0.6ex}{$\stackrel{\textstyle>}{\textstyle\sim}$}\,}
\shorttitle{Lyman Alpha and {\MgII} as Probes of Galaxies and their Environment}
\shortauthors{Barnes, Garel and Kacprzak}

\begin{document}


\title{Lyman Alpha and {\MgII} as Probes of \\
    Galaxies and their Environment}

\author{Luke A. Barnes}
\affil{Super Science Fellow, Sydney Institute for Astronomy, \\
    School of Physics, University of Sydney, Australia}
\email{L.Barnes@physics.usyd.edu.au}

\and

\author{Thibault Garel and Glenn G. Kacprzak}
\affil{Super Science Fellow, Center for Astrophysics \& Supercomputing, \\
    Swinburne University of Technology, Hawthorn, VIC 3122, Australia}

\begin{abstract}
\lya emission, \lya absorption and {\MgII} absorption are powerful tracers of neutral hydrogen. Hydrogen is the most abundant element in the universe and plays a central role in galaxy formation via gas accretion and outflows, as well as being the precursor to molecular clouds, the sites of star formation. Since 21cm emission from neutral hydrogen can only be directly observed in the local universe, we rely on \lya emission, and \lya and {\MgII} absorption to probe the physics that drives galaxy evolution at higher redshifts. Furthermore, these tracers are sensitive to a range of hydrogen densities that cover the interstellar medium, the circumgalactic medium and the intergalactic medium, providing an invaluable means of studying gas physics in regimes where it is poorly understood. At high redshift, \lya emission line searches have discovered thousands of star-forming galaxies out to $z = 7$. The large \lya scattering cross-section makes observations of this line sensitive to even very diffuse gas outside of galaxies. Several thousand more high-redshift galaxies are known from damped \lya absorption lines and absorption by the {\MgII} doublet in quasar and GRB spectra. {\MgII}, in particular, probes metal-enriched neutral gas inside galaxy haloes in a wide range of environments and redshifts ($0.1 < z < 6.3$), including the so-called redshift desert. Here we review what observations and theoretical models of \lya emission, \lya and {\MgII} absorption have told us about the interstellar, circumgalactic and intergalactic medium in the context of galaxy formation and evolution.

\end{abstract}

\keywords{galaxies: evolution --- galaxies: formation --- galaxies: ISM}

\section{Introduction}

In the modern picture of galaxy formation, primordial gas from the intergalactic medium (IGM) falling into the gravitational potential well of dark matter haloes can accrete onto galaxies to feed the interstellar medium (ISM) and fuel star formation. Feedback mechanisms powered by supernovae and active galactic nuclei are able to heat surrounding gas, and reprocess material from the galaxy into the circumgalactic medium (CGM) in the form of galactic winds.
 
As the most abundant element in the Universe, hydrogen is a unique tracer of the formation and evolution of galaxies, and its neutral atomic form (\hi) can probe gas at various scales (ISM, CGM and IGM). Much of what we know about neutral hydrogen in the ISM comes from 21cm observations of the Milky Way, local galaxies, and the low-redshift Universe ($z \lesssim 0.05$). Future surveys with the Square Kilometer Array (SKA) pathfinders, such as the Australian SKA Pathfinder \citep[ASKAP][]{2008ExA....22..151J}, the Karoo Array Telescope \citep[MeerKAT][]{2009arXiv0910.2935B} and APERture Tile In Focus \citep[APERTIF][]{2008AIPC.1035..265V}, will be restricted to redshifts less than 1, which will not allow to carry out \hi surveys at earlier epochs, i.e. when galaxies were forming stars at the highest rate ($z \gtrsim 1.5-2$). Until the large field of view and high sensitivity of the SKA become available, other tracers of the hydrogen gas are then needed at high redshift.
 
The Lyman Alpha (\lya) line offers an invaluable insight into the gas physics that drives galaxy formation. The \lya line results from a transition between the $2~^2P$ state and the $1~^2S$ (ground) state in hydrogen atoms ($\lambda=1215.67$\AA), that can be observed from the ground at $z \gtrsim 2$, both in emission or in absorption.
 
In star-forming galaxies, intense \lya emission lines can be produced in the ISM as the result of the photoionisation and the subsequent recombination of hydrogen by massive, short-lived stars. \lya emission features are commonly observed in high-redshift galaxies, and thousands of so-called Lyman-Alpha emitters (LAEs) have been detected at $z = 2-7$ \citep[e.g.][]{1998ApJ...502L..99H,2000ApJ...545L..85R,2008ApJS..176..301O,2011AandA...525A.143C}. A major uncertainty in the interpretation of the data comes from the resonant scattering of \lya photons by \hi atoms in the interstellar, circumgalatic, and intergalatic medium. Indeed, it has been observationally shown that the emergent \lya line profile and \lya spatial extent can be highly affected by the kinematics, geometry, and ionisation state of the gas in and around galaxies \citep[e.g.][]{2003ApJ...588...65S,2009AJ....138..923O,2010ApJ...717..289S}.

In absorption, the large cross-section for the \lya transition makes it the most sensitive method for detecting baryons at any redshift \citep[e.g.][]{1998ARAandA..36..267R}. Close to 7000 absorption systems with $z > 2.15$ have been detected with \hi column densities $N_\hi > 2 \times 10^{20} \cm$, suggestive of the ISM of galaxies.

In addition to \lya, metal-lines, such as the {\MgIIdblt} doublet, provide a direct tracer of neutral hydrogen column densities, $10^{16} \lesssim \nhi \lesssim 10^{22} \cm$ \citep{archiveI,weakII}, and can be observed in a redshift range much of which cannot be accessed in \lya from the ground ($0.1 < z < 6.3$). These metal-lines can be used to study the \hi when it is not directly detected or can be used in tandem to determine the origins of the gas.

A particularly important motivation for this review is the ability of \lya and the {\MgIIdblt} doublet to shed light on the circumgalactic medium, roughly defined as the 100--300 kpc region around a galaxy, distinct from the stellar system but within the virial radius of its halo. The CGM has been the subject of intense theoretical study in recent years, as it is where accretion (cold and hot) and feedback (supernovae and AGN) meet. \citet{1969ApJ...156L..63B} were the first to suggest that quasar absorption lines are caused by ``tenuous gas in extended haloes of normal galaxies''. Together, \lya scattering and emission, and \lya and metal-line absorption hold great potential to learn about gas in galaxies and their surroundings, and the mechanisms that govern their formation and evolution.

Here, we review observations and models of neutral hydrogen in and around galaxies in the high-redshift universe. Section \ref{S:lyaAbs} will examine \lya in absorption, focusing on observations and simulations of damped \lya absorption systems (DLAs). Section \ref{S:MgII} will examine the {\MgIIdblt} doublet and how it is used to trace the processes that shape the CGM. Section \ref{S:lyaEmission} will consider \lya in emission: the physics of its production and scattering, observations and models of \lya emitting galaxies, and its illumination of the wider cosmological context of galaxy formation.

\section{Lyman Alpha in Absorption} \label{S:lyaAbs}

Lyman alpha photons are strongly scattered by \hi. There are no other states between the $2~^2P$ state and the ground state in an \hi atom; thus, when the probability of collisional de-excitation is negligible (as it almost always is in astrophysical contexts), the absorption of a \lya photon by an \hi atom is almost immediately ($A_\ro{21}^{-1} = 1.6 \times 10^{-9}$s) followed by the re-emission of a \lya photon. This can be thought of as a scattering process. A \lya photon (at line centre) passing through an \hi region with column density \nhi and temperature $T$ encounters an optical depth of,
\begin{equation} \label{eq:tauLya}
\tau_0 \approx 0.59 \left( \frac{T} {10^4 ~ \ro{K}} \right)^{-\frac{1}{2}} \left( \frac{\nhi} {10^{13} \cm} \right) .
\end{equation}

\begin{figure*}[t]
\centering
	\includegraphics[width=\textwidth]{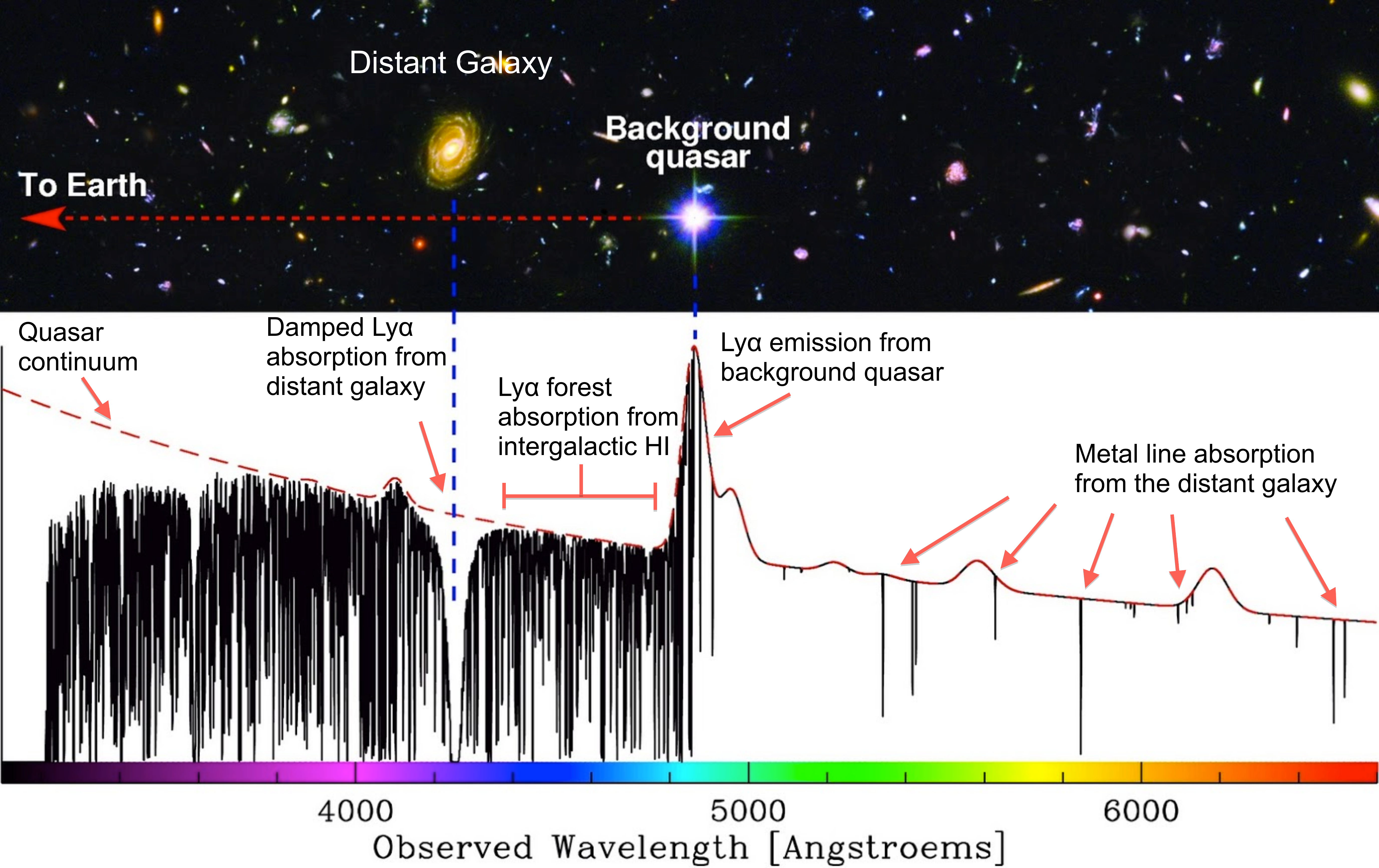}
	\caption{A schematic illustration of \lya absorption seen in the spectra of quasars. At wavelengths shorter than the (redshifted) \lya emission from the quasar, \lya is seen in absorption, as intervening \hi scatters the quasar continuum out of the line of sight. The large column of density \hi in intervening galaxies causes a broad, damped \lya absorption trough, along with associated metal absorption lines at the same redshift. Figure courtesy of Michael Murphy. An excellent movie of \lya absorption can be found on Andrew Pontzen's website: http://www.cosmocrunch.co.uk/media/dla\_credited.mov .}
	\label{fig:QSOlya}
\end{figure*}

Lyman alpha in absorption is observed in the spectra of quasars and gamma ray bursts (Figure \ref{fig:QSOlya}). Three classes of absorbers are distinguished by their neutral hydrogen column density, \nhi. \lya forest absorbers have $\nhi < 10^{17} \cm$, making them optically thin to ionizing radiation. Lyman limit systems (LLS) have $10^{17} \cm < \nhi < 2 \times 10^{20} \cm$. Damped \lya absorption systems (DLAs) are the highest column density systems, with $\nhi > 2 \times 10^{20} \cm$. Damped \lya absorption profiles are characterized by their Lorentz or damping wings: at such high column densities, unit optical depth occurs in the damping wings of the profile function, beyond the inner Doppler core. The equivalent width of the line is independent of the velocity and temperature structure of the absorber. The column density $\nhi = 2 \times 10^{20} \cm$ also fortuitously separates the predominantly ionized LLS population from DLAs, in which the hydrogen is mainly neutral due to self-shielding. We will focus here on the properties of DLAs, believed to probe galaxies and their immediate surroundings.

\subsection{Observed Properties of DLAs}
\begin{figure*}[t]
\centering
\begin{minipage}{0.45\textwidth}
\includegraphics[width=\textwidth]{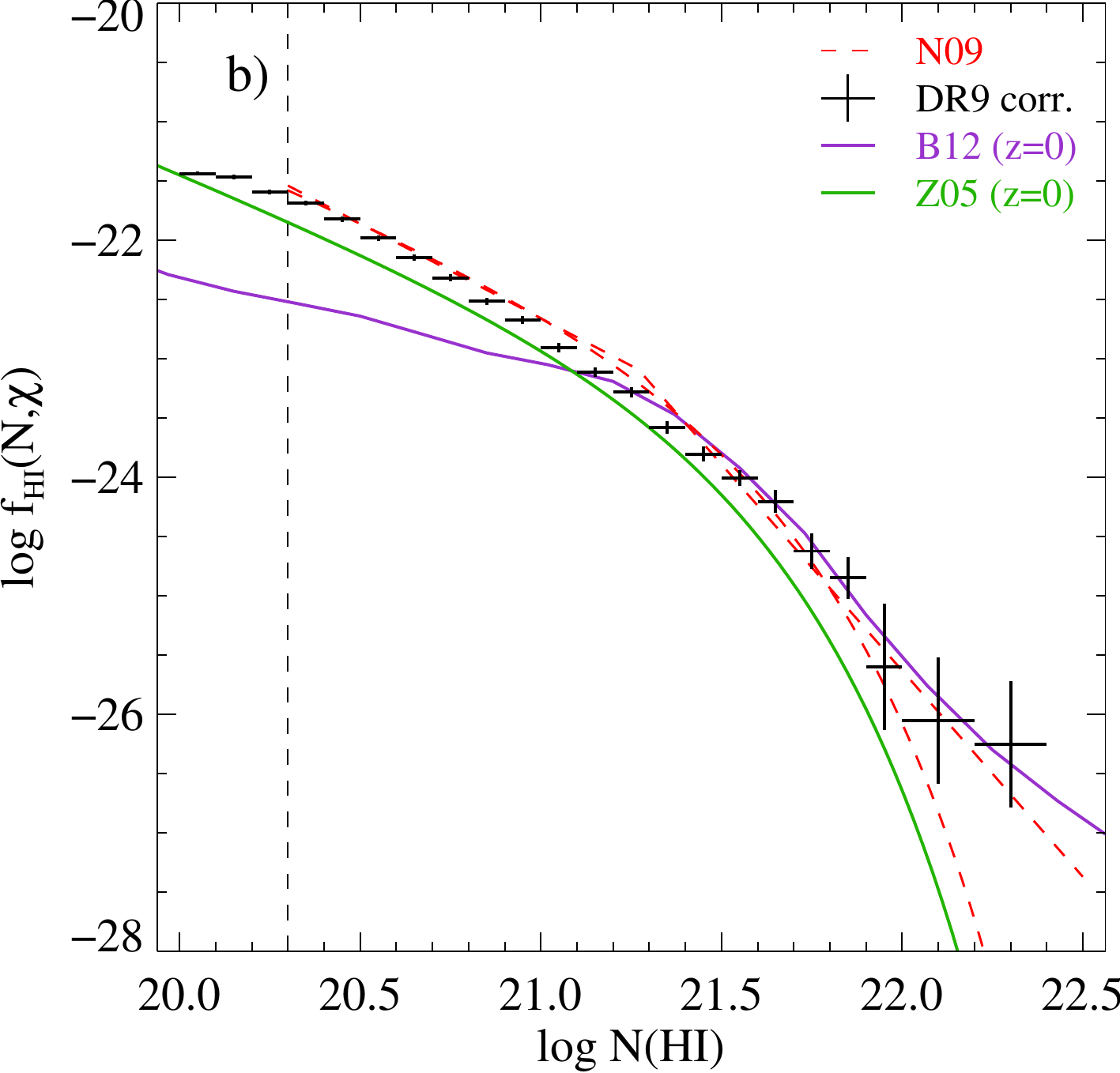}
\end{minipage}
\hspace{0.5cm}
\begin{minipage}{0.45\textwidth}
\includegraphics[width=\textwidth]{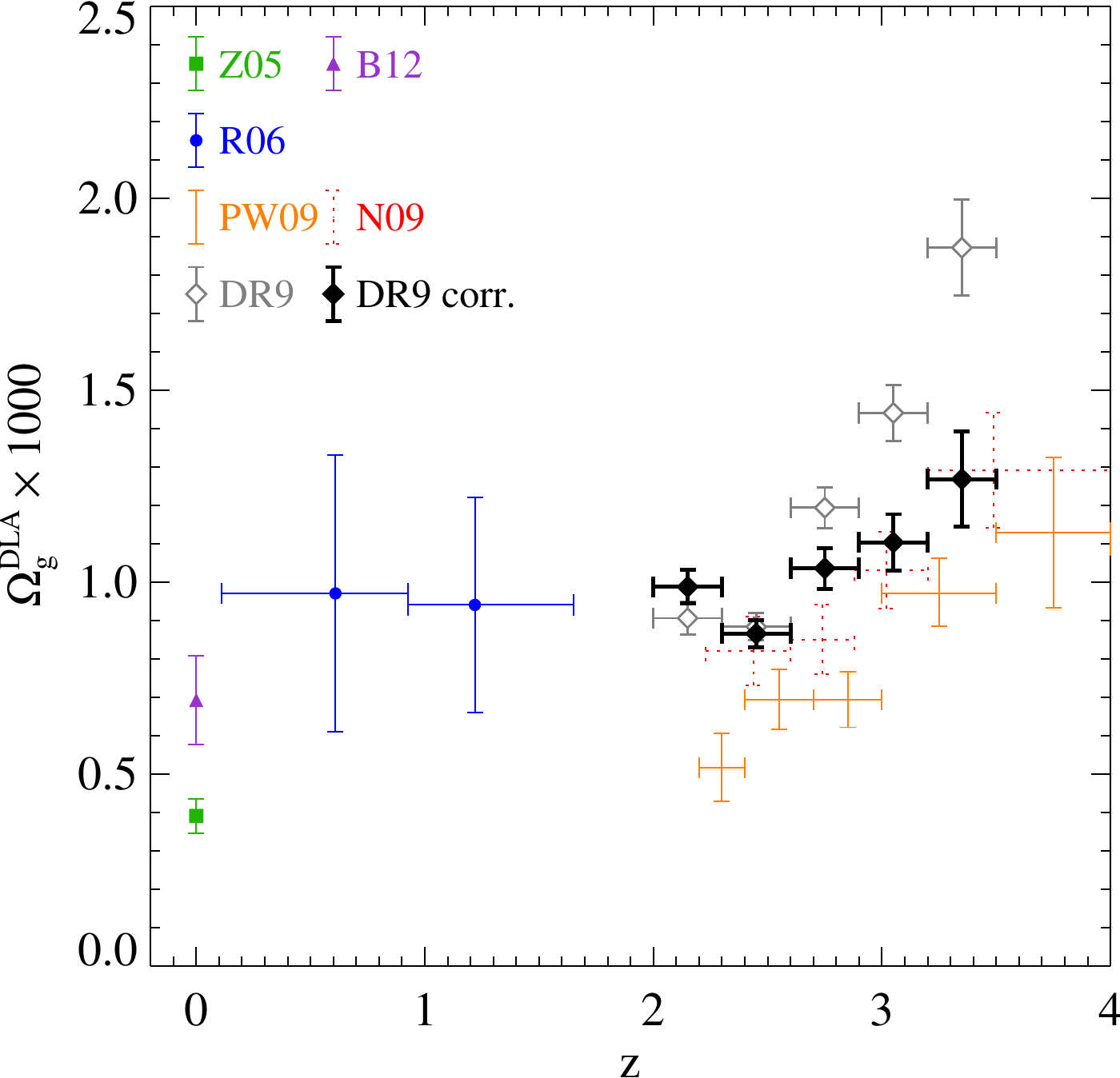}
\end{minipage}
\caption{\emph{Left:} Column density distribution function $f(\nhi,X)$ at $\langle z \rangle = 2.5$ from \citet{2012AandA...547L...1N}. Horizontal bars represent the bin over which $f$ is calculated and vertical error bars represent Poissonian uncertainty. The double power-law and $\Gamma$-function fits to the DR7 distribution of \citet{2009AandA...505.1087N} at $\langle z \rangle = 2.9$ are shown as red dashed lines. $f(\nhi,X)(z = 0)$ is taken from \citet[][purple]{2012ApJ...749...87B} and \citet[][green]{2005MNRAS.364.1467Z}. \emph{Right:} Cosmological mass density of neutral gas in DLAs as a function of redshift [Z05: \citet{2005MNRAS.364.1467Z}; B12: \citet{2012ApJ...749...87B}; R06: \citet{2006ApJ...636..610R}; PW09: \citet{2009ApJ...696.1543P}; DR9: \citet{2012AandA...547L...1N}]. Credit: \citeauthor{2012AandA...547L...1N}, A\&A, 547, L1, 2012, reproduced with permission \copyright ESO.}
\label{fig:fHIX_0}
\end{figure*}

\citet{1986RSPTA.320..503W} conducted the first DLA survey, using quasar spectra to detect neutral gas in absorption in the disks of high-redshift galaxies. To identify a DLA, one must distinguish a single absorption line, broadened by damping, from an absorption feature caused by the blending of Doppler-broadened, low column density systems. The \lya forest also generates confusion noise, contaminating the damping wings. The Sloan Digital Sky Survey \citep[SDSS][]{2004PASP..116..622P,2005ApJ...635..123P} overcame these problems using high-quality, good spectral resolution ($R \sim 2000$) that do not require higher-resolution follow-up spectroscopy to accurately fit Voigt profiles to the data. Using such methods, around ten thousand DLAs have been observed. The largest DLA survey to date is the Baryon Oscillation Spectroscopic Survey \citep[BOSS,][]{2013AJ....145...10D}, part of the Sloan Digital Sky Survey (SDSS) III \citep{2011AJ....142...72E}. The full sample, based on SDSS Data Release 9, contains over 150,000 quasar spectra over the redshift range $2.15 < z < 3.5$ and has discovered 6,839 DLAs, which is an order of magnitude larger than SDSS II.

The statistical properties of the DLA population can by summarised by its distribution function: the number of absorbers $(\dd^2 \mathcal{N})$ along a random line of sight in the redshift range $(z,z+\dd z)$ that have \hi column densities in the range $(\nhi, \nhi+\dd \nhi)$ is,
\begin{align}
	\dd^2 \mathcal{N} =& ~ n_{\nhi} (\nhi,z) \sigma_\ro{DLA}(z) (1+z)^3 \frac{\dd l_p}{\dd z} \df \nhi \df z \\
	 \equiv& f(\nhi,X) \df \nhi \df X ,
\end{align}
where $n_{\nhi} \dd \nhi$ is the comoving number density of DLAs within $(\nhi,\nhi+\dd \nhi)$, $\sigma_\ro{DLA}$ is the DLA absorption cross-section, $\frac{\dd l_p}{\dd z} = \frac{c} {H(z) (1+z)}$ is the ratio of proper distance interval to redshift interval, and the so-called absorption distance $X$ is defined by $\dd X \equiv \frac{H_0} {H(z)} (1+z)^2 \df z$. Figure \ref{fig:fHIX_0} shows the results of \citet{2012AandA...547L...1N}, derived from 5428 DLA systems observed as part of the BOSS survey.

$f(\nhi,X)$ can be approximated by a double power law. The distribution has a low-\nhi dependence of $f \propto \nhi^{-2}$, and drops more steeply above $\nhi = 10^{21.2} \cm$, which has be attributed to the  high-mass turnover of the halo mass function, the impact of stellar feedback, and the formation of molecular hydrogen in the highest column density clouds. A similar turnover is seen in the \hi column density profiles of local THINGS galaxies, which \citet{2012ApJ...761...54E} show is not strongly correlated with gas metallicity. This suggests that the \hi-$H_2$ transition is not the primary driver of this feature of $f(\nhi,X)$. 

The zeroth moment of $f(\nhi,X)$ gives the number of DLAs encountered along a line of sight per unit absorption distance $\df X$, i.e. the line density of DLAs,
\begin{equation}
l_{\ro{DLA}}(X) \df X \equiv \int^\infty_{N_{\ro{DLA}}} f(\nhi,X) \df \nhi \df X ~,
\end{equation}
where $N_{\ro{DLA}} = 2 \times 10^{20} \cm$. Alternatively, we can define this quantity in terms of $\df z$; $l_{\ro{DLA}}(z) \equiv \dd \mathcal{N} / \dd z$ is often referred to as the DLA covering fraction per unit redshift. \citet{2009ApJ...696.1543P}, using SDSS Data Release 5, show that $l_{\ro{DLA}}(X)$ drops from 0.08 at $z \sim 3-4$ to 0.05 at $z \sim 2.2$, at which point it is consistent with the present-day value estimated from 21cm observations \citep[][note however \citealt{2012ApJ...749...87B}]{2003MNRAS.343.1195R,2005MNRAS.359L..30Z}. This indicates surprisingly little evolution in galactic HI gas for the last $\sim 10$ Gyr.

% (1 + 2.3)^2 / sqrt(0.285*(1 + 2.3)^3 + (1 - 0.285))*0.055

The first moment of $f(\nhi,X)$ can be related to the mass density of gas associated with DLAs, 
\begin{equation}
\Omega^{DLA}_g (X)= \frac{\mu m_{\ro{H}} H_0}{c \rho_\ro{crit}} \int^\infty_{N_{\ro{DLA}}} \nhi ~ f(\nhi,X) \df N ~,
\end{equation}
where $\mu$ is the mean molecular mass of the gas; see \citet{2005ApJ...635..123P} for a discussion of the limits of the integral, and the contribution of LLSs to the mass density of gas associated with \hi atoms. DLAs dominate the neutral gas content of the Universe in the redshift interval $z = 0-5$. Further, DLAs between $z \sim 3.0-4.5$ contain sufficient neutral hydrogen to account for a significant fraction of the gas mass in stars in present-day galaxies. DLAs plausibly contain or are related to the reservoirs of gas needed to fuel star formation for much of the history of the Universe \citep{2005ARAandA..43..861W}.

\subsection{DLA Kinematics}

An important clue as to the nature of DLAs comes from the velocity profiles of associated metal lines. Because unit  optical depth occurs outside the Doppler core, the observed \lya absorption profiles of DLAs contain no information about the velocity structure of the gas. Observers have used high-resolution spectroscopy to study so-called low-ionisation metals (or low-ions). For example, Mg\textsc{i} (Mg\textsc{ii}) has an ionisation potential of 7.65 eV (15.04 eV) \citep{2003adu..book.....D}. Thus, in a region where neutral hydrogen is self-shielded from photons with energies above its ionisation potential (13.6 eV), the dominant ion of magnesium will be Mg\textsc{ii}. Other low-ions include Si\textsc{ii}, Fe\textsc{ii} and Ni\textsc{ii}, while N\textsc{i} and O\textsc{i} have ionisation potentials that are greater than 13.6 eV and are thus predominantly neutral in a self-shielded \hi region \citep{1995MNRAS.276..268V}. Thus, absorption features from low-ionisation metals at the same redshift as the DLA give us crucial information about the kinematics of the neutral gas.

\begin{figure}[t]
\centering
	\includegraphics[width=0.5\textwidth]{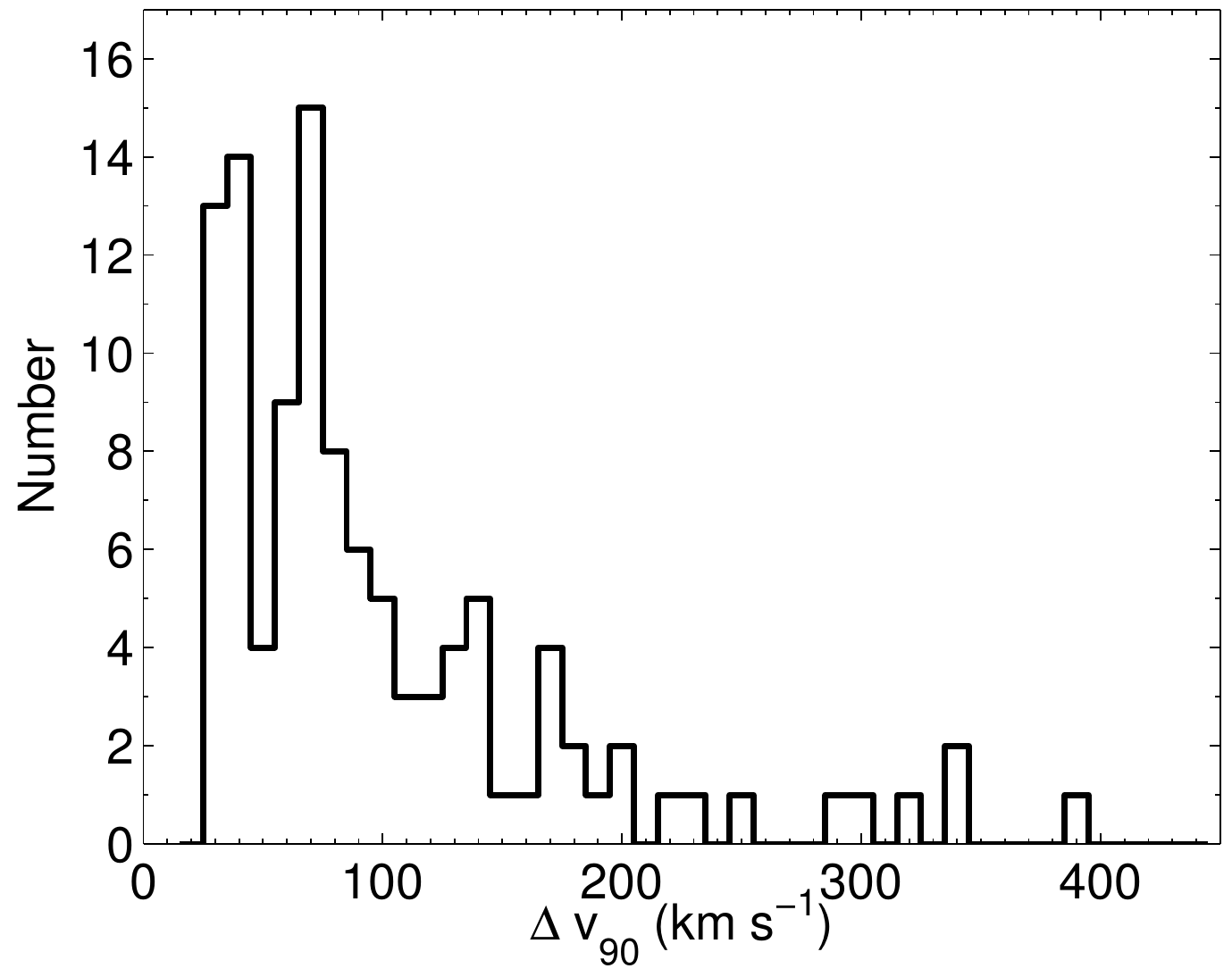}
	\caption[Velocity width distribution of low-ions associated with DLAs]{Histogram of the velocity width, $\Delta v_{90}$, of low-ions associated with the sample of DLAs found in \citet{2008ApJ...672...59P}. The sample shows velocity widths ranging from 15 \kmsec to several hundred \kmsec with a median of $\approx 80$ \kmsec. The mean redshift of the DLAs is $\langle z \rangle = 3$; \citet{2008MNRAS.390.1349P} and \citet{2013ApJ...769...54N} note that there is negligible redshift	evolution of the distribution.} \label{fig:Nvw}
\end{figure}

The most common statistic used to characterise the width of a low-ion metal absorption feature is $\Delta v_{90}$, defined by \citet{1997ApJ...487...73P} in their pioneering survey as the velocity interval encompassing 90\% of the total integrated optical depth. (We will often use the more compact notation $v_\ro{w}$.) Figure \ref{fig:Nvw} shows the observations of \citet{2008ApJ...672...59P}. The mean redshift of the DLA sample is $\langle z \rangle = 3$; \citet{2008MNRAS.390.1349P} and \citet{2013ApJ...769...54N} note that there is negligible redshift
 evolution of the distribution. DLAs velocity widths range from 15 \kmsec to several hundred \kmsec with a median of $\approx 80$ \kmsec. The high-velocity tail has been the subject of much theoretical attention, following the claim by \citet{1997ApJ...487...73P} that it conflicts with the predictions of hierarchical structure growth within CDM cosmologies (see Section \ref{Ss:DLAtheory}). 

An important low-redshift insight into DLA kinematics comes from 21cm emission from \hi in local galaxy disks. \citet{2008AJ....136.2886Z} use high-quality \hi 21cm data to measure the distribution of $\Delta v_{90}$ for local galaxies with $\nhi > 2 \times 10^{20} \cm$. The observed distribution peaks sharply at around 30 \kmsec, with a FWHM of $\sim 20$ \kmsec and a shallow tail out to $\sim 200$ \kmsec. This distribution is very different to that shown in Figure \ref{fig:Nvw} for high-redshift DLAs --- the median is smaller by more than a factor of two. \citet{2008AJ....136.2886Z} conclude that gas kinematics at high redshifts must be increasingly influenced by gas that does not participate in ordered rotation in cold disks.

\subsection{DLA Metallicity}
The metallicity of DLAs can shed light on their expected connection to star formation. The elemental abundances of DLAs are the most accurate measurements in the high-redshift Universe of the chemical enrichment of gas by stars. Furthermore, regardless of their exact identity, the mean metallicity of DLAs is the best measure we have of the amount of metal enrichment of neutral gas in the Universe at a given epoch. This subsection will follow the reviews of \citet{2004cmpe.conf..257P,2006fdg..conf..319P,2005ARAandA..43..861W}.

DLAs are generally metal-poor at all redshifts. This points to DLAs as arising in gas that is in its earliest stages of star formation. However, DLAs are not primordial clouds. No DLAs have been found without significant metal absorption. \citet{2003ApJ...595L...9P} report a ``metallicity floor'': though their observations are sensitive to $[M/H] < -4$, none were found with $[M/H] < -3$. \citet{2012ApJ...755...89R} show that this metallicity floor continues out to $z \sim 5$. Within the population of DLAs at a particular redshift, there is a wide scatter of metallicities, indicating different rates and stages of star formation within the DLA population. 

DLAs can trace the cosmic evolution of metallicity. \citet{2005ARAandA..43..861W}, \citet{2013ApJ...769...54N} and \citet{2013arXiv1310.6042R} report a statistically significant increase in the cosmic metallicity in DLAs with decreasing redshift, in contrast to the earlier conclusions of \citet{2004cmpe.conf..257P}. This is in keeping with the expectation that star formation will pollute the neutral gas with metals via supernovae and stellar winds. \citet{2013arXiv1310.6042R} further report evidence of a rapid increase in DLA metallicity between $z = 5$ and $z = 4.5$.

DLAs can also trace the large-scale distribution of metals. In particular, \emph{Proximate} DLAs (PDLAs), defined to have a velocity offset $< 3,000 \kmsec$ relative to their background quasar, are known to have higher metallicities than intervening DLAs, and higher metallicities as the quasar is approached \citep{2010MNRAS.406.1435E,2011MNRAS.412..448E}. This suggests PDLAs trace high-mass galaxies in the overdense regions of the universe that host quasars. Indeed, quasars with higher UV luminosities are assocated with lower-metallicity PDLAs. This is evidence of the direct impact of the quasar on surrounding galaxies, possibly involving the shutting-off of star-formation.

\citet{2006A&A...457...71L} find evidence of a velocity-metallicity correlation in DLAs, $[X/H] = 1.55 \log(v_w) - 4.33$, with X = Zn, S or Si. This has been confirmed by \citet{2013ApJ...769...54N} and \citet{2013MNRAS.430.2680M}, though with shallower slopes: 0.74 and 1.12 respectively. This correlation can be used to constrain the DLA mass-metallicity relationship \citep{2014MNRAS.440.2313B}, who conclude that metallicity increases with DLA host-halo mass as $[M/H]_\ro{mean} = 0.7 \log \left( \frac{M_v}{10^{11} \Msol} \right) ~- 1.8$. There is some evidence, too, that at a fixed halo mass, DLAs with above average velocity width (and thus presumably above average star formation rate) have above average metallicity.

%\citet{2006MNRAS.371.1519J} present a model that uses a physically motivated prescription for stellar feedback to follow the metallicity evolution of DLAs. Their model reproduces the low mean metallicities in DLAs, concluding that DLAs probe the outer gaseous parts of dwarf galaxies ($v_c \lesssim 70$ \kmsec). They further conclude that the galaxies responsible for DLAs make only a small contribution to the total star formation rate of the Universe.

An intriguing clue to the nature of DLA comes from observations of so-called \emph{sub-DLAs}, with $10^{19} \cm < \nhi < 2 \times 10^{20} \cm$. Imaging searches for DLA and sub-DLA galaxies show that DLAs probe smaller sightlines: median impact parameter is 17.4 kpc for the DLA galaxy sample and 33.3 kpc for the sub-DLA sample of \citet{2011MNRAS.416.1215R}. Given the expectation that DLAs to probe the densest, inner-most,  star-forming regions of galaxies, we might expect the sub-DLA population to probe the more pristine, outer regions of galaxies and their surroundings. Somewhat surprisingly, then, sub-DLAs are more metal-rich than DLAs, particularly at low redshift $(z < 1.5)$ \citep{2010NewA...15..735K}. The metallicities of DLAs and sub-DLAs decrease with increasing $\nhi$ \citep{2007A&A...464..487K}. Given the mass-metallicity relation, this plausibly implies that sub-DLAs are associated with higher-mass systems. One must keep in mind that the properties of DLAs and sub-DLAs vary both with the properties of its host system and the properties of the sightline through said system.

Recently, \citet{2011MNRAS.412.1047C,2011MNRAS.417.1534C} have reported observations of extremely metal-poor DLAs as probes of near-pristine clouds of star-formation fuelling gas. The abundance patterns of these systems are consistent with predictions of Population III supernovae, suggesting that these DLAs retain the signature of the earliest episodes of star-formation. Intriguingly, some of these systems also show an enhancement of carbon relative to other metals, reminiscent of carbon-enhanced-metal-poor (CEMP) stars --- a population of galactic halo stars whose abundances suggest that they were born from gas enriched by the first stars \citep{2005ARA&A..43..531B,2006ApJ...652L..37L}.

\subsection{The Search for DLA Host Galaxies}

Searches for the galaxies that are responsible for damped absorption in QSO spectra have a long history. The most common technique is to search adjacent to quasar sightlines with known absorption systems \citep{1999MNRAS.305..849F,1999MNRAS.309..875B,2000ApJ...536...36K, 2001ApJ...551...37K,2001MNRAS.326..759W,2007AandA...468..587C}. This is a difficult task, as the light of the extremely bright quasar must be accurately subtracted in order to study the light from the galaxy, which is very faint in comparison. \citet{2000ApJ...536...36K} and others caution that a given emission feature could be a Point Source Function (PSF) artefact rather than a real source. For many years, this search resulted in mostly non-detections \citep{1995ApJ...451..484L,1999MNRAS.309..875B,2000ApJ...536...36K,
2001ApJ...551...37K,2006ApJ...636...30K,2009AandA...505.1007C}.  \citet{2001MNRAS.326..759W} searched around 23 high-redshift, high-column density \lya absorbers with NICMOS, finding 41 candidates.  \citet{2007AandA...468..587C} report that, for $z > 2$, six DLA galaxies have been confirmed through spectroscopic observation of \lya emission, with other techniques producing a few additional candidates. \citeauthor{2007AandA...468..587C} added another six \lya emission candidates to this group. A few DLA counterparts have been discovered by searching in the damped \lya trough for \lya emission, beginning with \citet{1990ApJ...356...23H}, although that particular system failed to be detected by \citet{1992ApJ...385..151W} and \citet{1995ApJ...451..484L}, only to reappear in \citet{1997ApJ...486..665P}. Further \lya detections were reported by \citet{1999MNRAS.305..849F,2002ApJ...574...51M,2004AandA...422L..33M}. 

Recent success has followed from improved methods. For example, \citet{2010MNRAS.408..362F} use higher-redshift absorbers to do the QSO subtraction for lower redshift systems, detecting one DLA in rest-frame
FUV continuum emission. \citet{2010MNRAS.408.2128F,2011MNRAS.413.2481F}, noting theoretical predictions and tentative observational evidence of a luminosity-metallicity relation in DLAs, target high-metallicity DLAs and detect two systems in emission. \citet{2011MNRAS.410.2237P,2011MNRAS.410.2251P,2012MNRAS.419.3060P} used SINFONI integral field spectroscopy to detect faint DLA and LLS galaxies near bright quasars. They detect 5 galaxies (3 DLAs, 2 LLSs) in H$\alpha$ emission, 4 at $z \lesssim 1$, and the other at $z = 2.35$. \citet{2012AandA...540A..63N} and \citet{2013MNRAS.433.3091K} detect \lya, [O\textsc{iii}], and H$\alpha$ emission from DLAs at $z \sim 2.2$. (We will discuss these systems further in Section \ref{Ss:LyaCGM}.) \citet{2010MNRAS.409L..59R} stack the spectra from 341 DLAs observed in the Sloan Digital Sky Survey at $\langle z \rangle = 2.86$, searching for \lya emission from the DLA host. They report a non-detection of emission at line centre, which ignores the possible effects of \lya radiative transfer. \citet{2011MNRAS.412L..55R}, by relaxing this assumption, report a $2.7 \sigma$ detection. DLA host searches at lower redshift have had more success: \citet{2011MNRAS.416.1215R}  use photometric redshifts and colours to detect 27 DLAs (selected as Mg\textsc{ii} absorbers) in emission at $0.1 \leq z \leq 1.0$.

With relatively few DLA host galaxies known at $z \gtrsim 2$,  conclusions drawn about DLAs in emission must be tentative. \citet{2012MNRAS.424L...1K} use a sample of 10 DLAs in emission to argue for two correlations. Firstly, the DLA impact parameter (the distance between the centre of the host and the QSO line of sight) decreases with increasing \hi column density. This reflects the decrease in column density with distance from the centre of the host galaxy, as predicted by simulations \citep{2008MNRAS.390.1349P,2012arXiv1212.2965H}. Second, DLA metallicity increases with impact parameter, which is interpreted as a corollary of the correlation between DLA cross section and mass, and between mass and metallicity, though selection biases make comparison with observations difficult.

\citet{2014MNRAS.438..529R} provide a table of 13 confirmed $z \sim 2 - 3$ DLA-galaxy pairs from the literature. Proper impact parameter varies from $\sim 1 - 25$kpc, with a median of 8kpc. Noting that observations are typically only able to detect galaxies with SFR $\gtrsim 1 - 10 \Msol$ yr$^{-1}$, the observed star formation rates vary from $\sim 3 - 70\Msol$ yr$^{-1}$, with a median of $\sim 20 \Msol$ yr$^{-1}$. Note that the SFR's are often based on \lya emission, which is difficult to correct for dust extinction. The associated \lya luminosities (from Equation \ref{eq:LlyaSFR}, Section \ref{Ss:LAEs}) range from $3 - 80 \times 10^{42}$ erg s$^{-1}$. The two galaxies with measured stellar masses have $M_* = 2 - 12.6 \times 10^9 \Msol$. The simulations of \citet{2014MNRAS.438..529R}, which incorporate the physics of radiative transfer of the UVB and recombination radiation, show good agreement with these observations, though the problem of small number statistics is exacerbated by a lack of information about non-detections and detection limits (in both luminosity and impact parameter). Such simulations are discussed further in Section \ref{Ss:DLAtheory}.

\subsection{DLAs in Gamma-Ray Burst Afterglows}

Seeing \hi in absorption at cosmological distances requires an extremely bright background source. Quasars are very useful to this end, as we have seen. DLAs can also be seen in the afterglow of Gamma-Ray Bursts (GRBs). Importantly, the absorption seen in GRB-DLAs is intrinsic rather than intervening, that is, the absorbing gas is inside the GRB host galaxy. Thus, GRB-DLAs provide detailed information on the kinematics, chemical abundances, and physical state of the gaseous component of their host galaxy \citep{2001ASPC..251..168F,2001A&A...373..796F,2003ApJ...585..638S,2004A&A...419..927V,2004ApJ...611..200P,2004A&A...427..785J,2006A&A...460L..13J,2006A&A...451L..47F,2006ApJ...648...95P,2006ApJ...652.1011W}. Here, we will briefly review the properties of GRB-DLAs, highlighting clues as to their relationship to QSO-DLAs and other high-redshift galaxies.

\begin{figure}[t]
\centering
\includegraphics[width=0.46\textwidth]{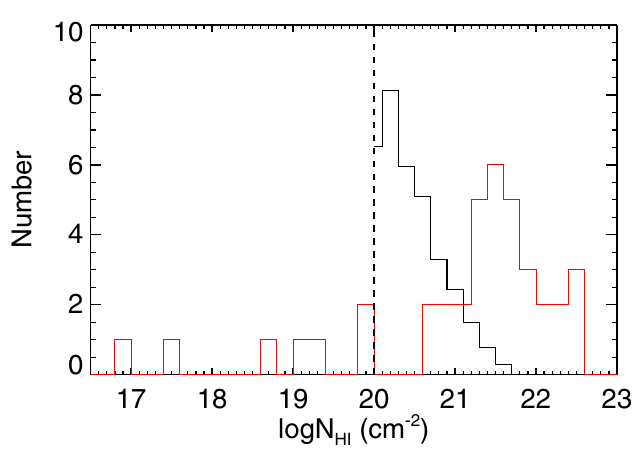}
	\caption{A comparison of the \hi column density distributions of QSO-DLAs (black) and GRB-DLAs (red). The numbers of QSO-DLAs have been renormalized to the same number of objects as the GRB absorbers. Figure from \citet{2009ApJS..185..526F}; reproduced by permission of the AAS.} \label{fig:GRBDLA_Fynbo}
\end{figure}

\begin{itemize}  \setlength{\itemsep}{-2pt}
\item GRB-DLAs have a dramatically different column density distribution to QSO-DLAs, as shown in Figure \ref{fig:GRBDLA_Fynbo}, from \citet{2009ApJS..185..526F}. While QSO-DLA column density distribution is a monotonically decreasing function of $\nhi$, the GRB-DLA distribution displays a prominent peak at $\nhi = 10^{21.5}\cm$. Only 1\% of QSO-DLAs display such high columns \citep{2010MNRAS.402.1523P}.
\item At $z > 2$, GRB-DLAs display a range of metallicities, from $1/100$ solar \citep{2010ApJ...720..862R} to super-solar \citep{2012MNRAS.420..627S}. Unlike QSO-DLAs, there is no clear trend of mean metallicity with redshift. \citet{2012MNRAS.420..627S} has shown that the mean GRB-DLA metallicity for $z = 2-4.5$ is somewhat higher ($\log(Z/Z_\odot) = -0.83 \pm 0.76$) compared to QSO-DLAs ($\log (Z/Z_\odot) = -1.39 \pm 0.61$) \citep[see also][]{2006NJPh....8..195S,2006A&A...451L..47F,2008ApJ...683..321F,2007ApJ...666..267P}. Metal absorption features in GRB-DLAs tend to be, on average, 2.5 times stronger and slightly more ionized that those of QSO-DLAs \citep{2012A&A...548A..11D}.
\item The velocity width ($v_w$) distribution of GRB-DLAs is consistent with that of QSO-DLAs, with a median of $\approx 80\kmsec$ and a high-velocity tail out to $200-300 \kmsec$ \citep{2008ApJ...672...59P}, though there are only 9 GRBs in the GRB-DLA $v_w$ sample. This suggests that the two populations inhabit parent haloes of comparable mass.
\item There is some evidence that GRB-DLAs are not significantly statistically distinct from the high-column density tail of QSO-DLAs. GRB-DLAs seem to smoothly extend the dust properties and metallicity of QSO-DLAs to larger column densities \citep{2009ApJS..185..526F,2012AJ....143..147G,2013arXiv1305.1153D}.
\end{itemize}

Given that GRB-DLAs are plausibly selected by their intense star formation, as opposed to the \hi cross-section selection of QSO-DLAs \citep{2008ApJ...683..321F}, \citet{2010MNRAS.402.1523P} modelled GRB sight-lines in a cosmological simulation by associating them with young star particles. The result was a successful prediction of the GRB-DLA column density distribution, and broad agreement with the range of metallicities. In the simulations, GRB-DLAs occupy dark matter host haloes that are an order of magnitude larger than their QSO counterparts, and probe regions typically four times closer to the centre of their host haloes. This is in agreement with the schematic picture of GRB-DLAs as resulting from sight lines that probe the denser, central regions of their host galaxies, as opposed to intervening absorbers where most of the cross-sectional area is in the outskirts of the galaxy \citep{2007ApJ...666..267P}. This picture also explains why GRB-DLA metallicities are systematically higher than the QSO-DLA metallicities \citep{2008ApJ...683..321F}. By simultaneously probing star-formation and ISM gas physics of galaxies, GRB-DLAs provide a much-needed test of galaxy formation simulations.

\subsection{Theoretical models of DLAs} \label{Ss:DLAtheory}

\subsubsection{Early Models of DLAs: From Galaxy Disks to $\Lambda$CDM}
In the local Universe, most \hi atoms are found in the disks of $L_*$-type galaxies. Thus, DLAs have traditionally been considered to be high-redshift galactic disks. This motivated \citet{1986RSPTA.320..503W} to search for the disks of such galaxies at high redshift by looking for high column density \hi in absorption in the spectra of quasars.  A variety of models have been proposed, as discussed in this section.

\citet{1972ApJ...172..553A} was the first to suggest that absorption lines in high-redshift QSO spectra could be \lya absorption from \hi in protogalaxies, though in those models the gas is highly ionized and thus not directly relevant to DLAs. \citet{york86} suggested that some DLAs could be associated with gas-rich dwarf galaxies, which would explain the complexity of the metal line profiles. \citet{1990ApJ...365..439S} used hydrodynamic simulations of the collapse of a gaseous corona into a ``\lya disk'' via radiative cooling to test the idea that DLAs originate in large, massive disks of gas that are the progenitors of present-day galaxies. They conclude that such a scenario is plausible, provided there is sufficient metallicity to allow for rapid cooling. \citet{1996ApJS..107..475L}, by considering elemental abundances, found that DLAs are much less metal-enriched than the Galactic disk in its past. They concluded that DLAs are not high-redshift spiral disks in the traditional sense, postulating thick disks or spheroidal components of (dwarf) galaxies as more likely scenarios.

Within the CDM model of cosmic structure formation, \citet{1994ApJ...430L..25M} modelled DLAs as gaseous disks within dark matter haloes. Such models were refined and extended by \citet{1996MNRAS.281..475K}, who based a disk-formation model on the paradigm of \citet{1978MNRAS.183..341W}: galaxies form by the continuous cooling and accretion of gas within a merging hierarchy of dark matter haloes. The model also incorporates star formation, with the gas supply regulated by infall from the surrounding halo. Chemical enrichment occurs through the ejection of metals back into the hot IGM by supernovae; this gas then cools back onto the disk. This model predicts, with reasonable success, the redshift dependence of $\Omega_{\hi}$ as well as $f(\nhi,X)$. DLAs are predicted to be smaller, more compact, and less luminous than today's galaxies.

\citet{1998MNRAS.295..319M} placed rotationally-supported disks within haloes with an NFW density profile \citep{1996ApJ...462..563N}, allowing the spin parameter $\lambda$ to vary over a lognormal distribution. These models successfully reproduce $\dd \cN / \dd z$ at $z=2.5$ by including the contribution of disks with rotation velocities down to 50-100 \kmsec. An important unknown in this and later models is the smallest halo that can host a DLA. Higher resolution simulations of smaller cosmological volumes by \citet{1996MNRAS.278L..49Q} indicated that haloes as small as 35 \kmsec can host DLAs.

\citet{1997ApJ...484...31G} were among the first to study DLAs using numerical simulations of cosmological structure formation. While extending the simulations of \citet{1996ApJS..105...19K,1996ApJ...457L..57K}, their simulations could not resolve dark matter haloes below 100 \kmsec. They used the \ps formalism to extrapolate the results of their simulation to smaller circular velocities. The results for $\dd N / \dd z$ are in good agreement with observations for $z = 2-4$ in an $\Omega_m = 1$ universe. \citet{1997ApJ...486...42G}, however, applied this method to other cosmological models, showing that in a $\Lambda$CDM universe (with $\sigma_8 = 0.79$), absorbers are underproduced at $2 \lesssim z \lesssim 3$ by a factor of 3. Further simulations \citep{2001ApJ...559..131G} produced a more adequate fit to the data, though precise predictions were affected by the uncertainty in determining the smallest halo capable of hosting a DLA.

\citet{1997ApJ...487...73P} showed that kinematical data present a considerable challenge to theoretical models of DLAs. They traced sightlines through disk galaxies, their favoured model being a single thick, cold disk ($h = 0.4 R_d$, $\sigma_{cc} = 10 \kmsec \ll v_\ro{rot}$, $v_\ro{rot} = 250$ \kmsec, where $\sigma_{cc}$ is the isotropic velocity dispersion of absorbing clouds within the disk). A CDM-inspired model faired much worse. In this model, $v_\ro{rot}$ for the thick disk is chosen from the distribution $P(v_\ro{rot})$ calculated by \citet{1996MNRAS.281..475K} for a CDM cosmology. This model fails to reproduce the velocity width distribution because of the predominance of small haloes with slowly rotating disks. The simulated profiles are also significantly more symmetric than is observed.

\citet{1998ApJ...495..647H} challenged the assumption that the velocity widths of high-redshift galaxies are due solely to rotational motion in disks. In hierarchical structure formation, galaxies are built up from the merging of protogalactic clumps, often moving along filaments. \citet{1998ApJ...495..647H} thus considered absorption arising in more realistic, irregular protogalactic clumps, whose velocity field is a mixture of rotation, infall, merging, and random motion \citep[Similarly complex velocity fields are seen in recent IFS observations of DLA galaxies, e.g.][]{bouche13,2013MNRAS.436.2650P,2014ApJ...785...16J}. Their simulations, which resolved the substructure of the clumps, produced a population of DLAs that was able to reproduce the kinematic data of \citet{1997ApJ...487...73P}.

%\citet{1999ApJ...519..486M} considered an analytical model that resembled the scenario of \citet{1998ApJ...495..647H}. They considered randomly moving clouds within a spherical halo formed in the context of a CDM universe, and found that such a model was able to reproduce the kinematic data of \citet{1997ApJ...487...73P}. They also identified the issue of energy dissipation: randomly moving clouds will collide often, producing shocks in the absorbing gas. These shocks seem to dissipate energy at a much higher rate than it can be supplied from the gravitational energy of merging haloes. This could suggest another energy source capable of maintaining gas motions in DLAs.

\subsubsection{The CGM in Cosmological Simulations} \label{S:DLA_CGM}

In light of the realisation that DLAs cannot be explained within a $\Lambda$CDM cosmology as isolated systems, the next generation of cosmological simulations sought to place DLAs within their full cosmological context, and in particular within the immediate surroundings of galaxies: the circumgalactic medium.  The CGM is roughly defined as the 100--300 kpc region around a galaxy, distinct from the stellar system but within the virial radius of its halo. Before we discuss the particular place of DLAs in cosmological simulations, we will here give a brief overview of the CGM in cosmological simulations. This is particularly important as recent theoretical studies have markedly changed our picture of how galaxies get their gas.

The cooling of gas in dark matter haloes has long been recognised as a crucial ingredient in galaxy formation \citep{1977ApJ...215..483B,1977ApJ...211..638S,1978MNRAS.183..341W}. In the classic picture of White and Rees, gas that falls into a dark matter halo is shock-heated at the virial radius to the virial temperature ($\sim 10^6$ K). The gas then accretes quasi-spherically when cooling (primarily via bremsstrahlung) removes the pressure-support of the gas against gravity. This process is inefficient above a halo mass of $5 \times 10^{12} \Msol$ \citep{2009MNRAS.395..160K}, because of low densities and long associated cooling times. This mode of accretion is known as ``hot mode'' accretion, to contrast it with another mode of accretion identified by \citet{2003MNRAS.345..349B} and \citet{2005MNRAS.363....2K}. In ``cold mode'' accretion, gas that has never been shock heated ($T \ll T_\ro{vir}$) radiates its potential energy and falls onto the central galaxy along thin filaments. This mode of accretion dominates in low-mass haloes ($M_\textrm{halo} \lesssim 2 -3 \times 10^{11} \Msol$), and is therefore the main accretion mechanism at high redshift. Both modes can coexist in massive haloes at high redshift, where cold, dense filaments are able to penetrate hot virialized haloes \citep{dekel06,2008MNRAS.390.1326O,dekel09,2013arXiv1308.1669F}\footnote{While the conclusions of \citet{2005MNRAS.363....2K} have been reproduced in higher resolution SPH simulations \citep{2009ApJ...694..396B,2009MNRAS.395..160K} and AMR simulations \citep{2008MNRAS.390.1326O,2009MNRAS.397L..64A,dekel09}, they have recently been challenged by the moving mesh simulations of the AREPO code \citep{2013MNRAS.429.3353N}. While there is still a low vs. high mass dichotomy for cold vs. hot accretion, AREPO concludes that more massive haloes $(M_\ro{halo} \gtrsim 10^{10.5} \Msol)$ have a much smaller cold fraction than is seen in SPH simulations. This is due to both an order of magnitude higher hot accretion rate, and a factor-of-two lower cold accretion rate. For cold streams, the AREPO flows are disrupted in massive haloes at 0.25 - 0.5 $r_\ro{vir}$. \citet{2013MNRAS.429.3353N} attribute the difference to numerical blobs of cold gas in SPH simulations. The situation, and the implications for the observable properties of gas accretion, remains unclear.}.

The gas in cold streams penetrates deep inside the virial radius \citep{dekel09}. The interaction of cold streams with hot halo gas as it settles into a rotating disks has been used to model a population of large, clumpy, rapidly star-forming disk galaxies at high redshift \citep{2009ApJ...703..785D,2009MNRAS.397L..64A}. The newly-accreted gas will fragment to form large ($10^7 - 10^9 \Msol$) star-forming clumps in an extended  $\sim 10$ kpc disc. While the inner parts of the disc have on average solar metallicity, the clump-forming region is only 0.1 $Z_\odot$. \citet{2009ApJ...694..396B} note that star formation in stellar disks of galaxies up to Milky Way mass is primarily fuelled at all times by gas that has accreted cold; such gas can dominate the supply of the stellar disk even in high mass haloes where most of the gas flowing through the virial radius is hot. The dominant gas supply at all masses is from smoothly accreted gas that has never belonged to another galaxy halo.

Inflowing gas is only half the story, however. Galactic winds, powered by some combination of stellar winds, supernovae, cosmic ray heating and AGN, are ubiquitous in both the local and distant universe \citep[see][for a comprehensive review]{2005ARA&A..43..769V}.  As \citet{2010ApJ...717..289S} notes, virtually every $z > 2$ galaxy bright enough to be observed spectroscopically is driving out material at velocities of at least several hundred \kmsec.  Such winds are the primary mechanism by which star-formation feeds back upon itself and energy and metals are circulated within galaxies and their environment. This feedback is widely believed to explain why stellar mass fails to follow halo mass in a $\Lambda$CDM cosmology, especially at low masses \citep[$M_\ro{halo} \lesssim 10^{11} \Msol$; see][]{2012RAA....12..917S}. Hydrodynamical simulations that fail to model feedback inevitably form galaxies with too many baryons.

Simulations of galactic winds, particularly those in a cosmological context, must rely on unresolved subgrid physics to approximate the combined effect of stellar and SN feedback. A range of prescriptions are available \citep[e.g.][]{2003MNRAS.339..289S,2006MNRAS.373.1265O,2008MNRAS.387..577O,2007MNRAS.374.1479G,2011MNRAS.417..950H,2012MNRAS.421.3522H,2012arXiv1210.3582B}; all are poorly constrained by observations and `first principles' simulations. The simulations of \citet{2012arXiv1205.0270S} show inflowing and outflowing gas coexisting in the CGM: about one third of all the gas within the virial radius is outflowing. This outflowing gas is enriched with the products of star-formation, so that at the virial radius inflowing gas has metallicity $0.05 Z_\odot$, while the mean metallicity of outflowing gas is $0.56 Z_\odot$. Outflows tend to be bipolar, taking the path of least resistance perpendicular to the plane of the disk, in agreement with observations at both low \citep[e.g. M82]{1963ApJ...137.1005L,1988Natur.334...43B,2002PASJ...54..891O} and high \citep{2011ApJ...743...10B,bordoloi12,bouche12} redshift. The launching of a large-scale galactic wind is not guaranteed, as the energy input may only pressurise the ISM \citep{2009MNRAS.395..160K}. \citet{2010MNRAS.406.2325O} find that the mean recycling time decreases with halo mass, meaning that recycled wind material is an important mode of accretion (distinct from cold and hot modes) in high mass haloes. In particular, in their favoured momentum-driven wind model, gas accreted via the fountain mode fuels over 50 per cent of the global star formation density at $z \lesssim 1.7$.

\subsubsection{DLAs in Cosmological Simulations}

Returning to DLAs within cosmological simulations, \citet{2004MNRAS.348..421N} used Smoothed-Particle Hydrodynamics (SPH) simulations in the context of the $\Lambda$CDM model. Their simulations included the effects of radiative cooling, the Ultra Violet Background (UVB), star formation, supernovae feedback, and in particular considered the effect of galactic winds using a simple phenomenological model that involves giving gas particles a ``kick'' in a random direction to drive them out of dense star-forming regions. They used the \ps formalism to extend their results below the resolution limit of their simulations. The result was a reasonable agreement with $\Omega_{\hi}(z)$ so long as ``strong'' winds were invoked; otherwise, there was too much gas left in the DLAs. Their prediction for $\dd N/\dd z (z)$ was also reasonably successful. Finally, $f(\nhi,X)$ is slightly underpredicted, with strong winds necessary to prevent overprediction at high-$N$. \citet{2007ApJ...660..945N} refined the previous models with a more careful consideration of winds, concluding that DLAs are hosted by small ($M_\ro{halo} < 10^{12} h^{-1} \Msol$), faint galaxies.

\begin{figure*}[t]
\centering
\begin{minipage}{0.6\textwidth}
\includegraphics[width=\textwidth]{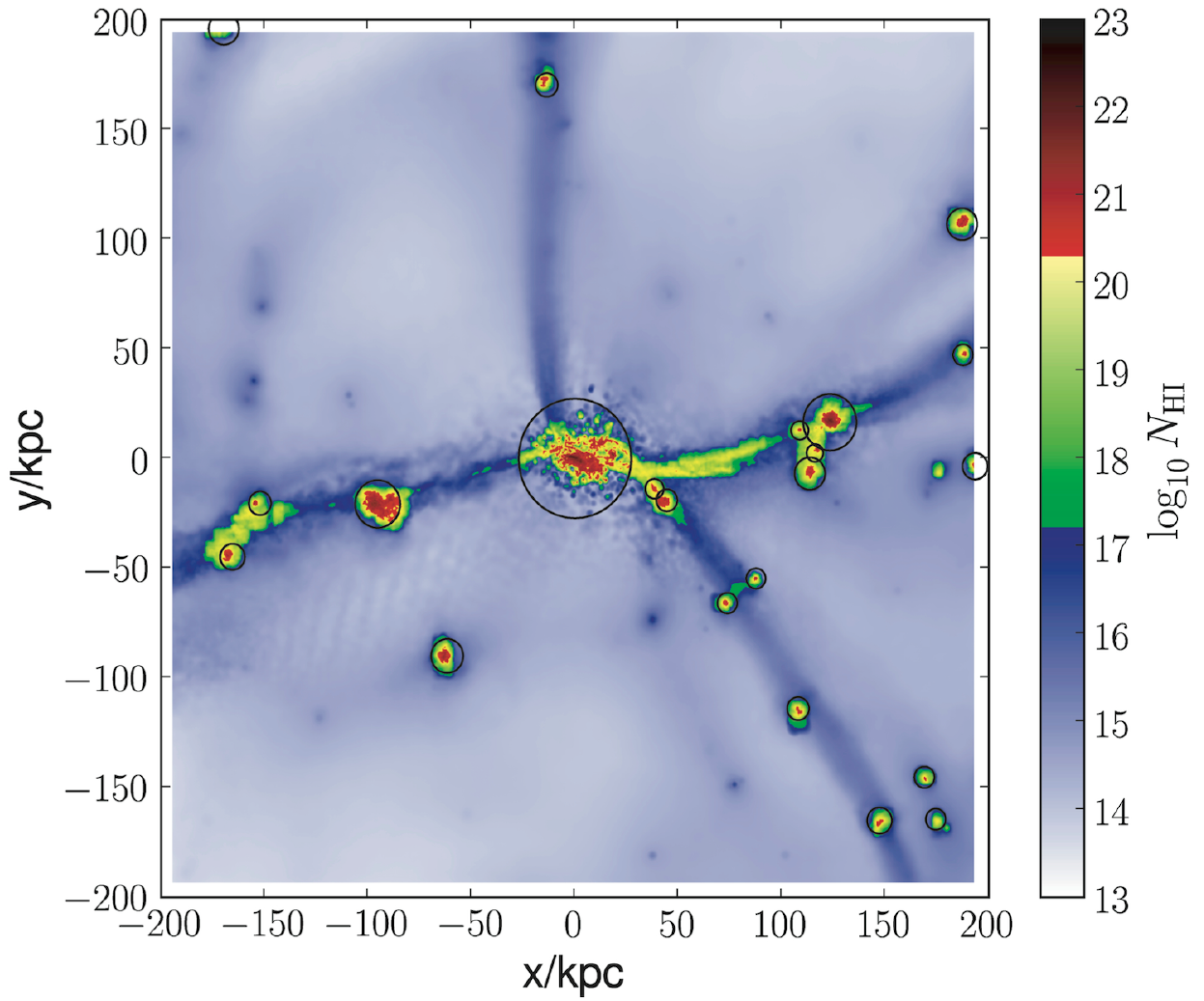}
\end{minipage}
\hspace{0.4cm}
\begin{minipage}{0.35\textwidth}
	\caption{The $z=3$ neutral column density of \hi in a 400kpc cube centred on the major progenitor to a $z=0$ Milky Way type galaxy in the simulations of \citet{2008MNRAS.390.1349P}.The colours are such that DLAs ($\nhi > 10^{20.3} \cm$) appear in dark red and Lyman limit systems ($10^{20.3} \cm > \nhi > 10^{17.2} \cm$) appear in green and yellow. The circles indicate the projected positions and virial radii of all dark-matter haloes with $M_\ro{halo} > 5 \times 10^8 \Msol$. All units are physical. In these and later simulations, almost all DLAs arise inside dark matter haloes, probing the ISM and CGM on typical scales of $\sim$10\% of the virial radius \citep{2013arXiv1310.3317R}. Figure from \citet[][Figure 2]{2008MNRAS.390.1349P}; used with permission.} \label{fig:Pontzen_DLA}
\end{minipage}
\end{figure*}

Further cosmological simulations have been able to model the DLA population without using analytic extensions based on the \ps formalism. The first to accomplish this was \citet{2006ApJ...645...55R}, who used AMR cosmological simulations to address both the neutral gas cross-section and the gas kinematics. Their results show that $f(\nhi,X)$ is overpredicted for $\nhi > 10^{21} \cm$, which may be due to the effects of grid resolution or the absence of a model for the formation of H$_2$, which will affect the highest density systems. The velocity width distribution is dramatically underpredicted at high-$v_w$, even when star-formation is taken into account \citep{2008ApJ...683..149R}. Another surprising conclusion of these simulations was the abundance of DLAs that are not associated with any halo --- intergalactic DLAs were found in tidal tails and quasi-filamentary structures.

\citet{2008MNRAS.390.1349P} analysed DLAs in the galaxy formation simulations of \citet{2007MNRAS.374.1479G,2008ASPC..396..453G} and \citet{2007ApJ...655L..17B}. These simulations produce impressively realistic disk galaxies at $z = 0$. It is thus a major success that the simulations, with no further tweaking of free parameters, are able also to match $f(\nhi,X)$ at $z = 3$, apart from a slight overprediction at high-\nhi. Good agreement is also found with the distribution of metallicities in DLAs, something previous simulations were unable to do. However, as before, DLA velocity width data is not reproduced, with too few high-$v_w$ systems. \citet{2009MNRAS.397..411T} followed a familiar pattern ---  success with $f(\nhi,X)$ but too few large $v_w$ systems, possibly due to inadequate mixing of wind-phase metals.

\citet{2010arXiv1008.4242H} mimicked metal diffusion by tying metals directly to \hi, and had somewhat more success in producing large $v_w$ systems. \citet{2012ApJ...748..121C} used an adaptive mesh-refinement code to study DLAs in two small cosmological volumes (for resolution reasons), one centred on a $1.8 \sigma$ overdensity and one on a $1 \sigma$ underdensity. These simulations bracket the observed column density and velocity width distributions. The OverWhelmingly Large Simulations (OWLS) project has been used to study QSO absorption systems, often in conjunction with post-processing to account for H$_2$ formation and ionizing radiation. The simulations have studied the observed $z = 3$ abundance of \lya forest, Lyman limit, and DLA absorption systems probed by quasar sight lines over 10 orders of magnitude in column density \citep{2011ApJ...737L..37A}, the lack of evolution in $f(\nhi, X)$ below $z = 3$ \citep{2012MNRAS.427.2889Y}, and the effect of subgrid feedback prescriptions \citep{2013MNRAS.tmp.2469A} and ionizing radiation from the UV background and local stellar sources \citep{rahmati13} on the DLA population. There is some tension between \citet{rahmati13}, who conclude that their predictions are highly sensitive to assumptions about the ISM, especially for $N_{\hi} > 10^{21}\cm$, and \citet{2012MNRAS.427.2889Y}, who conclude that ionizing radiation from stars makes little difference to DLAs as it is absorbed in very dense, compact clouds that contribute little to the DLA cross section.

\citet{2013arXiv1308.2598B} studied DLAs using semi-analytical models of galaxy formation. They found that  they needed to ``extend'' their gas disks by giving them more than their share of the halo angular momentum in order to reproduce the observed properties of DLAs at $z < 3$, while all models failed at $z > 3$.

An important constraint on models of DLAs comes from the use of cross-correlation to study host halo masses. For example, \citet{2001ApJ...562..628G,2003ApJ...584...45A,2004ApJ...609..513B,2006ApJ...636L...9C} measured the cross-correlation of DLAs with Lyman Break Galaxies. These observations, together with the modelling of \citet{2011MNRAS.411...54L}, suggest that DLAs inhabit smaller haloes than LBGs, with average halo masses of DLAs of $10^{11} \Msol$ compared to $10^{11.5-12}\Msol$ for LBGs.

More recently, however, \citet{2012JCAP...11..059F} calculated the cross-correlation of DLAs in the BOSS survey with the \lya forest, concluding that the derived DLA bias indicated a significantly larger typical halo mass of $\sim 10^{12} \Msol$. This conclusion was supported by \citet{2014MNRAS.440.2313B}. They used a simple analytical model of DLAs, which reproduces the column density and velocity width distribution of DLAs, to argue that the observed DLA bias presents a significant challenge to existing DLA models and galaxy formation simulations more generally, plausibly indicating that photoionisation and stellar feedback are more efficient in ionizing and/or ejecting \hi from 50 - 90 \kmsec haloes than is accounted for in current subgrid models.

The physical picture of DLAs that emerges from these simulations can be summarised as follows.
\begin{itemize}  \setlength{\itemsep}{-2pt}
\item DLAs arise in the ISM and CGM of low-mass, gas-rich galaxies. Most absorbers with $\nhi > 10^{17} \cm$ at $z = 3$ reside inside galaxy haloes, and either have been ejected from the ISM or will become part of the ISM by $z = 2$ \citep{2012MNRAS.421.2809V}. For DLAs, ISM absorption is most likely for systems with $\nhi > 10^{21} \cm$, with a significant fraction of DLAs not arising from the gaseous disks of spiral galaxies but rather from filaments, streams and clumps \citep{2011MNRAS.418.1796F}. An example is shown in Figure \ref{fig:Pontzen_DLA}. DLAs typically probe galaxies on $\sim$kpc scales, or the inner $\sim$10\% of the virial radius \citep{2013arXiv1310.3317R}.
\item DLAs arise in dark matter haloes in the mass range $10^{10} - 10^{12} \Msol$, which have very low stellar masses $< 10^8\Msol$ \citep{2013arXiv1310.3317R}. DLAs host haloes are generally smaller than the host haloes of LBGs. In this mass range, gas accretion is predicted to be via cold streams \citep{2003MNRAS.345..349B,2005MNRAS.363....2K}, that is, gas which enters dark matter haloes without being shock heated above $\sim 10^6$ K. \Citet{2012MNRAS.421.2809V} show that most of the gas in LLSs and DLAs has never been shock-heated, though the smooth component of cold streams is highly ionized \citep{2011MNRAS.418.1796F}. LLSs and DLAs at high redshift thus plausibly include the elusive cold streams of cosmological simulations.
\item DLA kinematics, derived from associated metal line absorbers, remain a significant hurdle for simulations. To reproduce the high-$v_w$ systems, significant ``non-gravitational'' sources of motion are required, that is, velocity components beyond that of a purely virialised halo of gas. Plausible sources include the motion of satellite haloes inside a larger parent halo, and supernovae-driven galactic winds.
\item The relatively low metallicity and low internal star-formation rates of DLA gas are explained by the fact that, at high redshift, they arise in gas which is on the outskirts of or outside of the ISM of small galaxies.
\end{itemize}
The typical limitations of cosmological hydrodynamical simulations --- resolution, box size, sub-grid physics --- all apply here. In particular, the importance of stellar and supernovae feedback to the properties of the ISM and CGM, and hence DLAs, makes theoretical predictions dependent on the details of processes that no simulation can yet resolve. The ability of Lyman alpha absorption to probe ISM and CGM gas at high redshift makes it one of the most important tests of galaxy formation simulations.

% % % % % % % % % % % % % % % % %% % % % % % % % % % % % % % % % %
% % % % % % % % % % % % % % % % %% % % % % % % % % % % % % % % % %
% % % % % % % % % % % % % % % % %% % % % % % % % % % % % % % % % %

\section{MgII Absorption Tracing HI} \label{S:MgII}

% comment on redshift difference of HI and MgII

The {\MgIIdblt} doublet probes a large range of neutral hydrogen
column densities, $10^{16}$ $\lesssim \hbox{N(\HI)} \lesssim$
$10^{22}$~{\cmsq} \citep{archiveI,weakII}, having typical temperatures
of 30,000 $-$ 40,000~K and average total hydrogen densities of
$\sim0.1$~atoms~cm$^{-3}$ \citep{cvc03,ding05}. Such conditions yield
a large range of {\MgII} rest-frame equivalent widths, $0.02$ $\lesssim
W_r(2796)\lesssim$ $10$~\AA, that are detectable in the spectra of
background quasars and galaxies probing intervening foreground
objects. It has been convincingly demonstrated that {\MgII} absorption
is produced within gaseous haloes surrounding galaxies and is not
produced within the diffuse intergalactic medium
\citep[see review by][]{cwc-china}.  

A significant quantity of {\HI} is probed by {\MgII} absorption,
comparable to roughly 15\% of the gas residing in DLAs and 5\% of the
total hydrogen in stars \citep{kacprzak11c,menard12}.  Roughly 20\% [50\%] of
all {\MgII} absorption systems with 0.6$\leq W_r(2796) \leq$1.7~{\AA}
[$W_r(2796)>1.7$~{\AA}] are associated with DLAs.  Furthermore,
\citet{menard09b} demonstrated a direct relation between the {\MgII}
rest equivalent width and the {\HI} column density such that the
geometric mean column density is $N({\HI})= A W_r(2796)^{\beta}$,
where $A=(3.06 \pm 0.55) \times 10^{19}$ cm$^{-2}$~{\AA}$^{-\beta}$
and $\beta=1.73\pm 0.26$ and is valid for $0.5 \leq W_r(2796) \leq
3$~{\AA} and $0.5 \leq z \leq 1.4$.

\subsection{MgII Absorption Statistics}

Current surveys place the number of known {\MgII} absorbers at around
$\sim$41,000 systems spanning a redshift range of $0.1<z<2.3$
\citep[e.g.,][]{sargent88,steidel92,weakI,narayanan07,barton09,chen10a,quider11,werk12,zhu13,seyffert13}
and $\sim$100 {\MgII} systems spanning a redshift range of $1.9<z<6.3$
\citep{matejek12}. The distribution of {\MgII} rest-frame equivalent
widths consists of two distinct populations: Strong systems,
$W_r(2796)>0.3$~{\AA}, are well described by an exponential
distribution \citep{nestor05,zhu13,seyffert13}, while weaker systems
follow a power-law distribution \citep{weakI,narayanan07}.  These two
distinct populations can be simultaneously described with a Schechter
function \citep{kacprzak11b}.

The incidence rate $dN/dz$ of absorption systems (number per unit
redshift and rest equivalent width) provides detailed information on
their cross-section and number density as a function of redshift.
\citet{seyffert13} found that the physical cross-section of {\MgII}
absorbers increases by a factor of three between $0.4<z<2.3$ for
$W_r(2796) \geq 0.8$~{\AA}.  Over the redshift range $0.4<z<5.5$, the
rate of incidence of weaker absorbers ($0.6< W_r(2796)< 1.0${\AA}) is
roughly constant, which suggests that these absorbers are established
early and are being constantly replenished as a function of time
\citep{zhu13,matejek12}.  Meanwhile, the incidence rate for stronger
absorbers ($W_r(2796)$$>$ 1.0~{\AA}) increases towards $z\sim2$ and
then decreases towards $z\sim5$, similarly to the cosmic star
formation history \citep[e.g.,][]{bouwens11}.  The strikingly similar
shapes of these two quantities points to a direct connection between
strong absorbers and star formation \citep{zhu13,matejek12}.

\subsection{Association with Galaxies}

The pioneering work of \citet{bergeron88} and \citet{bb91} identified
the first galaxies in close proximity to a quasar sight-line and at
the same redshift as absorbing gas traced by {\MgII} absorption.  Of
the $\sim41,000$ known {\MgII} absorbers, there are $\sim 300$
galaxies that are spectroscopically confirmed to reside at same
redshift as the absorber \citep[e.g.,][]{nielsen12,werk12}.  These
galaxies span a redshift range of $z=0.1$ \citep{barton09,kacprzak11a}
to $z=2$ \citep{bouche12b,lundgren12}. The small galaxy sample sizes
reflect only the difficulty of obtaining large amounts of telescope
time required to complete these spectroscopic surveys around
quasars. In addition, stacking techniques have been used to probe
1000's of galaxy haloes \citep{zibetti07,bordoloi11,menard09}.

Absorption by {\MgII} around galaxies extends to projected
galactocentric distances $(D)$ of $\sim$200~kpc. Host halo size scales
with galaxy luminosity and halo mass with some dependence on galaxy
colour
\citep{steidel95,zibetti07,kacprzak08,chen10a,nielsen12,cwc13b}.
Using clever stacking and cross-correlation techniques between {\MgII}
absorption systems and massive red galaxies, \citet{zhu14} and
\citet{Perez-Rafols14} detect absorption out to 20~Mpc around galaxies
with $<M_h>=10^{13.5}$M$_{\odot}$. These extended {\MgII} profiles
appear to change slope on scales of $\sim 1$~Mpc, which is the
expected transition from being dominated by the dark matter halo to
where it is dominated by halo-halo correlations \citep{zhu14}.

Within these {\MgII} haloes, the average gas covering fraction is
estimated to be $\sim 50-90$\% for galaxies at $z\sim 0.6$
\citep{tripp-china,chen08,kacprzak08,chen10a,nielsen12,cwc13a},
possibly decreasing to $25$\% at $z\sim0.1$ \citep{barton09}.  The gas
covering fraction has a radial and azimuthal dependence that increases
towards the host galaxy center and also along its major and minor axes
\citep{nielsen12,kacprzak12a}.  Furthermore, from halo abundance matching \citep[see][]{trujillo-gomez11}, we know that the  covering fraction within a given $D$ is constant with halo mass, $M_h$, over the range $10.4 \leq$ log$M_h \leq 13.3$  \citep{cwc13a}.

It has been firmly established that rest-frame equivalent width is
anti-correlated with $D$ at the 8$\sigma$ level
\citep{lanzetta90,steidel95,archiveII,kacprzak08,chen10a,2011MNRAS.416.1215R,nielsen12,kacprzak13}. The
anti-correlation can be represented by a single log-linear relation
log$[W_r(2796)]=(- 0.015\pm0.002)\times D+(0.27\pm0.11)$ that is found
to be valid for {\MgII} absorption around galaxies within
$0<D<200$~kpc \citep{nielsen12,kacprzak13}.  This anti-correlation
could be interpreted as a radially decreasing gas density profile
surrounding galaxies, although the complex velocity structure of the
absorption systems, which is independent of $D$, suggests that the
absorption arises from a variety of ongoing dynamical events within
the galaxy and halo \citep{csv96}. Recent work has shown that host
galaxy virial mass determines the extent and strength of the {\MgII}
absorption: the mean equivalent width increases with virial mass at
fixed $D$, and decreases with $D$ for fixed virial mass with the
majority of the absorption produced within 0.3 viral radii
\citep{cwc13b}.  On average, it appears that the {\MgII} equivalent
width is not dependent on, or is weakly correlated with, galaxy halo
mass \citep{bouche06,gauthier09,lundgren09,cwc13b,gauthier14}.

\subsection{Theory}

Semi-analytical models and isolated galaxy simulations
\citep[e.g.,][]{mo96,burkert00,lin00,maller04,chen08,kaufmann08,tinker08}
have been invoked to study isolated galaxy haloes.  In these models,
{\MgII} absorption typically arises from condensed, infalling,
pressure confined gas clouds within the cooling radius of the hot
halo. These models reproduce the general statistical
properties of the absorber population. However, they lack the dynamic
influences of cosmic structures and local environments.

Currently, one of the limitations of cosmological simulations is that
local sources of ionizing radiation are not properly accounted for,
although some attempts have been made \citep{goerdt12}. Ionizing radiation
can have a significant effect on high column density absorbers with
N(\HI)$>10^{17}$~{\cmsq} where {\MgII} absorption is typically found
\citep{rahmati13}.  None-the-less, there have been some attempts to
study {\MgII} absorption in simulations. Using simulated quasar
absorption-line sightlines through cosmological simulations,
\citet{kacprzak10a} demonstrated that {\MgII} absorption arises in
filaments and tidal streams with the gas infalling with velocities in
the range of the rotation velocity of the simulated galaxy. This is
consistent with simulations of \citet{stewart11b} who showed that
these cold-flows produce a circumgalactic co-rotating gaseous disk
that is infalling towards the galaxy. In absorption these structures
are expected to produce $\sim 100$\kmsec velocity offsets relative to
the host galaxy in the same direction of galaxy rotation, which is
consistent with observations \citep{steidel02,kacprzak10a}. The
accreting gas spends only 1$-$2 dynamical times in this disk before
accreting onto the host galaxy \citep{stewart13}.

Simulations have yet to consistently reproduce the observed {\MgII}
covering fractions \citep[][though some recent simulations are very
close]{ford13}, which are typically underestimated by a factor of two
or more. However, \citet{stewart11b} showed that the {\MgII} covering
fraction drops dramatically for galaxies with a halo mass above
$M_h>$$10^{12}$~M$_{\odot}$ where cold mode accretion is predicted to
be quenched when the halo is massive enough to support a stable shock
near the virial radius
\citep[e.g.,][]{dekel06,dekel09,keres09,freeke11a}. This phenomena may
have been directly observed for one galaxy \citep{cwc12}, however,
\citet{cwc13a} has shown statistically that the covering fraction does
not precipitously drop for $M_h>10^{12}$~M$_{\odot}$, which is in
direct conflict with theoretical expectations of a mass dependant
truncation of cold-mode accretion.

\subsection{Sources of MgII Absorption}

The goal of absorption-line studies is to determine the exact source
of the absorption so we can study specific phenomena/physics in
detail.  However, we have yet to determine an efficient method of
conclusively determining the origin of individual systems. In general,
there are five likely sources of {\MgII} absorption: 1) ISM, 2) HVCs,
3) tidal debris, 4) galactic outflows, and 5) filamentary accretion.

\subsubsection*{ISM}

A significant fraction of cool gas is located within the ISM of
galaxies. {\MgII} absorption is always detected along quasar
sight-lines that probe the ISM of our Milky Way
\citep[e.g.,][]{savage00,bowen95}. The majority of absorption seen in
galaxy spectra arises from the ISM; only a small fraction of the
absorbing gas appears to be outflowing/infalling \citep[see][and
references therein]{martin12}.  It is increasingly difficult to find
extremely low impact parameter systems as the quasar either outshines
faint foreground galaxies or becomes more obscured by large low
redshift foreground galaxies.  However, several studies report {\MgII}
absorption systems detected within $\sim$5~kpc of local
\citep{bowen96} and low-redshift galaxies \citep{kacprzak13} that have
a similar $W_r(2796)$ distribution to that of the Milky Way.

\subsubsection*{HVCs}

HVCs are expected to contribute to the covering fraction of {\MgII}
absorbers because they are sources of {\MgII} absorption around
local galaxies \citep[e.g.,][]{savage00,herenz13}.  This
contribution is difficult to compute, because we can only currently
measure it for the local galaxy population. It has been estimated that
roughly 30-50\% of known strong {\MgII} absorbers arise from
HVCs \citep{richter12,herenz13}.  This suggests that the majority of
absorbers have an alternative origin.

\subsubsection*{Tidal Debris}

Low-redshift {\HI} surveys show that galaxies having undergone a minor
merger/interaction typically exhibit a perturbed/warped {\HI} disk
that is more extended than those of isolated galaxies
\citep[e.g.,][]{fraternali02,chynoweth08,sancisi08}. It is common to
detect {\MgII} absorption near galaxy groups. In most cases, the
absorbing gas tends to have lower metallicity than the nearby host
galaxies, and the absorption-line kinematics tend to be more complex. In
addition, stellar tidal features are commonly seen for galaxy group
members. All of the above suggests that the gas has a tidal origin
within the intragroup medium
\citep{york86,kacprzak07,kacprzak10b,rubin10,meiring11,battisti12,gauthier13}.
\citet{chen10a} demonstrates that group galaxies do not follow the
well known anti-correlation between $W_r(2796)$ and $D$ but exhibit a
more random distribution of $W_r(2796)$ as a function of
$D$\citep[also see][]{2013ApJ...772L..29Y}. These results suggest that galaxy
environment plays a role in the metal distribution within galaxy
haloes.

The extreme role of environment is further evident in galaxy clusters
where {\HI} disks are truncated \citep[e.g.,][]{chung09} and the CGM
covering fraction is significantly reduced \citep{2013ApJ...772L..29Y}. In addition,
{\MgII} haloes in galaxy clusters are only $\sim$10~kpc in size
\citep{lopez08,padilla09}.

\subsubsection*{Galactic Outflows}

High equivalent width {\MgII} absorption has been observed to directly
trace 100$-$1000 \kmsec galactic-scale outflows originating from their
host galaxies
\citep{tremonti07,weiner09,martin09,rubin10,chelouche10,coil11,lundgren12,martin12,rubin13,bordoloi13}. These
outflows can extend out to $\sim$50~kpc and are orientated along the
galaxy minor axis \citep{bordoloi11,
  bouche12,gauthier12,kacprzak12a,bordoloi12, kacprzak14}, having
opening angles of $\sim 100$ degrees
\citep{kacprzak12a,bordoloi12,martin12}.  The outflows observed in
disk galaxies produce {\MgII} equivalent widths and velocities that
are higher for the face-on systems compared to the edge-on ones, which
is consistent with a primarily bipolar outflow geometry
\citep{bordoloi13}.  Furthermore, the outflowing absorbing gas
velocities originating from the host galaxies appear to correlate with
the host galaxy star formation rates (SFRs) and B-band magnitude/mass
\citep{weiner09,rubin10,martin12,kornei12,bordoloi13}.  Similarly,
correlations between host galaxy colours and SFRs with {\MgII}
equivalent widths detected along quasar sight-lines probing their
haloes also indirectly suggest that absorption is produced in outflows
\citep{zibetti07,noterdaeme10,nestor11}.

Additional evidence of outflows is provided by a handful of absorption
systems with measured metallicities near solar that exist near
galaxies
\citep{2012AandA...540A..63N,2013MNRAS.433.3091K,peroux13,crighton14,kacprzak14}.  It is
expected that these high metallicity systems are outflowing from their
host galaxies. This is supported by the fact that some of these quasar
sight-lines tend to probe galaxies along their projected minor axis,
which is where outflows are expected to exist
\citep{peroux13,kacprzak14}. Further evidence supporting outflows
comes from combined kinematic, geometric and metallicity arguments
showing that the velocities of gas outflowing directly from a galaxy
at $z=0.2$ can reproduce the observed transverse absorption velocities
found at 58~kpc along the galaxy's projected minor axis
\citep{kacprzak14}.

\begin{figure}[t]
\centering
\includegraphics[scale=0.66]{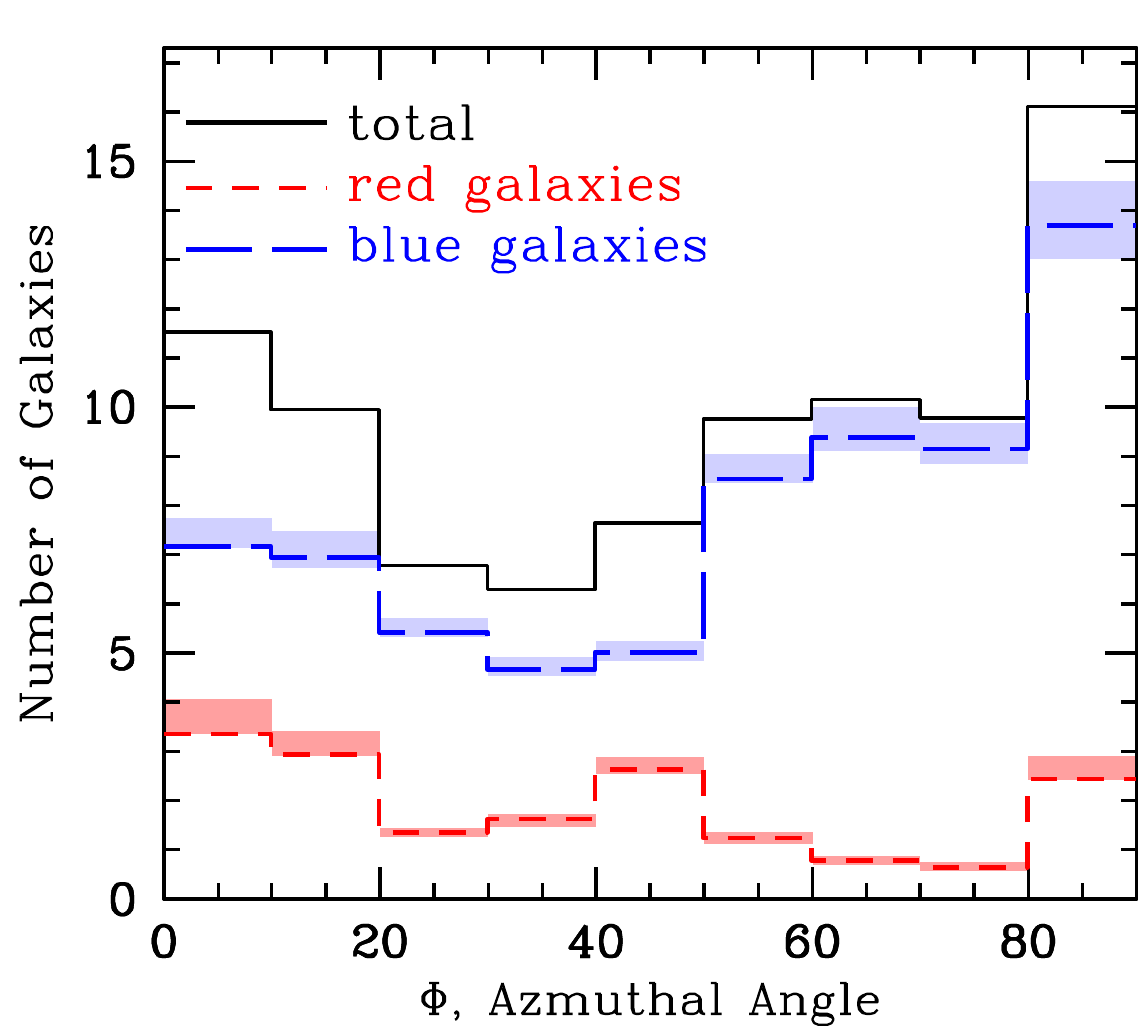}
\caption{Binned azimuthal angle mean probability distribution function
  for all absorbing galaxies (solid line) and for blue (dash line) and
  red (dotted line) absorbing galaxies from \citet{kacprzak12a}. The
  areas of the histograms are normalized to the total number of
  galaxies in each sub-sample. The shaded regions are $1\sigma$
  confidence intervals.  The bimodal distribution suggests a
  preference for {\MgII} absorbing gas toward the galaxy projected
  major axis ($\Phi=0^{\circ}$) and along the projected minor axis
  ($\Phi=90^{\circ}$).  Blue galaxies dominate the bimodal
  distribution while red galaxies tend to have a flatter
  distribution.}
%\vspace*{0.3cm}}
\label{GGKf1}
\end{figure}

\subsubsection*{Filamentary Accretion}

MgII absorption has been observed infalling \citep{martin12} onto
highly inclined galaxies with velocities of $\sim$200 \kmsec
\citep{rubin12}. This is consistent with \citet{kacprzak11b} who
showed that $W_r(2796)$ is correlated with galaxy inclination,
implying a significant fraction of absorption systems are coplanar and
likely accreting toward the galaxy.  The bimodal azimuthal angle
distribution of quasar sight-lines with detected {\MgII} absorption
around host galaxies, shown in Figure~\ref{GGKf1}, also suggests that
infall occurs along the projected galaxy major axis
\citep{bouche12,kacprzak12a}. 

The co-rotating and infalling accretion seen in the simulations of
\citet{stewart11b} is consistent with observations of
\citet{steidel02} and \citet{kacprzak10a} that show {\MgII} absorption
residing fully to one side of the galaxy systemic velocity and aligned
with expected galaxy rotation direction --- mimicking the rotation
curve out into the halo. 

Additional key evidence is provided by a handful of systems with
measured absorption-line metallicities ranging between [M/H] $<-1.8$
to $-1$ that exist near galaxies that have nearly solar metallicities
\citep{zonak04,chen05, tripp05,
  cooksey08,kacprzak10b,ribaudo11,thom11,bouche13,crighton2013}. It is
expected that these low metallicity systems are accreting onto their
host galaxies and possibly trace cold mode accretion.

\subsection{Future Progress}

A large body of evidence suggests that the majority of {\MgII}
absorption traces both outflows from star-forming galaxies and
accretion onto host galaxies. While is clear both processes are
occurring, it remains difficult to disentangle which absorption
systems may be uniquely associated with either process.

\begin{figure}[t]
\centering
\includegraphics[scale=0.35]{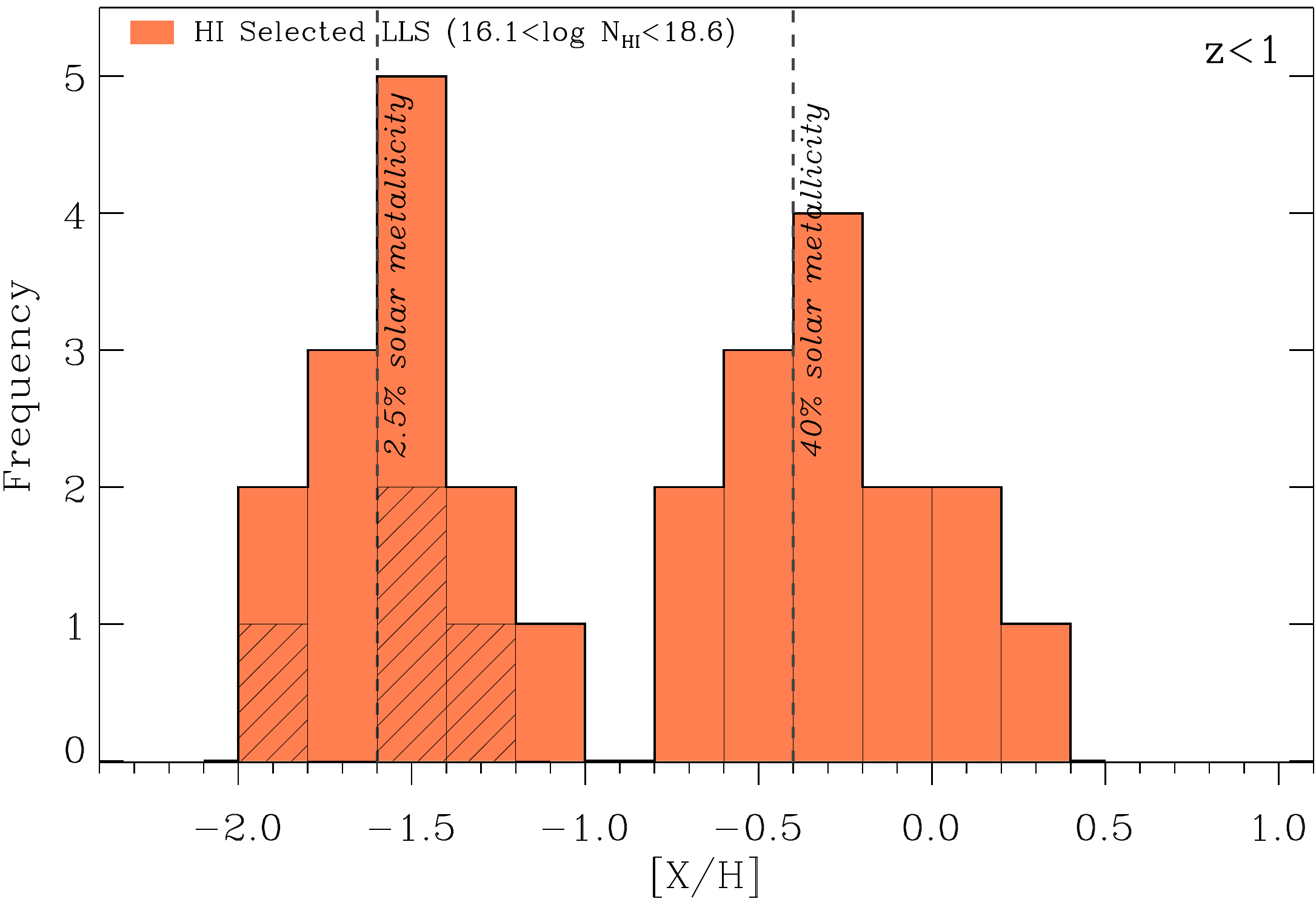}
\caption{ Metallicity distribution function of absorption systems with
  column densities of $16.2 \leq$N(\HI)$\leq19$ at $z\leq1$ from
  \citet{lehner13}. The hashed histograms highlight values that are
  upper limits. The metallicity distribution is bimodal where the peak
  values are indicated. These results are suggestive of two dominant
  sources giving rise to the majority of absorption-line systems: The
  metal-rich population tracing winds, recycled outflows, and tidally
  stripped gas and the metal-poor population tracing cold accretion.}
%\vspace*{0.3cm}}
\label{GGKf2}
\end{figure}

Two measurements may aid in identifying which absorption systems are
associated with either outflows or accretion: galaxy orientation with
respect to the quasar sight-line and the absorption-line metallicity.
The {\MgII} spatial distribution relative to the host galaxy provides
a promising test since outflows are expected to extend along the host
galaxy minor axis \citep[e.g.,][]{strickland04}, while accretion
should progress along filaments towards the galaxy major axis
\citep[e.g.,][]{stewart11b}.  In Figure~\ref{GGKf1},
\citet{kacprzak12a} shows that {\MgII} absorption is primarily
concentrated around the projected major and minor axis of host
galaxies. Galaxies with no measurable absorption have a random
distribution of quasar line-of-sight position angles. Furthermore,
star-forming galaxies drive the bimodal distribution while red
galaxies show some signs of gas accretion along the projected major
axis \citep{kacprzak12a}. This is consistent with previous works and
recent models \citep{bordoloi11,bouche12,bordoloi12}.

In addition, since outflows are expected to be metal-enriched and
accreting material metal-poor, comparing the absorption-line and host
galaxy metallicity provides another suitable test for discriminating
outflows from accretion \citep[e.g.,][and references
therein]{bouche13,crighton14}. Recent work by \citet{lehner13} shows
that the metallicity distribution of Lyman limit systems, $16.2
\leq$N(\HI)$\leq19$, is bimodal with metal-poor and metal-rich
populations which peak at 2.5\% and 40\% of solar metallicity 
(Figure~\ref{GGKf2}). Thus, similarly to the galaxy orientation
bimodality, it is suggestive of two dominant sources of the
majority of absorption-line systems: the metal-rich population likely
traces winds, recycled outflows, and tidally stripped gas while the
metal-poor population is consistent with cold accretion.

Measuring the metallicity and relative orientations, combined with
relative kinematics, yields the most promising selection criteria in
isolating individual sources producing {\MgII} absorption
\citep[see][]{kacprzak14}. Future {\HI} surveys, such as WALLABY
(targeting $600,000$ galaxies at $z=0-0.25$) and DINGO (targeting
$100,000$ galaxies at $z=0-0.40$), using the Australian Square
Kilometer Array Pathfinder (ASKAP) will provide complementary data
that will aid in our understanding of galaxy evolution and feedback
processes.

% % % % % % % % % % % % % % % % %% % % % % % % % % % % % % % % % %
% % % % % % % % % % % % % % % % %% % % % % % % % % % % % % % % % %
% % % % % % % % % % % % % % % % %% % % % % % % % % % % % % % % % %

\section{Lyman Alpha in Emission} \label{S:lyaEmission}

\subsection{Physics of Lyman Alpha Emission} \label{Ss:Lyaphysics}
\citet{1967ApJ...147..868P} predicted that \lya emission would be an excellent tracer of young galaxies. Unlike absorption lines studies, which give one-dimensional information, \lya emission can trace gas in three dimensions (two spatial and one wavelength). While it was three decades before observational searches for \lya emitters were successful, the detection of \lya emission is now an important window into the high-redshift Universe.

The physics of Lyman alpha emission is neatly summarised by the mechanisms by which a hydrogen atom is excited into the $2~^2P$ state.

\paragraph*{Recombination:} when an electron and a proton combine to make neutral hydrogen, the electron may pass through the $2~^2P$ state on its way to the ground state:
\begin{equation*}
\ro{e}^- + \ro{H}^+ \rightarrow \ro{H}(2~^2P) + \ro{photon(s)} \rightarrow \ro{H}(1~^2S) + \ro{Ly}\alpha ~.
\end{equation*}
For atoms at $T=10^4$ K, with only a weak dependence on temperature, $\sim 42$\% of recombinations will pass through the $2~^2P$ state on their way to the ground state and produce a \lya photon, $\sim 38$\% will go directly to the ground state and produce an ionizing photon, and $\sim 20$\% go to the $2~^2S$ state, producing 2 continuum photons in a forbidden transition to the ground state \citep{1996ApJ...468..462G}.  If the surrounding medium is optically thick to ionizing radiation, then ionizing photons will be reprocessed, while \lya photons are simply scattered. In this case, known as \emph{Case B}, \lya photons are emitted at $\sim 68$\% of the rate at which ionizing radiation is absorbed\footnote{Resonant absorption of a higher Lyman-series photon (Ly$\beta+$) will also result in a cascade back to the ground state, possibly resulting in Lyman alpha emission. The Ly$\beta+$ photon itself will result from recombination or excitation, and so can be included in the respective case.}.

Thus, the photoionisation of neutral hydrogen will produce Lyman alpha radiation. The \hi could surround the photoionising source, producing \emph{nebular} emission. In star-forming galaxies, for example, massive O and B stars (and PopIII stars) produce copious ionizing radiation \citep{2001ApJ...552..464B,2003AandA...397..527S,2010A&A...523A..64R}, which immediately encounters \hi in the surrounding ISM from which the stars formed. Upon recombination, \lya photons are emitted. Because the \lya line is narrow and strong, it should (in theory) provide a signature of primeval, high-redshift galaxies.

For the same reason, ionizing radiation from quasars should light up the surrounding gas in \lya. \citet{2001ApJ...556...87H} studied the effect of a quasar turning on within an assembling protogalaxy, and found that it would boost \lya emission in a spatially extended region dubbed ``\lya fuzz''. While, observations show that the fraction of LAEs associated with AGN is a few percent \citep{2006ApJ...642L..13G,2008ApJS..176..301O}, the brightest, most spatially extended emitters (``blobs'') could be driven by an obscured quasar \citep{2013ApJ...771...89O}.

An \hi cloud can also be illuminated by an external source of ionizing photons, such as the UV background or a nearby quasar. Observations of these Fluorescent \lya Emitters (FLEs) hold great potential --- they illuminate gas \emph{outside} of galaxies, where most of the baryons are at high redshift. Further, for clouds that are optically thick to ionizing radiation, the surface brightness of the fluorescent emission is set by the strength of the ionizing background \citep{1996ApJ...468..462G,2007ApJ...657..135C}.

\paragraph*{Excitation:} A \hi atom can be placed in the $2~^2P$ state via collisional excitation. Most collisions at the relevant temperatures place the \hi atom in an $n=2$ state: $\sim 25$\% go to the $2~^2S$ and $\sim 75$\% go to the $2~^2P$. Approximately 10\% of the energy is lost to bremsstrahlung, meaning that $\sim 68$\% of the thermal energy that is radiated away by collisional excitation is in the form of \lya photons \citep{1996ApJ...468..462G}. However, the \lya collisional emissivity is a strong function of temperature. For example, for gas in collisional ionisation equilibrium (e.g. self-shielded), the emissivity drops by $\gtrsim 3 \times 10^3$ between its peak at $1.8 \times 10^4$ K and $1 \times 10^4$K \citep{1995ApJ...442..480T}.

There is thus an important connection between \lya emission and cooling radiation, long recognised as a crucial ingredient in galaxy formation \citep{1977ApJ...215..483B,1977ApJ...211..638S,1978MNRAS.183..341W}.  \citet{2003MNRAS.345..349B} and \citet{2005MNRAS.363....2K} found that gas accretes onto galaxies in two modes: a hot mode, where particles are heated to $T_{\ro{vir}} \sim 10^6$ K before cooling via bremsstrahlung and accreting onto the galaxy quasi-spherically; and a cold mode, where particles that have never been heated above $\sim 10^5$ K are accreted along filaments, and cool primarily by \lya line emission. Thus, gas that is cooling within dark matter haloes is expected to radiate a substantial fraction of its gravitational energy via collisionally excited \lya emission \citep{2000ApJ...537L...5H,2001ApJ...562..605F}. 

We will examine these mechanisms and emitters in more detail in later sections. Next, we will survey the history and current status of observations of \lya emission from the high-redshift universe.

\subsection{Physics of Lyman Alpha Scattering} \label{SS:lyaRT}

Lyman alpha photons are strongly scattered by \hi, and so an understanding of radiative transfer effects is crucial to interpreting \lya observations. With quasar absorption spectra frequently revealing \hi regions with column densities \nhi of order $10^{16}-10^{22} \cm$ (Section \ref{S:lyaAbs}, extremely large optical depths of $\tau_0 = 10^3 - 10^9$ are likely to be encountered by emitted \lya photons (see Equation \ref{eq:tauLya}). This is particularly true when ionizing radiation is the source of \lya photons. At temperatures of $\sim 10^4$ K, the optical depth of \hi in \lya is about $10^4$ times larger than the optical depth at the Lyman limit \citep[][pg. 77]{1989agna.book.....O}. A Lyman limit photon that enters an \hi region will be absorbed at a depth of $\tau_{LL} \sim 1$. When recombination produces a \lya photon, this photon will find itself at an optical depth of $\tau_{\lya} \simeq 10^4$.

Radiative transfer calculations are necessary for any theoretical prediction of the spectra and spatial distribution of \lya emission. \lya photons will typically undergo many scatterings before escaping an \hi region. The properties of the emergent radiation depend sensitively on the spatial distribution, kinematics, temperature and dust content of the gas.

\begin{figure}[t]
\centering
	\includegraphics[width=0.5\textwidth]{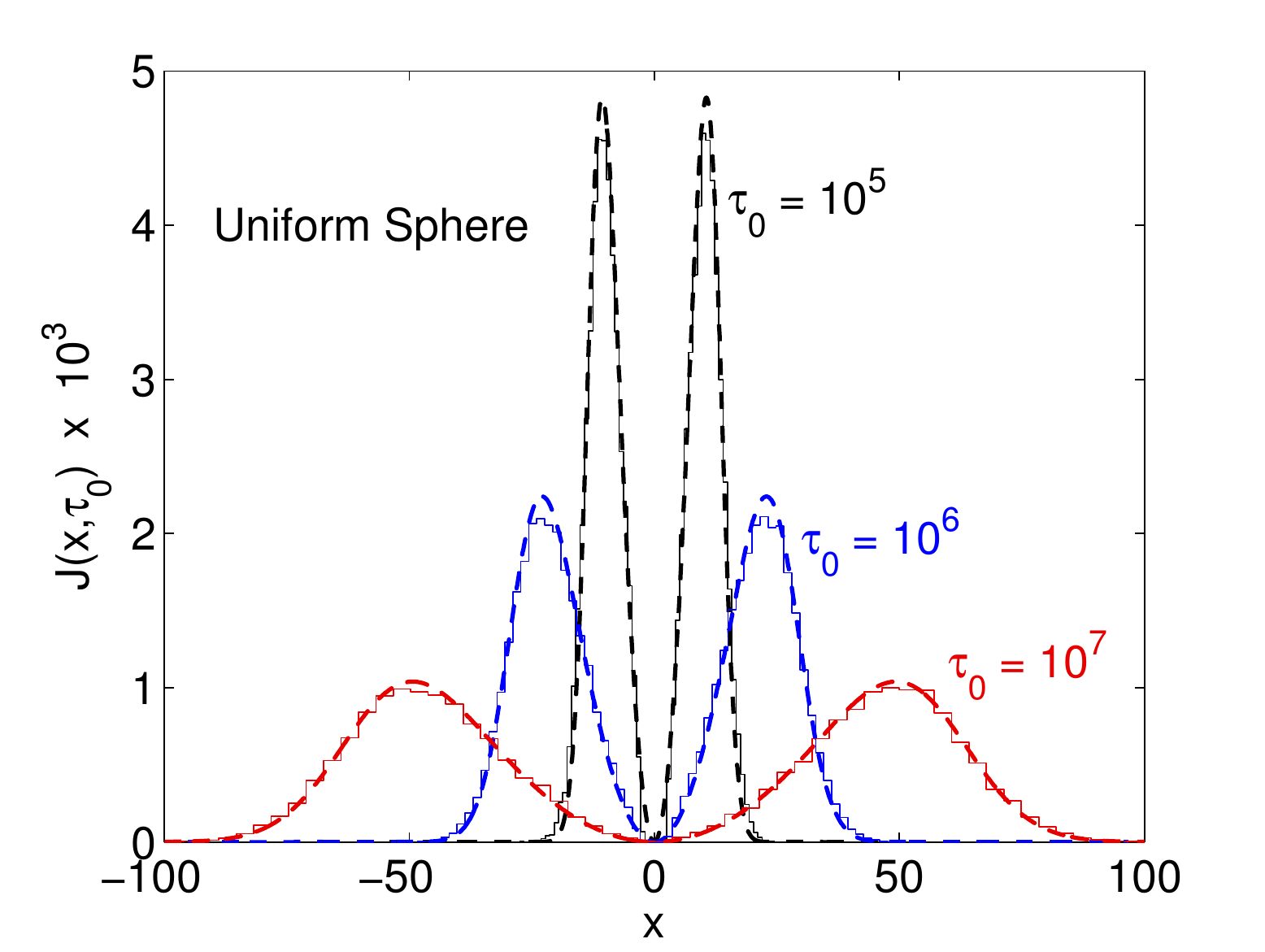}
	\caption{The emergent spectrum for an optically thick ($\sqrt{\pi} \tau_0 \apgt 10^3/a$), uniform, static sphere of neutral hydrogen, where line-centre photons are injected at the centre of the slab. The $x$-axis shows the frequency of the emergent radiation, relative to line centre, in units of the thermal Doppler frequency width. We set the temperature of the gas to $T = 10$ K, and $\tau_0$ is the line-centre, centre-to-edge optical depth as labelled on the plot. The dotted line shows the analytic solution of \citet{2006ApJ...649...14D}. This solution is used as a test for \lya Monte Carlo radiative transfer codes; the solid histogram shows the result of the code of \citet{2010MNRAS.403..870B}.} \label{fig:sphtest}
\end{figure}

We will begin by examining a simple scenario for which an analytic solution is available. \citet{1973MNRAS.162...43H} and \citet{1990ApJ...350..216N} derived an analytic expression for the spectrum $J(x,\tau_0)$ of radiation emerging from an optically thick ($\sqrt{\pi}\tau_0 \gtrsim 10^3/a$), uniform, static slab of neutral hydrogen, where line-centre photons are injected at the centre of the slab, atomic recoil is neglected, $\tau_0$ is the centre-to-edge optical depth at line-centre, $x$ is the frequency of \lya radiation relative to line centre and in units of the thermal Doppler frequency width, and $a$ is the ratio of the Lorentz to (twice) the Doppler width. The analogue of this solution for a uniform sphere is shown in Figure \ref{fig:sphtest}.

The key to understanding this spectrum is that the photon executes a random walk in \emph{both frequency and physical space}. When a photon is in the Doppler core, its mean free path is very short, so there is very little spatial diffusion. Most scatterings are with atoms that have the same velocity along its direction of motion as the atom that emitted the photon. Occasionally, however, the photon will collide with a very fast moving atom from the tail of the Maxwell-Boltzmann distribution, with large velocities perpendicular to the photon's direction. When this photon is re-emitted, it will be far from line-centre. The photon is now travelling through a slab that is comparatively optically thin. What happens next depends on the optical depth of the slab ($\tau_0$).

In the case of moderate optical depth ($a \tau_0 \lesssim 10^3$), a single ``catastrophic'' scattering into the wings is enough to render the slab optically thin to the photon. A rough estimate of the frequency of the escaping photons ($x_e$) is given by,
\begin{equation}
	\tau \approx \tau_0 e^{-x_e^2} \approx 1 \qquad \Rightarrow \qquad x_e \approx \pm \sqrt{\ln \tau_0} .
\end{equation}
This case is discussed in \citet{1962ApJ...135..195O}. \citet{1972ApJ...174..439A} introduced the term \emph{single longest flight} to describe this scenario.

For extremely optically thick media ($a \tau_0 \gtrsim 10^3$), however, the optical depth in the damping wings is enough to prevent the photon from escaping from the medium in a single long flight. Instead, the photon will execute a random walk in physical space with a relatively long mean free path. \citet{1962ApJ...135..195O} showed that during this `walk in the wings', there will also be a random walk in frequency space: the rms Doppler shift of each scatter is $x \sim 1$, with a mean shift per scatter of $1/|x|$, biased to return the photon to line-centre. Thus, after a large number of scatterings, the photon will return to the Doppler core and once again experience very little spatial diffusion. The cycle of an initial scatter to the wings followed by the random walk back to the core (in frequency space) is termed an \emph{excursion} \citep{1972ApJ...174..439A}.

It is at this point that \citet{1972ApJ...174..439A} points out a mistake in \citet{1962ApJ...135..195O}, the resolution of which is quite illuminating. \citet{1962ApJ...135..195O} assumes that, in extremely optically thick media, the distance travelled in any particular excursion will be small compared to the size of the region. Thus, each excursion can be considered to be a single step in an ordinary random walk. However, \citet{1972ApJ...174..439A} points out that we can test this assumption by asking which is more likely to happen first: the photon uses a large number of small excursions to random-walk out of the medium, or the photon uses one large excursion to escape? \citet{1972ApJ...174..439A} showed that it is the second option --- photons will escape the medium on their \emph{single longest excursion}.

We can again give a rough estimate of the escape frequency of the photons. If a photon is scattered to frequency $x$ in the wings, and each scattering sends the frequency on average $1/|x|$ back to the core, then each excursion will contain $\mathcal{N} \sim x^2$ scatterings. Between each scattering, the photon will travel a physical distance $\Delta s$ defined by $\sigma_x \dhi \Delta s \sim 1$. If this distance were travelled at line-centre, it would correspond to an optical depth of $\Delta \tau_0 = \sigma_0 \dhi \Delta s = \sigma_0 / \sigma_x$. Now, $\sigma_0 / \sigma_x \approx 1/H(a,x) \sim x^2 / a$, where we have (reasonably) assumed that we are in the damping wings of the Voigt function $H$. Thus, between each scattering, the photon will travel a line-centre optical depth of $\Delta \tau_0 \sim x^2 / a$. Further, after $\mathcal{N}$ scatterings, the photon will have travelled an rms line-centre optical depth of $\tau_0^{\ro{rms}} \sim \sqrt{\mathcal{N}} \Delta \tau_0 = |x|^3 / a$ \citep[see, e.g.][pg. 35]{1979rpa..book.....R}. 

The photon will escape when, in the course of an excursion, it can diffuse a distance comparable with the size of the medium, i.e. $\tau_0^{\ro{rms}} \sim \tau_0$. Putting the above equations together, we find that the critical escape frequency is $x_e \sim \pm (a \tau_0)^{1/3}$. This agrees very well with the analytic solution of \citet{1973MNRAS.162...43H} and \citet{1990ApJ...350..216N} for a static slab, which has its peak at $x_p = \pm 1.06 (a \tau_0)^{1/3}$. The static sphere solution of \citet{2006ApJ...649...14D} has its peak at $x_p = \pm 0.92 (a \tau_0)^{1/3}$. This means that if we modelled a DLA as a static sphere of \hi at temperature $T$ with maximum edge-to-edge column density \nhi, a \lya photon emitted at the centre of the system will emerge with a characteristic frequency (expressed as a velocity) of:
\begin{equation}
v_p = 165.7 ~ \kmsec \left( \frac{T}{10^4 \ro{ K}} \right)^\frac{1}{6} \left( \frac{\nhi}{2 \times 10^{20} \cm} \right)^\frac{1}{3} ~,
\end{equation}
even though the gas itself is static.

Given the complexity expected of the distribution and kinematics of \hi in the real universe, Monte Carlo simulations are the method of choice for \lya radiative transfer models. In full generality, a radiative transfer problem is specified by the number density of \hi (\dhi), the \lya emissivity ($\epsilon$, in photons/s/cm$^3$), the bulk velocity $\mathbf{v}_b$, the temperature $(T)$, and the dust number density $(n_d)$ and scattering/absorption properties, all as a function of position. The Monte Carlo algorithm involves creating a photon and propagating it in a random direction for a certain distance (that depends on the optical depth), at which point the photon will interact with an atom. After the interaction, the photon will have a new frequency and a new direction chosen from an appropriate distribution. We repeat until the photon escapes the system. Mock observations of the system are built up from the escaping photons. The algorithm is discussed in detail in \citet{BarnesPhDThesis}\footnote{Available at www.dspace.cam.ac.uk/handle/1810/224480.}. A plethora of \lya radiation transfer codes, based on the Monte-Carlo technique, have been published \citep{2001ApJ...554..604A,2002ApJ...578...33Z,2005ApJ...628...61C,2006AandA...460..397V,2006ApJ...649...14D,2006MNRAS.367..979H,2006ApJ...645..792T,2007ApJ...657L..69L,2011MNRAS.416.1723B,2012MNRAS.424..884Y}. Different geometrical and kinematic configurations can be tested with these models.  We will consider specific applications of such simulations in the following sections.

\subsection{Lyman Alpha Emitters (LAEs)} \label{Ss:LAEs}
\paragraph*{The search for \lya Emitters:}

For many years, attempts to detect the population of high-redshift \lya Emitters (LAE) proved mostly unsuccessful \citep[e.g.][]{1980PASP...92..537K,1981ApJ...246L.109M,1992IAUS..149..337D,1993AJ....105.1243D,1994PASP..106.1052P,1995AJ....110..963T}, with the discovery of only a handful of candidates  \citep{1992ApJ...385..151W,1993AandA...270...43M,1993ApJ...404..511M}. This was in strong disagreement with the predictions of \citet{1967ApJ...147..868P}, who predicted emission lines as intense as a few $10^{45}$ erg s$^{-1}$ (corresponding to fluxes of a few $10^{-14}$ erg s$^{-1}$ cm$^{-2}$ at $z =  3$), that should have been seen by the aforementioned surveys. For example, \citet{1995AJ....110..963T} conducted both  shallow/wide survey (F$_{\rm Ly\alpha} \gtrsim 10^{-16}$ erg s$^{-1}$ cm$^{-2}$ over $\approx 10^5$ Mpc$^3$) and a deeper/smaller one (F$_{\rm Ly\alpha} \gtrsim 10^{-17}$ erg s$^{-1}$ cm$^{-2}$ over $\approx 10^3$ Mpc$^3$) using narrow-band (NB) imaging at $\sim3-5$ that led to no detections.

Telescopes remained blind to the long-sought distant \lya galaxy population until the late 90's, and the advent of high-sensitivity and larger collecting-area instruments. 
\citet{1998ApJ...502L..99H} detected 10 LAE candidates with the 10m Keck II telescope in a deeper survey at $z =  3.4$ spanning only $46$ arcmin$^2$ over $\delta$z$ = 0.07$. Their sensitivity (L$_{\rm Ly\alpha} \gtrsim 10^{42}$ erg s$^{-1}$) enabled them to find a more common, faint, LAE population.
\citet{2000ApJ...545L..85R} found $\sim$150 bright LAEs (L$_{\rm Ly\alpha} \gtrsim 5.10^{42}$ erg s$^{-1}$) at $z =  4.5$ as part of 
the Large Area Lyman Alpha survey \citep{2000ApJ...545L..85R,2002ApJ...565L..71M}. Their observations took advantage of the $8192^2$ pixel CCD Mosaic camera at the 4 m Mayall telescope of Kitt Peak National Observatory to cover $0.72$ deg$^2$ in the redshift range $4.37 \leq$ $z  \leq 4.57$, corresponding to a volume of almost $10^6$ Mpc$^3$. The large surveyed volume allowed them to probe rarer, brighter LAEs.

Many ideas have been suggested to explain the previous failures to uncover high-redshift LAEs, e.g. short \lya emission duty cycle \citep{1993ApJ...415..580C}, stellar absorption \citep{1993ApJ...419....7V}, or line suppression due to metals and dust \citep{1981ApJ...246L.109M,1988ApJ...326..101H,1993ApJ...415..580C}. In addition, the models of \citet{1967ApJ...147..868P} were based on the monolithic collapse paradigm of galaxy formation, which overpredicts the \lya luminosities of galaxies in the early Universe.

\citet{1999ApJ...518..138H}, using a hierarchical galaxy formation formalism and a simple model of dust extinction, showed that the properties of the population observed by \citet{1998ApJ...502L..99H} could be roughly recovered. Moreover, the escape of \lya photons from galaxies is very sensitive to resonant scattering in \hi gas \citep[e.g.][]{1962MNRAS.125...21H,1968ApJ...153..783A,1973MNRAS.162...43H,2001ApJ...554..604A}. Indeed, even for low dust content the observed fluxes can be considerably reduced because of dust grain absorption along the increased path of resonantly scattered \lya photons \citep{1980ApJ...236..609H,1990ApJ...350..216N,1991ApJ...378..471C}. An extreme example of the impact of \lya resonant scattering is the strong \lya absorption profile observed in metal/dust-poor low-redshift galaxies \citep{1994AandA...282..709K,2009AJ....138..923O}. Similar processes are thought to alter \lya spectra of high-redshift galaxies, and therefore the visibility of LAEs (see Section \ref{SS:lyaRT}).

\lya radiative transfer is strongly affected by interstellar gas kinematics, geometry and ionisation state \citep{1998AandA...334...11K,1999MNRAS.309..332T,2003ApJ...588...65S,2003ApJ...598..858M,2006AandA...460..397V,2009ApJ...704.1640L,2012MNRAS.424..884Y}, dust distribution \citep{1991ApJ...370L..85N,2006MNRAS.367..979H}, intergalactic attenuation \citep{1995ApJ...441...18M,2007MNRAS.377.1175D}, making the interpretation of \lya observations difficult. We will next discuss such observations of LAEs, returning to theory in Section \ref{subsubsec:lya_ism}.

\paragraph*{Observational Methods:} \label{obslae}

\lya photons emitted at high redshift ($2  \lesssim z \lesssim 7$) can detected from the ground in the optical and near infrared. However, only certain spectral ranges are free from confusion due to night sky line emission. This translates into a series of redshift windows at which LAEs can be observed from earth.

At low redshift ($z \lesssim 2$), the \lya line is seen in the ultra-violet and one needs to utilise space instrumentation. Observations in the local universe are even more restricted due to (i) the strong geocoronal emission which blinds nearby \lya emission lines, and (ii) the damped \hi absorption produced by the Galactic centre, which suppresses many lines of sight. Nevertheless, low-redshift LAEs have been studied by space-based telescopes, e.g. with the HST \citep[\textit{Hubble Space Telescope, }][]{1998AandA...334...11K,2009AJ....138..923O}, IUE \citep[\textit{International Ultraviolet Observatory, }][]{1981ApJ...246L.109M,1993MNRAS.260....3T} and GALEX \citep[\textit{Galaxy Evolution Explorer, }][]{2008ApJ...680.1072D,2010ApJ...711..928C}. Observations of nearby LAEs  probe galaxies on small scales, providing detailed \lya and multi-wavelength mapping of the interstellar and circumgalactic media \citep[e.g.][]{2013ApJ...765L..27H}. Although samples of are still rather small, the analysis of well-resolved objects can serve as a useful benchmark for interpreting high-redshift data \citep{2009Ap&SS.320...35M}.

\begin{table*}
\scriptsize
\begin{adjustwidth}{-1.cm}{-1.cm}
\caption{(Non-exhaustive) Compilation of surveys of LAEs.}
\vspace{0.5cm}
\begin{tabular}{cccccccc}
\hline
 \hline
 \noalign{\vskip 0.1cm}    
\textbf{Reference} & \textbf{Redshift} & \textbf{N$_{\rm obj}$} & \textbf{EW$_{Ly\alpha}^{\rm lim}$} & \textbf{L$_{Ly\alpha, 42}^{\rm lim}$} & \textbf{Ident. technique} & \textbf{Instrument} & \textbf{Area} \\
 & (1) & (2) & (3) & (4) & (5) & (6) & (7) \\
\noalign{\vskip 0.1cm}
\hline
\noalign{\vskip 0.1cm}

\citet{2012ApJ...749..106B} & $ 0.67-1.16 $ & $ 28 $ & $20$ & $3$ & grism & GALEX & $2286$\\ % Barger 2012

\citet{2011ApJ...736...31B} & $1.9-3.8$ & $98$ & $20$ & $4$ (z$\approx$3) & IFS & VIRUS & $169$ \\

\citet{2010ApJ...714..255G} & $2.07^{\pm 0.02}$ & $250$ & $20$ & $0.64$ & NB & MOSAIC & $998$ \\

\citet{2011AandA...525A.143C} (Deep) & $2-6.62$ & $42$ & $-$ & $\approx$0.4 (z$\approx$3)& MOS & VIMOS & 2230 \\

\citet{2011AandA...525A.143C} (Ultra-Deep) & $2-4$ & $188$ & $-$ & $\approx$0.1 (z$\approx$3)& MOS & VIMOS & 576 \\

\citet{2010Natur.464..562H} & $2.2^{\pm 0.05}$ & $38$  & $20$ & $0.3$ & NB & FORS1 & $56$ \\

\citet{2009AandA...498...13N} & $2.26^{\pm 0.05}$ & $170$ & $20$ & $2.3$ & NB & WFI & $1190$ \\

\citet{2005MNRAS.359..895V} & $2.3-4.6$ & $14$ & $-$ & $1.1$ (z$\approx$3) & IFS & VIMOS & $1.4$ \\

\citet{2008ApJ...681..856R} & $2.67-3.75$ & $27$ & $-$ & $\approx$0.1 (z$\approx$3) & S-S & FORS2 & $0.252$ \\

\citet{2009AandA...497..689G} & $2.8-3.2$ & $83 (59)$ & $25$ & $\approx 0.4$ & NB-MOS & FORS1-2 & $133$\\

\citet{2005PASJ...57..881Y} & $3-5$ & $198$ & $20$ & $5$ (z$\approx$4) & IB & S-Cam & $944$ \\

\citet{2004AJ....128.2073H} & $3.09^{\pm 0.03}$ & $283$  & $37$ & $4.1$ & NB & S-Cam & $770$ \\

\citet{2012ApJ...744..110C} & $3.1^{\pm 0.04}$ & $141$ & $20$ & $2.1$ & NB & MOSAIC II & $\approx 1000$\\

\citet{2007ApJ...667...79G} & $3.11^{\pm 0.02}$ & $162$ & $20$ & $1.2$ & NB & MOSAIC II & $993$\\

\citet{2000ApJ...536...19K} & $3.13^{\pm 0.01}$ & $9$ & $-$ & $1.8$ & S-S & FORS & $50$ \\

\citet{2008ApJS..176..301O} & $3.13^{\pm 0.03}$ & $356 (41)$ & $64$ & $1$ & NB & S-Cam (FOCAS-VIMOS) & $3538$\\

\citet{1998AJ....115.1319C} & $3.44^{\pm 0.03}$ & $10$ & $17$ & $2$ & NB & LRIS & 46 \\

\citet{2008ApJS..176..301O} & $3.69^{\pm 0.03}$ & $101 (26)$ & $44$ & $4$ & NB & S-Cam (FOCAS-VIMOS) & $3474$ \\

\citet{2003AJ....125...13F} & $3.71^{\pm 0.12}$ & $6$  & $53$ & $5$ & IB & S-Cam & $132$ \\

\citet{2002ApJ...565L..71M} & $4.47^{\pm 0.10}$ & $194 (110)$ & $14$ & $5$ & NB & MOSAIC & $1116$ \\

\citet{2003ApJ...582...60O} & $4.86^{\pm 0.03}$ & $87$ & $14$ & $0.8$ & NB & S-Cam & $543$ \\

\citet{2009ApJ...696..546S} & $4.86^{\pm 0.03}$ & $79$ & $11$ & $3$ & NB & S-Cam & $6588$ \\

\citet{2004AJ....127..563H} & $5.68^{\pm 0.05}$ & $24 (19)$  & $17$ & $7.5$ & NB & S-Cam (DEIMOS) & $918$ \\

\citet{2007ApJS..172..523M} & $5.7^{\pm 0.05}$ & $119$ & $18$ & $6.3$ & NB & S-Cam & $6696$ \\

\citet{2003AJ....126.2091A} & $5.7^{\pm 0.05}$ & $20 (2)$ & $25$ & $7$ & NB & S-Cam (FOCAS-ESI) & $720$\\

\citet{2006ApJ...638..596A}& $5.7^{\pm 0.05}$ & $14$ & $17$ & $4.8$ & NB & S-Cam & $320$\\

\citet{2010ApJ...725..394H} & $5.7^{\pm 0.05}$ & $87$  & $-$ & $5.6$ & NB & S-Cam (DEIMOS) & $4168$ \\

\citet{2008ApJS..176..301O} & $5.7^{\pm 0.05}$ & $401 (17)$ & $27$ & $3$ & NB & S-Cam (FOCAS-VIMOS) & $3722$ \\

\citet{2006PASJ...58..313S} & $5.7^{\pm 0.05}$ & $89 (28)$ & $17$ & $2.2$ & NB & S-Cam (FOCAS-DEIMOS) & $725$ \\

\citet{2001ApJ...563L...5R} & $5.73^{\pm 0.06}$ & $18$ & $14$ & $5$ & NB & MOSAIC & $1116$ \\

\citet{2011ApJ...740...71D} & $5.75^{\pm 0.05}$ & $122$ & $-$ & $1$ & MNS & IMACS  & 110 \\

\citet{2010ApJ...725..394H} & $6.54^{\pm 0.08}$ & $ 27 $ & $-$. & $6.7$ & NB & S-Cam (DEIMOS) & $4168$ \\

\citet{2006ApJ...648....7K} & $6.56^{\pm 0.05}$ & $75 (17)$ & $17$ & $2$ & NB & S-Cam (FOCAS-DEIMOS) & $876$ \\

\citet{2010ApJ...723..869O} & $6.56^{\pm 0.05}$ & $207 (24)$ & $14$ & $2.5$ & NB & S-Cam (DEIMOS) & $3238$\\

\citet{2012ApJ...744...89H} & $ 7^{\pm 0.04} $ & $ 7 $ & $-$ & $9$ & NB & S-Cam & $2340$\\ % Hibon 2012

\citet{2012ApJ...752..114S} & $7.26^{\pm 0.08}$ & $4 (1)$ & $0$ & $10$ & NB & S-Cam (FOCAS-DEIMOS) & $1718$ \\

\citet{2010AandA...515A..97H} & $ 7.7^{\pm 0.01} $ & $ 7 $ & $-$ & $6$ & NB & WIRCAM & $390$\\ % Hibon 2010

\citet{2010ApJ...721.1853T} & $ 7.7^{\pm 0.005} $ & $ 4 $ & $-$ & $5$ & NB & NEWFIRM & $784$\\ % Tilvi 2010

\citet{2012ApJ...745..122K} & $ 7.7^{\pm 0.01} $ & $ 4 $ & $-$ & $5$ & NB & NEWFIRM & $760$\\ % Krug 2012

\noalign{\vskip 0.1cm}
\hline 
\noalign{\vskip 0.1cm}
\end{tabular}

\footnotesize{\textbf{Columns:} Col. (1): Redshift.
Col. (2): Number of LAE detections (Number of candidates confirmed with spectroscopy).
Col. (3): \lya{} rest-frame Equivalent Width threshold (\AA).
Col. (4): \lya{} luminosity threshold in units of $10^{42}$ erg s$^{-1}$ (we assume $h=0.7$, $\Omega_m=0.3$ and $\Omega_\Lambda=0.7$).
Col. (5): Observational technique to identify \lya sources.
Col. (6): Instrument used for detection (Instrument used for follow-up observation).
Col. (7): Field size (arcmin$^2$).}\\
\footnotesize{\textbf{Acronyms}: NB: \textit{Narrow-band imaging.} IB: \textit{Intermediate-band imaging.} S-S: \textit{Slit-Spectroscopy.} IFS: \textit{Integral Field Spectroscopy.} MOS: \textit{Multi-Object Spectroscopy.} S-Cam: \textit{Suprime-Cam.} MNS: \textit{Multi-slit Narrow-band Spectroscopy.}}\\
Some flux/EW limits and survey areas/depths given in this table are only approximate values; the reader should refer to the original articles for full details on the surveys. Similarly, the number of LAEs that we quote, \textbf{N$_{\rm obj}$}, can either correspond to the total number of detections, or to the number of objects used to compute the luminosity functions.
\label{table_surveys}
\end{adjustwidth}
\end{table*}

\begin{figure}[t]
\centerline{\includegraphics[width=9.cm,height=8.7cm]{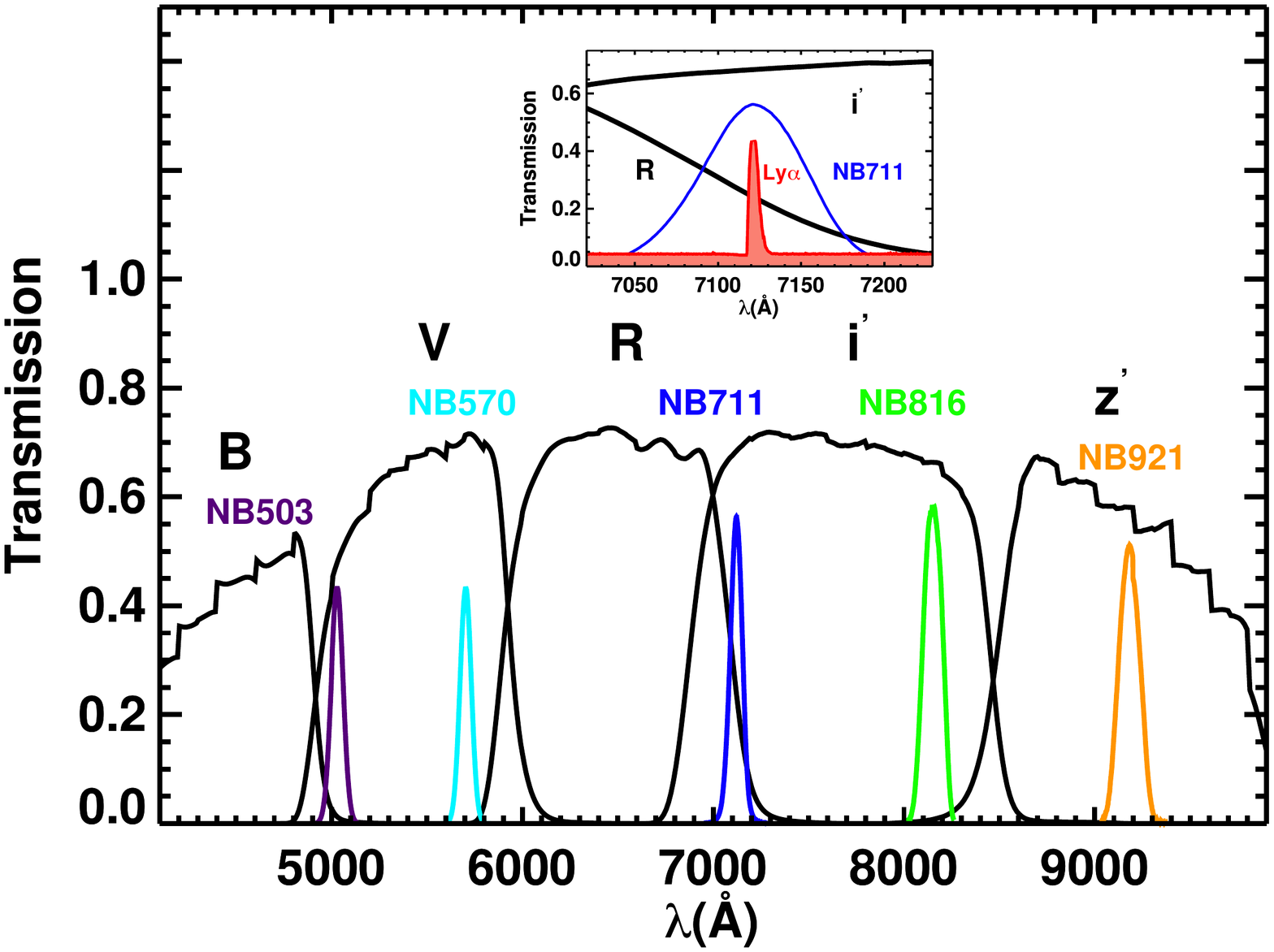}}
\caption{\textit{Main panel}: Set of broad-band (black curves: B, V, R, i$^\prime$ and z$^\prime$) and narrow-band filters (coloured curves: NB503, 570, 711, 816, 921) used in \citet{2003ApJ...582...60O,2008ApJS..176..301O,2010ApJ...723..869O} with Subaru/SuprimeCam to detect LAEs at $z =  3.1, 3.7, 4.5, 4.9, 5.7$ and $6.6$. \textit{Inner panel}: Illustration of a \lya emission line (red curve) interpreted as a z = 5.7 LAE.}
\label{nbfilters}
\end{figure}

At high redshift, thousands of galaxies have been found via the \lya emission line; Table \ref{table_surveys} presents a compilation of \lya surveys. The most popular imaging technique to detect LAEs requires a set of broad- and narrow-band (NB) filters to detect the emission line at the redshifted \lya wavelength \citep[e.g.][ and references in Table \ref{table_surveys}]{2008ApJS..176..301O}. Figure \ref{nbfilters} shows examples of the various filters used with the Subaru/SuprimeCam to detect LAEs at $z =  $ 3.1, 3.7, 4.5, 4.9, 5.7 and 6.6. Colour-magnitude selections are usually applied to the NB detections to ensure the reliability of the LAE candidates and remove contaminants from the sample, such as lower redshift H$\alpha$, O\textsc{ii} or O\textsc{iii} emitters. These criteria can introduce a bias if they preferentially select LAEs with large equivalent widths, or if they eliminate objects without a signature of IGM attenuation shortward 1216 \AA, hence missing galaxies located on \textit{clean lines of sight}. Although the number of clear sightlines is reduced at high redshift as IGM opacity increases \citep{2011ApJ...728...52L}, ionizing emission from the galaxy itself is expected to photoionise the surrounding HI gas.

\setlength{\abovecaptionskip}{3pt}
\setlength{\belowcaptionskip}{-10pt} 

Alternatively, \lya emission lines can be identified spectroscopically in blind surveys, or as part of a follow-up of NB candidates. Spectroscopic observations have been carried out by several teams in the last years, e.g. \citet{2008ApJ...681..856R} with VLT/FORS2, \citet{2011ApJ...734..119K} with Subaru/FOCAS and Keck II/DEIMOS, and \citet{2011AandA...525A.143C} with VLT/VIMOS. Parameters such as \lya equivalent width are usually measured more accurately with spectroscopy than with narrow-band imaging, but slit losses and sky noise can still introduce significant errors in the measurement of the flux.

Another efficient way to search for line emitters is Integral-Field Spectroscopy (IFS): the spectra of many sources are acquired simultaneously in a large field-of-view. At fixed sensitivity, this method considerably reduces the integration time required to observe a given number of galaxies. The first attempt to find LAEs with IFS was by \citet{2005MNRAS.359..895V} with the VLT/VIMOS integral field unit. More recently,
\citet{2011ApJ...736...31B} used the Visible Integral-field Replicable Unit Spectrograph \citep[VIRUS;][]{2010SPIE.7735E..20H} in the HETDEX Pilot survey to observe about a hundred LAEs at $1.9 <$ $z < 3.8$. The Multi Unit Spectroscopic Explorer \citep[MUSE;][]{2006Msngr.124....5B}, recently installed at the VLT, will probe LAEs at redshifts between 2.8 and 6.7, reaching \lya surface brightness limits of $2 \times 10^{-19} \ergsca$ in $\sim 80$ hour integrations. The Keck Cosmic Web Imager \citep[KCWI;][]{2010SPIE.7735E..21M} aims to detect faint \lya emission from the CGM at $2 < z < 6$ to similar limiting surface brightness limits. %http://arxiv.org/pdf/1008.1791.pdf

Another method is Multi-slit Narrow-band Spectroscopy (MNS), in which a narrow or intermediate bandpass is used to select an air-glow free spectral region to find LAE candidates. The MNS technique, pioneered by \citet{1999ASPC..191..229C} and \citet{2004ApJ...603..414M}, was used by \citet{2008ApJ...679..942M} and \citet{2011ApJ...740...71D} with the Inamori-Magellan Areal Camera \& Spectrograph \citep[IMACS;][]{2006SPIE.6269E..13D} to detect LAEs.

All methods to identify LAEs must face the problem of false positives, interlopers that can be confused with \lya emitting galaxies. In addition to low-redshift emission line objects (galaxies or AGN), AGN located at the same redshift as the targeted galaxies can contaminate LAE samples at the level of a few percent \citep{2008ApJS..176..301O}.

In addition to \lya-selected surveys, \lya emission is also often seen in Lyman Break Galaxies \citep[hereafter, LBG;][]{1993AJ....105.2017S,1999ApJ...519....1S}, which are selected via their a) intense UV magnitude, and b) the discontinuity caused by photoelectric absorption ($\lambda < 912$ \AA) in the interstellar medium and line blanketing by the intervening \lya forest  ($912 > \lambda > 1216$ \AA). The Lyman break technique (or dropout technique) efficiently selects high-redshift, star-forming, galaxies using a set of broad-band UV and optical filters \citep[][and references therein]{2003ApJ...592..728S,2004AandA...421...41G,2007ApJ...670..928B,2009MNRAS.395.2196M,2011ApJ...737...90B}. 

\paragraph*{Statistical properties of \lya Emitters:}
\label{statprop}

LAEs observed in current surveys span a range of \lya luminosities from $\sim 10^{42}$ erg s$^{-1}$ to a few times 10$^{43}$ erg s$^{-1}$ \citep[e.g.][]{2006PASJ...58..313S,2007ApJ...667...79G,2008ApJS..176..301O}. The typical \lya luminosity reached by wide-field surveys probes star-formation rates (SFR) greater than $\sim 1 \Msol$ yr$^{-1}$, according to,
\begin{equation} \label{eq:LlyaSFR}
{\rm L}_{{\rm Ly}\alpha} = 1.1 \times 10^{42} \left(\frac{{\rm SFR}}{ \Msol {\rm yr}^{-1}}\right) \: {\rm erg s}^{-1} \: .
\end{equation}
This equation is derived from the SFR-H$\alpha$ relation for constant star formation rate \citep{1983ApJ...272...54K,1998ApJ...498..541K} and the Ly$\alpha$-H$\alpha$ emissivity ratio under Case B recombination \citep{1971MNRAS.153..471B,2006agna.book.....O}. The following assumptions are made to compute the coefficient in Eq. \eqref{eq:LlyaSFR}: (i) Salpeter IMF, (ii) solar metallicity, (iii) and ionisation bound nebula (no ionizing photon can escape the medium). The conversion factor can vary significantly if we modify these assumptions, especially for very low metallicities or extreme IMF cut-off \citep{2003AandA...397..527S,2010A&A...523A..64R}.

The \lya Luminosity Function (LF) at various redshifts has been used to characterise the evolution of LAEs as a population. It does not seem to evolve significantly from $z  =6$ to 3 as the characteristic number density ($\Phi^{*} \sim 10^{-3}$ Mpc$^{-3}$)  and luminosity of LAEs (L$^{*} \sim {\rm{a ~ few}} \: 10^{42}$ erg s$^{-1}$) appear to remain almost unchanged over this redshift range \citep{1998ApJ...502L..99H,2008ApJS..176..301O,2011AandA...525A.143C}\footnote{Although the \lya LFs reported by \citet{2008ApJS..176..301O}, \citet{2007ApJ...667...79G} and \citet{2011AandA...525A.143C} at $z \sim$ 3 are in good agreement, these surveys have applied very different LAE selection criteria. \citet{2008ApJS..176..301O} select only strong line emitters (EW $> 64$ \AA), whereas \citet{2007ApJ...667...79G} use EW $> 20$ \AA. The spectroscopic survey of \citet{2011AandA...525A.143C} does not apply any EW selection. A large fraction of the LAEs reported by \citet{2007ApJ...667...79G} and \citet{2011AandA...525A.143C} have EW $< 64$ \AA, which might have been missed by \citet{2008ApJS..176..301O}. The apparent agreement between $z =3$ LFs may be affected by observational uncertainties and cosmic variance.}.
It is worth noticing that, unlike the \lya LF, the UV LF of LBGs evolves significantly (as expected in a hierarchical growth scenario), its characteristic luminosity being about ten times larger at $z = 3$ than at $z = 6$ \citep{2004AandA...421...41G,2007ApJ...670..928B,2009MNRAS.395.2196M,2011ApJ...737...90B}. 

The non-evolution of the \lya LF is a subject of debate. \citet{2010ApJ...725..394H} find a significantly lower abundance of LAEs at $z = $ 5.7 than \citet{2008ApJS..176..301O} and \citet{2006PASJ...58..313S}. The former argue that the photometric sample of \citet{2008ApJS..176..301O} at the same redshift might contain a large fraction of contaminants, leading them to strongly overestimate the $z = 5.7$ LF. However, the recent spectroscopic follow-up of the \citet{2006PASJ...58..313S} sample at the same redshift by \citet{2011ApJ...734..119K} seems to favour a low contamination rate and a value of $\Phi^{*}$ consistent with \citet{2008ApJS..176..301O}, though slightly smaller. Additional and more homogeneous datasets with spectroscopic confirmations will refine constraints on the \lya LFs. 

\begin{figure}[t]
\includegraphics[width=7.8cm,height=5.5cm]{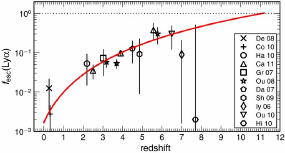}
\caption{Redshift evolution of the \textit{sampled-average volumetric} \lya escape fraction reported by \citet{2011ApJ...730....8H}. Symbols correspond to values derived from observations (\lya, H$\alpha$ and UV), and the red solid line is the best-fitting power-law to the data points ($\propto (1+z)^{2.6}$). The dotted line denotes the $100\%$ escape fraction limit, which is reached at $z = 11$ according to the fitting relation. (Reproduced by permission of the AAS.)}
\label{hayes11}
\end{figure}

A differing evolution of the \lya and UV LFs would imply a variation with redshift of the mechanisms that power the \lya emission, or more probably the ability of \lya photons to escape galaxies. \citet{2011ApJ...730....8H} have quantified this evolution in terms of the \textit{sampled-average volumetric} \lya escape fraction, defined as the ratio of observed to intrinsic \lya luminosity density. Figure \ref{hayes11} (from \citet{2011ApJ...730....8H}) illustrates the rise of this volumetric escape fraction from $z \approx 0 - 6$. They find an evolution of the escape fraction that scales like $(1+z)^{2.6}$ over this redshift range (red curve). Although it is not shown on this figure, we note that the $f_\ro{esc} (\lya)$ value derived by \citet{2014ApJ...783..119W} at $z = 1$ is fully consistent with this relation.

\citet{2013MNRAS.435.3333D} derive an effective \lya escape fraction from SFR and \lya luminosity functions, instead of luminosity densities \citep[see also][]{2011ApJ...736...31B}. Their results are in good agreement with those of \citet{2011ApJ...730....8H}, except at $z=0.35$. Indeed, the value quoted by \citet{2011ApJ...730....8H} at this redshift may be underestimated, because $f_\ro{esc} (\lya)$ is computed by comparing the \lya luminosity density (as observed above a given detection threshold) to the total SFR density.

Regarding the faint-end slope of the \lya LF, and its evolution with redshift, low luminosity sources have been found in very deep surveys down to a few 10$^{-18}$ erg s$^{-1}$ \cm \citep{2008ApJ...681..856R,2011AandA...525A.143C,2011ApJ...740...71D}. The data seem to favour steeper slopes (and especially a steepening towards higher redshift). However, the number of detections at such low fluxes is limited; larger samples are required to draw robust conclusions.

A valuable observable to tackle questions of \lya emission and radiative transfer is the \lya equivalent width (EW), which measures the intensity of the line with respect to the adjacent continuum. Most LAEs have (\textit{rest-frame}) \lya EW between 0 and 100 \AA; the tail of the distribution extends to 250 \AA \citep[][]{2006PASJ...58..313S,2007ApJ...667...79G,2008ApJS..176..301O,2011AandA...525A.143C}. ￼Higher EW values have been claimed \citep[e.g.][]{2002ApJ...565L..71M,2007ApJ...671.1227D,2011ApJS..192....5A}, and recently \citet{2012ApJ...761...85K} reported a z=6.5 LAE with EW $= 436^{+422}_ {149}$\AA.
Such large EW are surprising because they exceed the standard limit of 240 \AA{} set by theoretical models of stellar-powered \lya emission \citep{1993ApJ...415..580C}. Assuming that these extreme EW are real \lya sources, they might hint at a top-heavy IMF, very-low metallicity stars \citep{2003AandA...397..527S} or radiative transfer effects in neutral hydrogen \citep[see Section \ref{subsubsec:lya_ism}; and also][]{1991ApJ...370L..85N}. They could also imply that \lya emission is not powered by star formation alone. Indeed, such measurements are often lower limits with no obvious continuum detection, so other processes like cooling radiation or fluorescence may contribute significantly to the \lya emission.

\begin{figure}[t]
\centerline{\includegraphics[width=8.9cm,height=6.9cm]{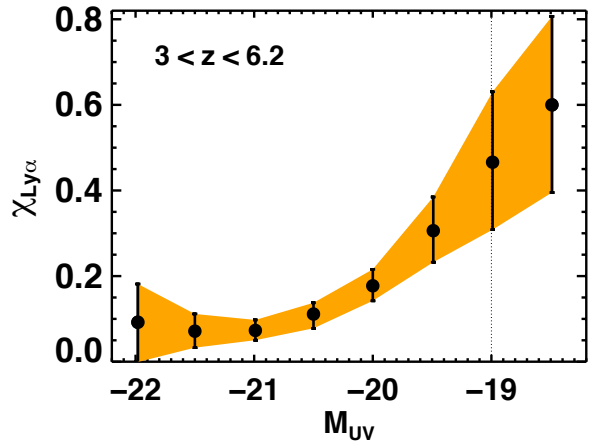}}
\caption{Evolution of the fraction \protect \raisebox{2pt}{$\chi^{}_{{\rm Ly}\alpha}$} of strong \lya emitters (EW$>50$\AA) as a function of rest-frame absolute UV magnitude for Lyman-Break galaxies between $z = 3$ and $6.2$ observed by \citet{2010MNRAS.408.1628S}. Black circles correspond to the fraction \protect \raisebox{2pt}{$\chi^{}_{{\rm Ly}\alpha}$} per magnitude bin, and the orange region encloses the error bars. We see that \protect \raisebox{2pt}{$\chi^{}_{{\rm Ly}\alpha}$} increases gently from 10 to 50 $\%$ between M$^{*}_{\rm UV}=-22$ to $-19$. The dotted vertical line gives the  90$\%$ completeness limit of the dropout galaxies sample.}
\label{2010MNRAS.408.1628S}
\end{figure}

In recent years, several groups have studied the link between LAEs and LBGs. \citet{2003ApJ...588...65S} find that half of $z = $ 3 LBGs have detectable \lya emission, and only $25\%$ display a \lya equivalent width larger than $20$\AA. As part of a spectroscopic survey of faint LBGs, \citet{2011ApJ...728L...2S} have characterized the fraction of \lya emitting galaxies within UV continuum-selected dropout galaxies between $z = 3$ and 7. They find that the fraction of strong emitters (EW$>50$\AA) increases towards fainter UV luminosities. As shown in Figure \ref{2010MNRAS.408.1628S} (black points and error bars), the LAE fraction increases from $\approx 10\%$ at M$_{\rm UV}=-21$ to about $50\%$ for galaxies as faint as M$_{\rm UV}=-19$ at $3<$ $z <6.2$. This result echoes the observed trend between UV magnitude and \lya EW: UV bright galaxies have lower \lya EW, while fainter ones seem to be stronger \lya emitters on average \citep{2006ApJ...645L...9A,2008ApJS..176..301O}. Further, the fraction of \lya emitting galaxies at fixed UV luminosity is found to increase from $z \approx$ 3 to 6 \citep{2010MNRAS.408.1628S,2011ApJ...728L...2S,2012MNRAS.422.1425C}. The rise is even more significant when considering weaker \lya line emitters \citep[EW$>25$\AA;][]{2012ApJ...744...83O}. \\

Untangling the relationship between the LAE and LBG populations is complicated by their respective selection criteria. \citet{2006ApJ...642L..13G} points out that up to 80\% of narrow-band selected LAEs have the right colours to be detected by typical LBG surveys at z = 3, but only 10\% are bright enough in the UV continuum (assuming a detection limit of $R_{AB} \leq 25.5$). Indeed, narrow-band surveys of LAEs preferentially pick up low-continuum galaxies. Nevertheless, the fraction of LAEs among samples of dropout galaxies increases as fainter LBG surveys are carried out \citep[e.g.][]{2010MNRAS.408.1628S}. \lya emission from high-redshift sources, and its variation with respect to the physical properties of galaxies, remains to be fully understood.

\paragraph*{Physical properties of \lya Emitters:}

The \hi gas distribution of LAEs can be constrained by comparing the spatial extent of \lya and UV emission. \citet{2011ApJ...729...48B} and \citet{2011ApJ...743....9G} conducted HST imaging on a subsample of $z = 3$ LAEs previously \lya-selected by \citet{2007ApJ...667...79G}. They find that the \lya emission of these objects is typically compact ($< 1.5$ kpc), and almost coincides with the far-UV ($< 1$ kpc) that traces the young stars. This contrasts with similar observations at lower redshift, which suggest that \lya is more extended than the UV continuum and H$\alpha$ emission by factors of 2-3 \citep{2013ApJ...765L..27H}. This could indicate an evolution of the morphology of LAEs with redshift, or at least a change in the neutral gas distribution surrounding those galaxies \citep{2009AJ....138..923O,2009AandA...498...13N,2012ApJ...753...95B}. Deeper observations by \citet{2011ApJ...736..160S} show a diffuse halo of \lya emission extending well beyond the galaxy; these observations will be discussed further in Section \ref{Ss:LyaCGM}.

Physical properties of LAEs have been studied with multiband photometry and stacking. These analyses reveal that these objects have moderate star formation rates (1 -- 10 $\Msol$ yr$^{-1}$) and low stellar masses $(10^7 - 10^{9} \Msol)$ \citep{2006ApJ...642L..13G,2007ApJ...660.1023F,2010ApJ...724.1524O,2013arXiv1309.6341V}, such that LAEs are thought to be the building blocks of local L$^{*}$ galaxies \citep{2007ApJ...671..278G}.  LAEs are also believed to host young stellar populations on average ($\lesssim 100$ Myr) and have low dust content, as suggested by their blue colours \citep{2008ApJS..176..301O,2010ApJ...724.1524O}. These characteristics are also typical of local galaxies with \lya in emission \citep{2013arXiv1308.6578H}. Note that higher stellar masses, older ages, and significant dust extinction have also been reported by \citet{2009ApJ...691..465F}, \citet{2009AandA...494..553P} and \citet{2010MNRAS.402.1580O}, which may highlight an inherent spread in the population of LAEs. Further uncertainties arise from SED fitting techniques and their assorted assumptions: star formation history, IMF, dust model, nebular emission, etc \citep{2009ApJ...691..465F,2009AandA...502..423S}.\\

Compared to LAEs, LBGs are on average larger (a few kpc), more massive ($10^{10}-10^{11}$ $\Msol$), highly star-forming ($10-1000$ $\Msol$ yr$^{-1}$), old ($\approx 1$ Gyr), and dusty (E(B-V)$\approx 0.3$) \citep{2000ApJ...544..218A,2001ApJ...559..620P,2001ApJ...562...95S,2002ARAandA..40..579G}. The rest frame UV sizes of LAEs and LBGs are similar at $z \sim 5$, and thereafter LBGs grow as $H(z)^{-1}$ while LAEs remain remarkably constant \citep{2012ApJ...750L..36M}. \citet{2001ApJ...562...95S} divided their sample of $z = 3$ LBGs in two categories, according to the age of the stellar populations. They find that the old sample ($\approx 1$ Gyr) contains stronger \lya emitters than the young galaxy sample ($<35$ Myr). While the two samples show rather similar stellar masses, older galaxies are more dusty and less star-forming. \citet{2001ApJ...562...95S} propose an evolutionary sequence for LBGs: young, starbursting galaxies quickly produce dust, extinguishing their \lya emission; later, when dust content has decreased (possibly ejected by Type II supernovae), \lya is seen in emission. While supported by the work of \citet{2010ApJ...711..693K} and GALEX observations of \citet{2012ApJ...751..139O} at $z \sim 0.3$, this picture contrasts with other observations reporting that \lya emission is preferentially found in younger galaxies \citep{2006ApJ...642L..13G,2007AandA...471..433P,2007ApJ...660.1023F,2007ApJ...667...49P,2011ApJ...738..136C}, and with the older idea that \lya emission is related to an early phase of galaxy formation \citep{1996Natur.382..231H,2007MNRAS.379.1589D}.

As an added complication, it has been shown that the relation between \lya emission and colour excess E(B-V) is  scattered. While \lya emission is often found to anti-correlate with dust extinction \citep{2003ApJ...588...65S}, young and dusty objects can also appear as \lya emitters \citep{2009ApJ...691..465F,2009AandA...494..553P,2010ApJ...720.1016Y}.

While \lya emission is an excellent tool for detecting high-redshift galaxies, questions remain about the nature of LAEs and their link with LBGs. The observed scatter in the relations between \lya emission and the physical properties of galaxies calls for a better understanding of the processes governing the production and the radiative transfer of \lya photons. Theoretical models are therefore crucial to decyphering the \lya signature of high-redshift galaxies.

\subsection{\lya Emission in Theory - Models of LAE galaxies}
\label{subsubsec:lya_ism}
The scattering of \lya photons through neutral hydrogen in galaxies implies that models must take into account, and thus can potentially constrain, the kinematics, structure and composition of the ISM of high-redshift galaxies.

\paragraph*{Importance of ISM kinematics:} \label{ism_kinem}

\begin{figure*}
\begin{minipage}{0.32\textwidth}
\includegraphics[width=\textwidth]{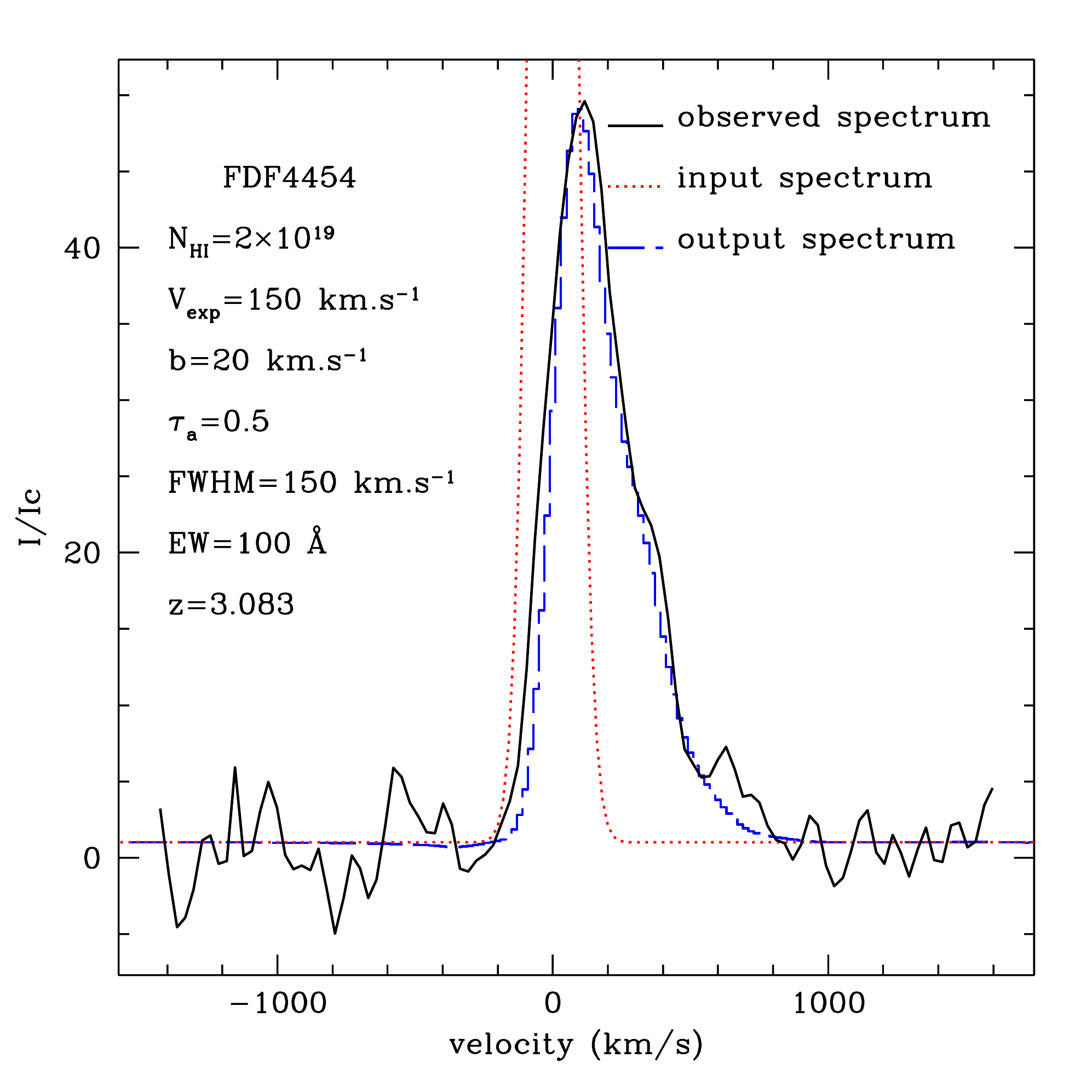}
\end{minipage}
\begin{minipage}{0.32\textwidth}
\includegraphics[width=\textwidth]{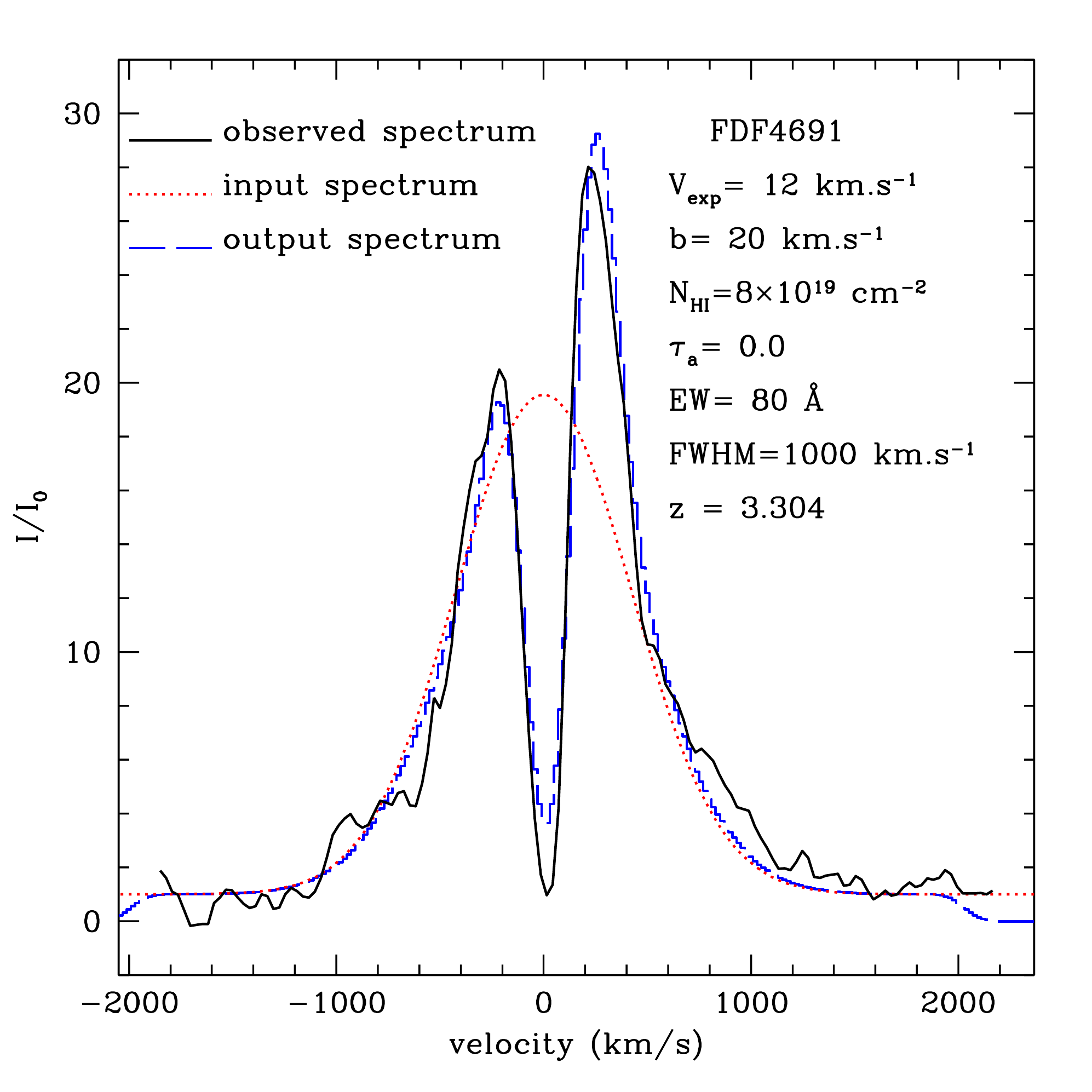}
\end{minipage}
\begin{minipage}{0.32\textwidth}
\includegraphics[width=\textwidth]{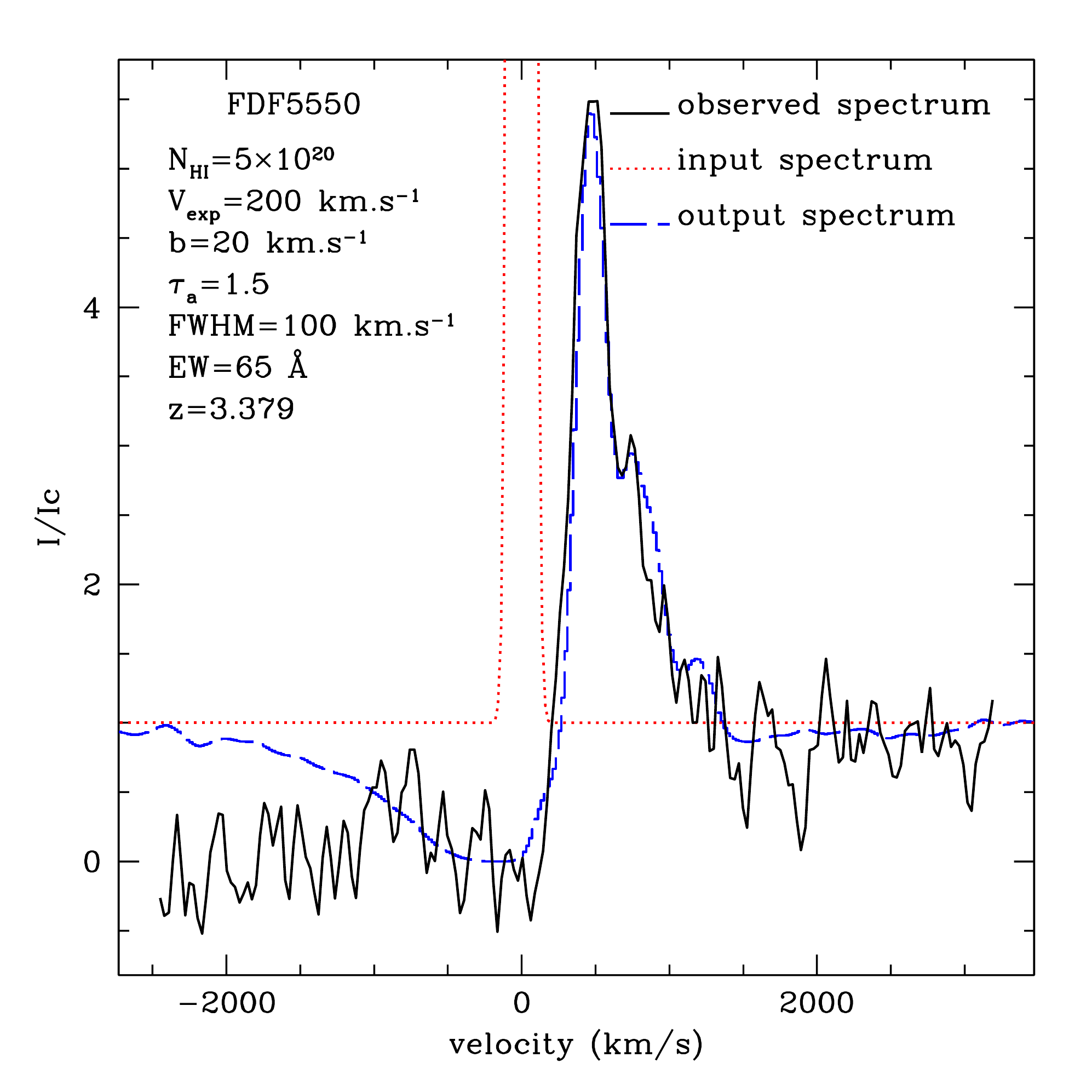}
\end{minipage}
\caption{Illustration of the variety of \lya line profiles observed at high redshift (black curves): asymmetric (left), double bump (middle) and P-Cygni (right). These medium-resolution ($R=2000$) spectra belong to the FORS Deep Field (FDF) sample presented by \citet{2007AandA...467...63T}. The figure appears in \citet{2008AandA...491...89V}, who show that these various profiles can be reproduced with numerical models (dashed blue line). The models compute the radiative transfer of an intrinsic \lya line (dotted red line) through an expanding shell. Six parameters are adjusted to fit the observed spectra: the width (FWHM) and equivalent width (EW) of the intrinsic line, the expansion velocity V$_{\rm exp}$, the \hi column density \nhi, the dust opacity $\tau$, and the thermal/turbulent velocity of the gas in the shell $b$. The values of the parameters are specified in each panel. Credit: \citet{2008AandA...491...89V}, reproduced with permission \copyright  ESO.}
\label{verh}
\end{figure*}

A striking feature of \lya emitting galaxies is the shape of their line profiles (see Figure \ref{verh}). Most show a broad, asymmetric red line; other spectral shapes are observed \citep[P-Cygni, double-peaked, damped, etc.;][]{2003ApJ...588...65S,2003ApJ...598..858M,2004AandA...416L...1T,2004ApJ...617..707D,2007AandA...467...63T}. Though IGM absorption might be partly responsible for the attenuation of the blue side of the \lya line, similar line shapes are observed in the local Universe \citep[e.g.][]{1998AandA...334...11K,2011ApJ...730....5H}. The diversity of \lya profiles is largely due to radiative transfer effects in the ISM of galaxies.

\citet{2000JKAS...33...29A,2001ApJ...554..604A,2002ApJ...567..922A}\footnote{See also  \citet{2002JKAS...35..175A,2003MNRAS.340..863A,2003JKAS...36..145A,2004ApJ...601L..25A}.} and \citet{2002ApJ...578...33Z} were amongst the first to use Monte Carlo radiative transfer codes to predict \lya spectra from simple models of protogalaxies. \citet{2006ApJ...649...14D} studied \lya transfer through collapsing clouds, representing the gas accretion experienced by primeval galaxies \citep[see also][]{2002ApJ...578...33Z}. They find a boost of the blue peak and a suppression of the red part of the \lya profile, at odds with what is usually observed.

Signatures of neutral gas outflows are detected ubiquitously at all redshifts in galaxies, with velocities ranging from a few tens to hundreds of \kmsec \citep[e.g.][]{2001ApJ...554..981P,2005ASPC..331..305M,2009ApJ...692..187W}. \citet{2003ApJ...588...65S} used more than 800 spectra of LBGs to construct a high-signal-to-noise composite spectrum. They find kinematic offsets between \lya emission and low-ionisation interstellar (LIS) absorption lines. The absorption lines associated with outflowing gas are blueshifted compared with the systemic redshift of the galaxies, such that $\Delta v_{\rm LIS}\approx -150$ \kmsec. In contrast, \lya emission appears to be redshifted, with a typical offset of $\Delta v_{\lya} \approx +360$ \kmsec. Comparable results have been reported in both LBGs and LAEs \citep{2010ApJ...717..289S,2011AAS...21733543M,2011ApJ...729..140F,2012ApJ...745...33K,2012ApJ...749....4B}. 

Doppler shifts induced by winds ease the escape of \lya, since the photons are scattered away from the \lya line centre \citep{1998AandA...334...11K,1999MNRAS.309..332T,2003ApJ...598..858M}. However, an anti-correlation between \lya EW and kinematic offset ($\Delta v_{\lya} - \Delta v_{\rm LIS}$) is reported \citep{2003ApJ...588...65S,2013ApJ...765...70H}: stronger \lya emission is found in cases where the velocity shift is smaller. Although these results are still debated \citep{2008AandA...491...89V,2012ApJ...749....4B}, it may indicate that the enhancement of dust extinction due to resonant scattering in \hi remains the main driver of \lya escape from galaxies. Alternatively, \citet{2006MNRAS.373..571F} interpret this trend as the time evolution of a galactic wind: after the starburst, the outflow slows down and the covering factor is reduced, which favours the escape of \lya photons.

\begin{figure}[t]
\centerline{\includegraphics[width=8.1cm,height=6.5cm]{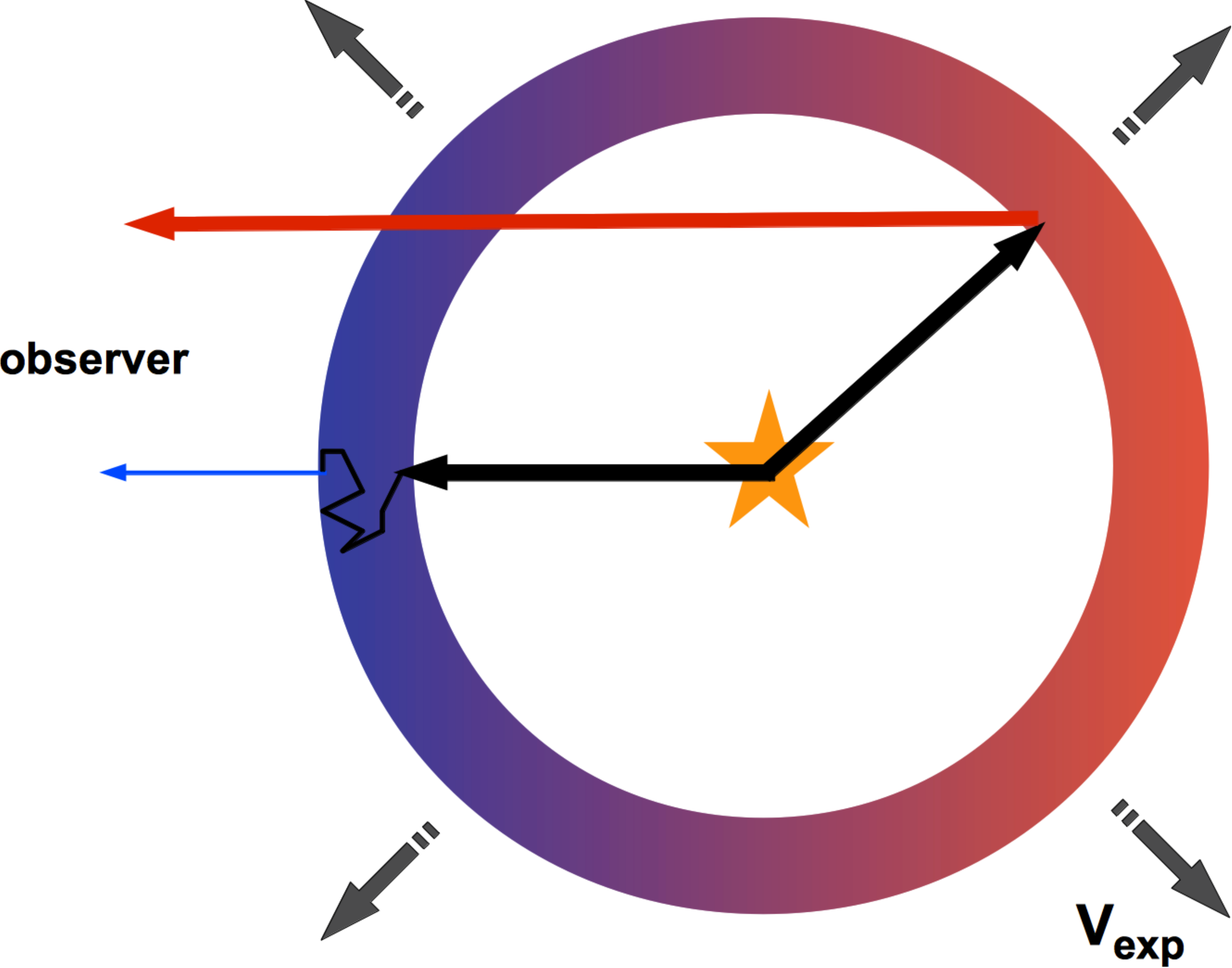}}
\caption{Illustration of an expanding shell model, as an ideal representation of a supernova-driven gas outflow around a source of \lya emission (i.e. a star forming region: yellow star). \lya photons are emitted isotropically at the centre of the shell (solid black arrows), which is expanding at speed V$_{\rm exp}$. Photons emitted directly towards the observer (left) may experience strong extinction due to resonant scattering in the gaseous dusty shell. Photons emitted towards the receding shell (right) can \textit{backscatter} towards the observer. These photons are redshifted relative to the front of the shell (and the observer), hence they have greater probability of escaping.}
\label{fig:shell}
\end{figure}

Radiative transfer through an expanding shell is also invoked to explain the line profiles of \lya emitting galaxies \citep{2003MNRAS.340..863A,2006AandA...460..397V,2008AandA...491...89V}. In this model (Figure \ref{fig:shell}), an isotropic  \lya source is located at the center of a thin, homogeneous, and spherical shell of \hi gas mixed with dust. The expanding shell mimics the gas outflow generated by strong stellar winds and supernovae during starburst events, associated with photoionisation and subsequent \lya emission \citep{1977ApJ...218..377W,1985Natur.317...44C,2006MNRAS.373..571F,2009MNRAS.396L..90N}. Using different values of the shell parameters (expansion velocity $V_{\rm exp}$, \hi column density, dust opacity and thermal/turbulent velocity of the gas within the shell), \citet{2008AandA...491...89V} are able to reproduce the wide variety of observed \lya profiles \citep[see also][]{2008AandA...480..369S,2012MNRAS.420.1946L}. As an example, we show in Figure \ref{verh} three \lya profiles of LBGs at $z = 3$ from \citet{2007AandA...467...63T} that are successfully fitted by the shell model of \citet{2008AandA...491...89V}. Typical red asymmetric profiles arise from photons \textit{backscattered} from the receding side of the shell (red arrow on Figure \ref{fig:shell}), while photons directly emitted towards the observer will be preferentially destroyed by dust due to a smaller Doppler shift (blue arrow on Figure \ref{fig:shell}).

Shell models provide a simple and physically motivated interpretation of \lya spectra, but they remain very idealised. In particular, the emitted spectra does not depend on the spatial extent of the \hi entrained in the wind. The observations of \citet{2010ApJ...717..289S,2011ApJ...736..160S} indicate clumpy outflows and the presence of scattering gas well outside the ISM of LAEs. More realistic models of the impact of a galactic wind on the circumgalactic medium of high-redshift galaxies are required, and are discussed in Section \ref{Ss:LyaCGM}.

\paragraph*{Importance of ISM structure and composition:} \label{ism_struct}

\citet{2006MNRAS.367..979H} simulated the transfer of \lya photons through clumpy media. Since the analytic model of \citet{1991ApJ...370L..85N}, it has been argued that a multi-phase ISM could explain the abnormally high observed \lya EW, as previously discussed in Section \ref{statprop}. Of particular importance is the distribution of dust, as interactions with dust can destroy \lya photons\footnote{It should be remembered that while \hi scattering does not destroy \lya photons, it can scatter them below the surface brightness limit of observations, so that \lya photons disappear in the noise, so to speak.}. \citet{1991ApJ...370L..85N} models an ISM in which dust is locked in dense \hi clouds that are embedded in a diffuse, dust-free, inter-cloud medium (ICM). \lya and UV continuum photons are generated at the center of a plane-parallel slab in the diffuse phase. While (non-resonant) continuum radiation passes through the clouds and is attenuated by dust, \lya photons are scattered off the surface of the clouds as a result of their high probability of interaction with hydrogen. The resulting \lya boost is called the \textit{Neufeld effect}, which \citet{2011ApJ...733..117F} quantify with the parameter $q$:
\begin{equation} \label{eq_q}
q=\tau_{{\rm Ly}\alpha} / \tau_{\rm cont} ~ ,
\end{equation}
where $\tau_{\lya}$ is the dust opacity of \lya photons, and $\tau_{\rm cont}$ is the dust opacity of UV-continuum photons. In the simplest case, one expects $q \ll 1$ for a clumpy ISM, and $q \gg 1$ for a homogeneous ISM. Note, however, the simulations of  \citet{2012arXiv1211.2833L}, discussed below. 

\begin{figure*}[t]
\centering
\begin{minipage}{0.46\textwidth}
\includegraphics[width=\textwidth]{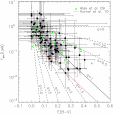}
\end{minipage}
\hspace{0.1cm}
\begin{minipage}{0.46\textwidth}
	\caption{Observed relation between the \lya escape fraction and the colour excess E(B-V) \citep{2011ApJ...736...31B}. The black circles and the green triangles are the measurements of \citet{2011ApJ...736...31B} and \citet{2009AandA...506L...1A} for Lyman-alpha emitters (LAE) at $z = 2-4$ and $0.3$ respectively. The escape of \lya photons is anti-correlated with dust extinction, despite a strong dispersion. The black lines show the expected correlation for different clumpiness parameters $q$. The $q=1$ model, for which \lya and continuum photons suffer a similar dust extinction, describes well the median distribution of the data. It suggests that the impact of \lya resonant scattering is partly suppressed, certainly due to geometry or kinematics effects. At given E(B-V), Lyman Break galaxies display lower f$_{\rm esc}$ than LAEs on average \citep[red solid line;][]{2010ApJ...711..693K}. Reproduced by permission of the AAS.}
	\label{blanc}
\end{minipage}
\end{figure*}

\citet{2011ApJ...733..117F} measured $q$ using the ratio of emission line fluxes for a sample of 12 GALEX LAEs at $z \sim 0.3$. They find that most of the sample lies at $q \sim 1-3$, indicating that the ISM neither enhances nor seriously attenuates \lya. Even for those LAEs with $q < 1$ they note an important \emph{caveat}, first discussed by \citet{2009ApJ...704L..98S}: if the dust is clumpy, then the observed line ratios do not follow the simple $ \exp \{-\tau_{\rm cont}\}$ law. Further, \citet{2011ApJ...736...31B} showed from the HETDEX pilot survey that the \lya escape fraction, estimated from \lya and (dust-corrected) UV luminosities, anti-correlates with dust extinction, as shown in Figure \ref{blanc} \citep[see also][]{2013arXiv1308.6577A}. The black circles correspond to the $z = 2-4$ sources observed by \citet{2011ApJ...736...31B}. The dearth of $q\ll1$ objects indicates that the Neufeld effect is not a common process in LAEs. In spite of some dispersion, the cloud of points is well fitted by a $q=1$ model (black solid line in Figure \ref{blanc}) which suggests that \lya and continuum photons suffer a very similar dust extinction. The green triangles indicate the objects found by \citet{2009AandA...506L...1A} at $z = 0.3$, and agree fairly well with the data of \citet{2011ApJ...736...31B}. The red solid line is the relation reported by \citet{2010ApJ...711..693K} for LBGs at $z = 3$. It lies slightly below the $q=1$ relation measured for LAEs at the same redshift, and is consistent with the idea that LAEs are a high f$_{\rm esc}$ subset of the LBG population.

The simulations of \citet{2006MNRAS.367..979H} suggest that the Neufeld effect will be strong if most of dust grains reside in optically thick \hi clouds. Recently, \citet{2012arXiv1211.2833L} and \citet{2013arXiv1302.7042D} revisited this issue, finding that a significant boost requires a relatively rare confluence of conditions: high metallicity, little outflow, very high cloud density, very low density of \hi in the ICM, and \lya and UV photons that originate from regions deprived of neutral gas. They conclude that an EW boost is unlikely to occur in real galaxies, with $q \ll 1$ no guarantee of a homogeneous ISM. The preferential escape of \lya compared with UV continuum is also not seen in the \lya RT hydrodynamic simulations of \citet{2012MNRAS.424..884Y} and \citet{2012AandA...546A.111V}.

Although many numerical \lya RT experiments have been published, only a few have focused on transfer through interstellar media in state-of-the-art hydrodynamic simulations. As they include a range of relevant physics, like gas cooling, star formation, feedback, metal enrichment or dust formation, such studies are essential to gaining insight into the complex mechanisms altering the \lya line. We emphasise that the \emph{\lya escape fraction}, as calculated by simulators, usually refers to the ratio of emitted to escaping \lya photons. Photons are lost only to absorption by dust. Observed escape fractions can also lose photons to scattering by the CGM/IGM, and beneath surface brightness detection limits. The observed quantity is termed ``effective escape fraction'' by \citet{2013MNRAS.435.3333D}.

\citet{2009ApJ...704.1640L} analysed the transfer of \lya photons in $z \approx3$ galaxies extracted from a cosmological N-body/hydrodynamical simulation. Although their sample contains only nine galaxies, there is a clear trend between \lya escape fraction and the mass of the host halo.  f$_{\rm esc}$ is of order unity for galaxies in $10^9-10^{10}\Msol$ haloes, but only a few percent for  $10^{11}-10^{12}\Msol$. There are two complementary reasons for this trend. First, galaxies in smaller haloes form fewer stars, hence limiting the production of metals/dust. Second, the shallower gravitational potential wells of low-mass galaxies makes supernovae feedback more efficient at disturbing the ISM, which could help \lya photons to escape from galaxies in smaller haloes. 

Unlike most of the other models who assume density/temperature cuts to determine the state of the gas (ionized vs neutral), the simulations of \citet{2012arXiv1209.5842Y} include the transfer of ionizing radiation. It allows them to compute (i) the recombination of hydrogen that powers the \lya emission, and (ii) the ionisation state of the ISM. However, the resolution of the zoom-in region investigated by these authors ($250$h$^{-1}$ comoving pc) is not sufficient to accurately model the ISM on small scales. They find that the median f$_{\rm esc}$ is almost constant ($\approx 25\%$) between $z = 0$ and 3, and increases towards higher redshifts to reach a value of about $90\%$ at $z = 10$, mainly due to galaxies being increasingly metal and dust poor at high redshift. They interpret the trend at low redshift as a mutual cancellation of the effects of increased metallicity and dust, and decrease ISM gas fraction. At all redshifts, the values of  f$_{\rm esc}$ are nonetheless very scattered around the median. \citet{2012arXiv1209.5842Y} find only weak correlations between \lya escape fraction and the galaxy properties. f$_{\rm esc}$ anti-correlates with host halo mass, galaxy mass, dust mass, but the dispersion in these relations is again quite significant. A clearer trend is that metal-enriched galaxies have lower f$_{\rm esc}$, in good agreement with the observations of \citet{2009AandA...506L...1A}, \citet{2010ApJ...711..693K}, and \citet{2011ApJ...736...31B}, discussed above.

\begin{figure}[t]
\includegraphics[width=7.5cm,height=3.3cm]{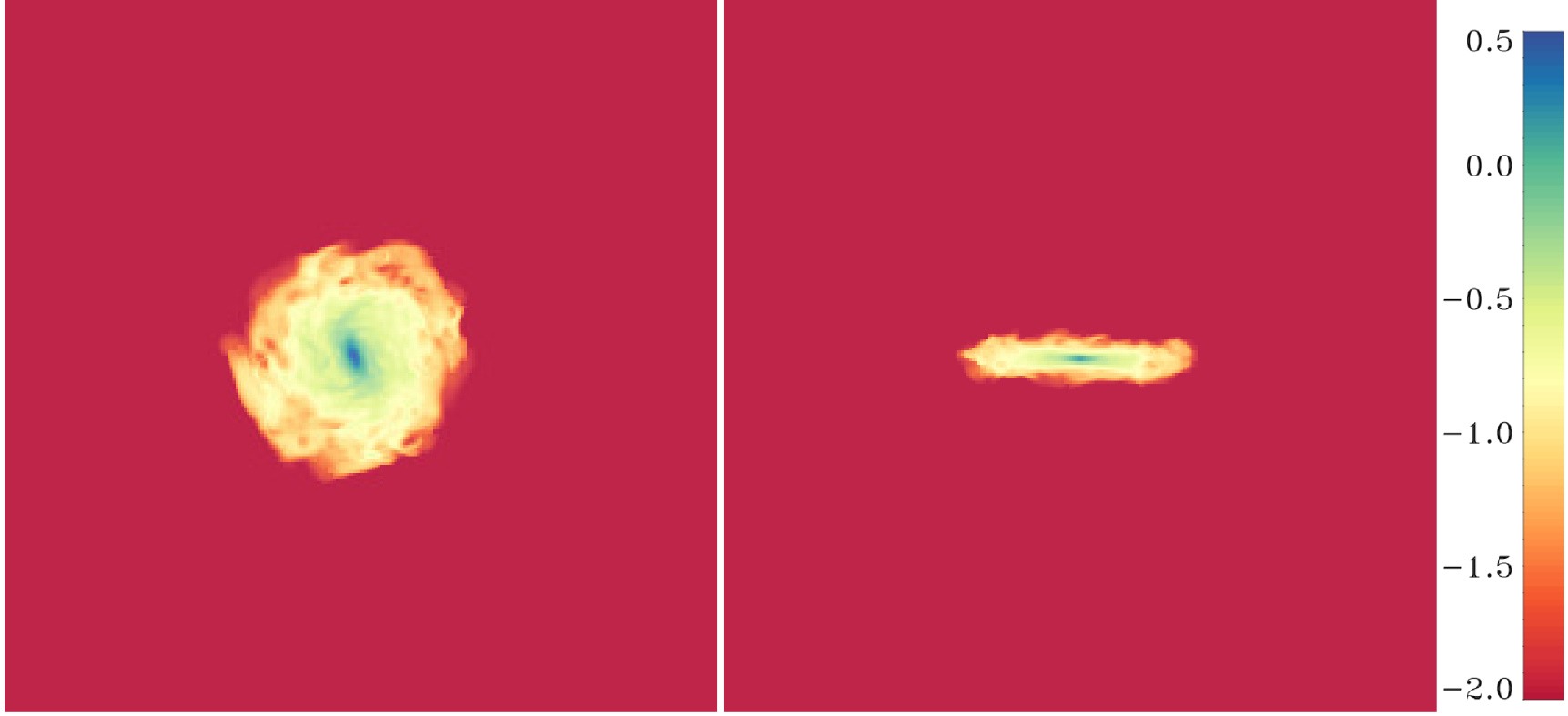}
\includegraphics[width=7.5cm,height=3.3cm]{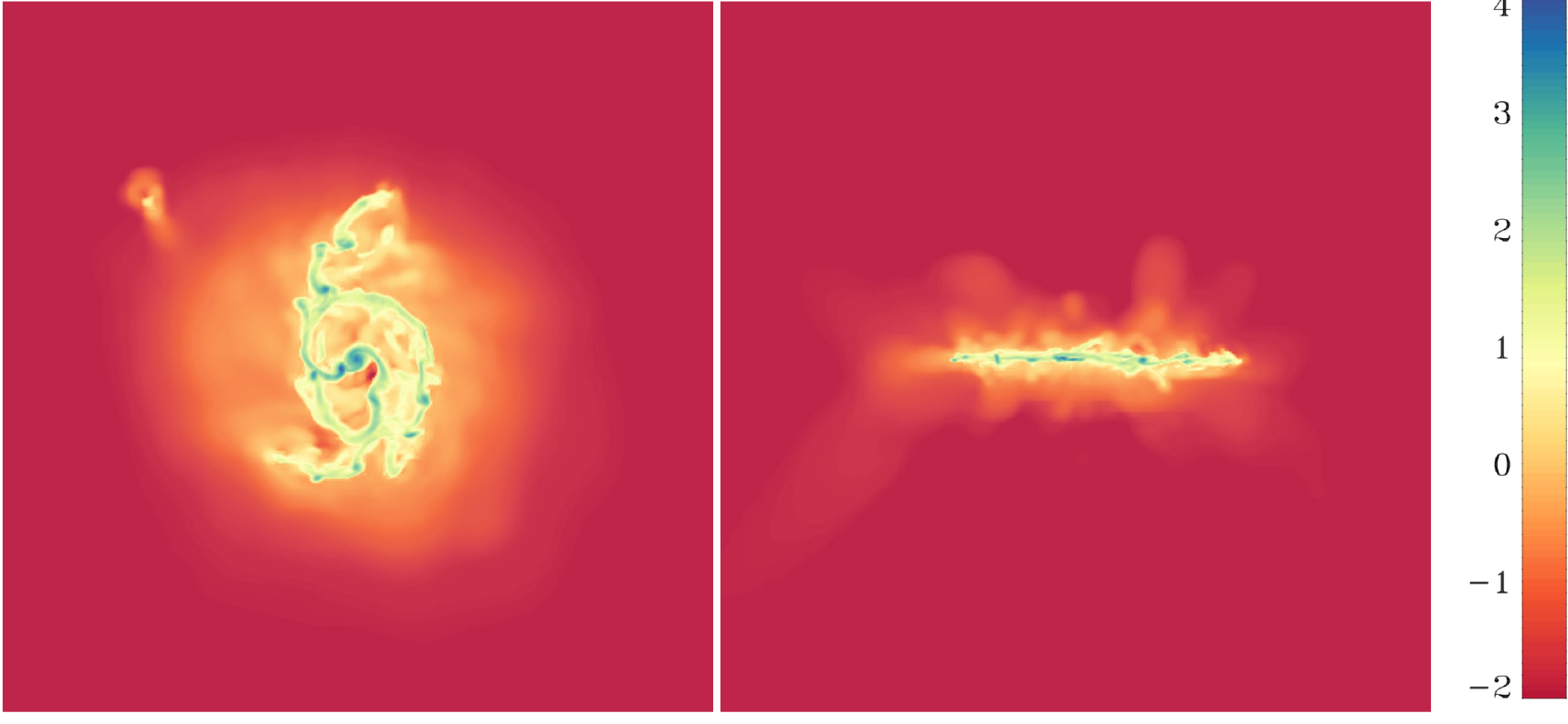}
\caption{Density maps of gas distribution for the two isolated galaxies, viewed face-on (left) and edge-on (right), used by \citet{2012AandA...546A.111V} to study \lya transfer in the interstellar medium (ISM). The colour scales with gas density (arbitrary log units). \textit{Top}: Gas is allowed to cool down to $10^4$ K in the ISM, leading to a rather smooth matter distribution. \textit{Bottom}: Extra energy loss due to metal cooling enables the gas temperature to drop down to 100K. In this more realistic case, smaller scale structures appear.  \citet{2012AandA...546A.111V} post-process the simulations with a \lya transfer algorithm and find that the \lya escape fraction is ten times lower in the inhomogeneous galaxy than in the smoother one. Credit: \citet{2012AandA...546A.111V}, reproduced with permission \copyright ESO.}
\label{verh12map}
\end{figure}

\citet{2012AandA...546A.111V} showed that the small-scale structure of the ISM has a strong effect on the escape of \lya photons. They performed \lya radiative transfer within two simulations of idealized dwarf galaxies at different resolution ($\sim 20$ and 150 pc respectively), run with the adaptive-mesh refinement code RAMSES \citep{2002AandA...385..337T}. Gas density maps of the two simulations are shown in Figure \ref{verh12map}. In the higher-resolution galaxy, the ISM can cool down to $100$K via metal cooling and fragment into smaller clumps in the disk. The \lya escape fraction  f$_{\rm esc}$ is as low as $5\%$ in this case, whereas it goes up to $50\%$ for the lower-resolution (with smoother gas distribution) galaxy. \citet{2012AandA...546A.111V} also report that the \lya line profile, escape fraction, and equivalent width vary with respect to the angle at which the galaxy is viewed; especially the latter is higher in the face-on direction. As pointed out by previous studies \citep{1993ApJ...415..580C,2009ApJ...704.1640L,2010ApJ...716..574Z,2011MNRAS.416.1723B}, orientation effects may introduce a bias in LAE observations where a \lya EW detection threshold is set, which could be partly responsible for the scatter in properties of LAEs and LBGs.

While our knowledge of the ISM of high-redshift galaxies and its effect on the \lya line transfer is still incomplete, the understanding of the mechanisms at play is improving. The detailed analysis of observational data and the comparison with realistic theoretical models has resulted in better constraints on the emission and escape  from the ISM of \lya photons. The gas/dust structure and kinematics appear to be key drivers to shape the \lya emission line, and more detailed descriptions of neutral outflows \citep{2012MNRAS.426..140D}, of the ionisation state of the ISM \citep{2012arXiv1211.0088Y}, and of its molecular hydrogen content \citep{2012MNRAS.421.3266V} will help refining the models.

\subsection{\lya Emission and the CGM} \label{Ss:LyaCGM}

The extremely large cross-section for \lya scattering off \hi makes \lya spectra sensitive to gas \emph{outside} of galaxies. In a hierarchical cosmology, the environment of a galaxy plays a crucial role in its formation. The CGM, introduced in Section \ref{S:DLA_CGM}, potentially includes inflowing cold streams, hot and shocked infalling gas, satellite galaxies, feedback from AGN, and outflowing galactic winds and recycling galactic fountains of hot wind fluid and entrained cool star-formation enriched gas.

%\citet{2010ApJ...717..289S} studied the CGM at $z > 2$ within $\sim 125$ kpc of Lyman Break galaxies using a sample of 512 close galaxy pairs. These galaxies predominantly show \lya emission (when present) that is strongly redshifted ($\Delta v_{\lya} \approx +445 \kmsec$), while the strong interstellar  absorption lines are strongly blueshifted ($\Delta v_\textrm{IS} \approx -160 \kmsec$), consistent with the presence of a galactic wind (Section \ref{subsubsec:lya_ism}). Absorption from \hi and metals is observed in the CGM, with absorber equivalent width declining with impact parameter. A very rapid decline in equivalent width is observed at large distances, beginning at 70-90 kpc for all transitions except \lya, which remains strong out to 250 kpc. Similarly, \citet{2012ApJ...750...67R} observe that \hi absorbers within $\sim 100$ kpc of galaxies at $z \sim 2.5$ have $10^3$ times higher median \nhi than random IGM absorbers; even at $1000$ kpc, \hi absorbers have a median \nhi twice as high as the IGM.

What light does \lya emission throw on the CGM? \lya emission is an ideal tracer of the CGM, being copiously produced by star-forming galaxies and strongly scattered by \hi. Further, \lya is the strongest emission line from cooling gas in the universe. \citet{2001ApJ...562..605F} estimate that \lya emission accounts for 57\% of cooling radiation, with just 2\% coming from bremsstrahlung. Note, however, that they assume primordial composition gas. More recent calculations by \citet{2013arXiv1301.5330B}, which take into account $\sim$ 2000 emission lines from 11 elements, show that while \lya carries about 50\% of the energy in H and He lines, it carries 13\% of the total energy emitted by diffuse gas ($n_H < 0.1$ cm$^{-3}$) at $z = 2$. Most of the diffuse emission comes from dense ($n_H \sim 10^{-3} - 10^{-1}$ cm$^{-3}$), cool ($\sim 10^4 - 10^{4.5}$ K), metal rich (0.1-1 solar) gas, 80\% in emission lines and 20\% in the continuum \citep[see Figure 14 of][]{2013arXiv1301.5330B}.

Spatially-resolved \lya spectra are sensitive to the distribution, kinematics and dust content of CGM gas. As discussed in Section \ref{subsubsec:lya_ism}, the vast majority of \lya emitting galaxies at high redshift have spectral lines shifted to the red by hundreds of \kmsec, in agreement with models of \lya radiative transfer through an expanding galactic wind \citep{2006AandA...460..397V}. Spectra alone, however, cannot constrain the spatial extent of the gas and so cannot say whether the gas simply puffs up the ISM or is blasted right out of the halo. We require ultra-deep, spatially-resolved observations.

\citet{2008ApJ...681..856R} performed an ultra-deep spectroscopic search for low surface brightness \lya emitters at redshift $z \sim 3$. A 92-hour long exposure with the ESO VLT FORS2 instrument yielded a sample of 27 faint line emitters with fluxes of a few times $10^{-18} $ erg s$^{-1}$cm$^{-2}$. Based on their large number density, the sample is likely dominated by \lya emitters, rather than low-redshift interlopers. A number of lines of evidence lead \citet{2008ApJ...681..856R} to claim that these emitters are the host galaxies of DLAs:
\begin{itemize}  \setlength{\itemsep}{-2pt}
\item Both must host extended, optically thick neutral hydrogen. 
\item The incidence rate ($\dd N / \dd z$) for the emitters and for DLAs is consistent. The combination of the large sizes and high space density of the emitters mean that they can account for the high incidence rate of DLAs. 
\item Both populations have a low star formation rate, which would explain the low success rate for direct searches for the counterparts of DLAs, and the low observed metallicity of DLAs. 
\item Both populations have low dust content, assuming that a high dust content would extinguish the line.
\item If the large sizes of the emitters are due to radiative transfer effects  \citep[which seems likely in light of the strict upper limits on \emph{extended} star formation in DLAs derived by][]{2006ApJ...652..981W}, then the emitters must contain significant amounts of neutral hydrogen. The majority of \hi at these redshifts resides in DLAs.
\end{itemize}

\citet{2009MNRAS.397..511B,2010MNRAS.403..870B} used an analytic model of neutral hydrogen in dark matter haloes and radiative transfer simulations to simultaneously reproduce the observed properties of DLAs and the faint \lya emitters of \citet{2008ApJ...681..856R}. These emitters are hosted by $10^{9.5} - 10^{12} \Msol$ haloes, with little contribution from haloes with virial velocities $\lesssim 50 - 70 \kmsec$. Their observed sizes are due to centrally concentrated star-formation at a few tenths $\Msol $ yr$^{-1}$ producing \lya photons that scatter through the surrounding \hi and are observable \citep[at the flux limit of][]{2008ApJ...681..856R} to $\sim 30-50$ kpc.

 \citet{2011ApJ...736..160S} used a sample of 92 continuum-selected galaxies at $\langle z \rangle = 2.65$ to study very faint \lya emission surrounding LBGs. The CGM of LBGs had previously been studied in absorption: \citet{2010ApJ...717..289S} studied the CGM at $z > 2$ within $\sim 125$ kpc of Lyman Break galaxies using a sample of 512 close galaxy pairs. These galaxies predominantly show \lya emission (when present) that is strongly redshifted ($\Delta v_{\lya} \approx +445 \kmsec$), while the strong interstellar  absorption lines are strongly blueshifted ($\Delta v_\textrm{IS} \approx -160 \kmsec$), consistent with the presence of a galactic wind (Section \ref{subsubsec:lya_ism}). Absorption from \hi and metals is observed in the CGM, with absorber equivalent width declining with impact parameter. A very rapid decline in equivalent width is observed at large distances, beginning at 70-90 kpc for all transitions except \lya, which remains strong out to 250 kpc. Similarly, \citet{2012ApJ...750...67R} observe that \hi absorbers within $\sim 100$ kpc of galaxies at $z \sim 2.5$ have $10^3$ times higher median \nhi than random IGM absorbers; even at $1000$ kpc, \hi absorbers have a median \nhi twice as high as the IGM.
 
In emission, \citet{2011ApJ...736..160S} stacked UV continuum and \lya line images to reach a surface brightness threshold of $\sim 10^{-19} \ergsca$. While the UV continuum drops with an exponential scale length of 3-4 kpc (proper) and is undetectable beyond 24 kpc, Lyman alpha emission is $\sim 5-10$ times more extended, with a scale length of $20-30$ kpc. \lya remains detectable out to $\sim 80$ kpc, comparable to the virial radius of the galaxies. Such signal is missed by NB observations and slit spectroscopy of individual objects, so that the \lya flux of LAEs are underestimated. They argue that this emission arises from resonantly scattered \lya photons in the CGM.
 
 \begin{figure*}
 \centering
 \begin{minipage}{0.45\textwidth}
 \includegraphics[width=\textwidth]{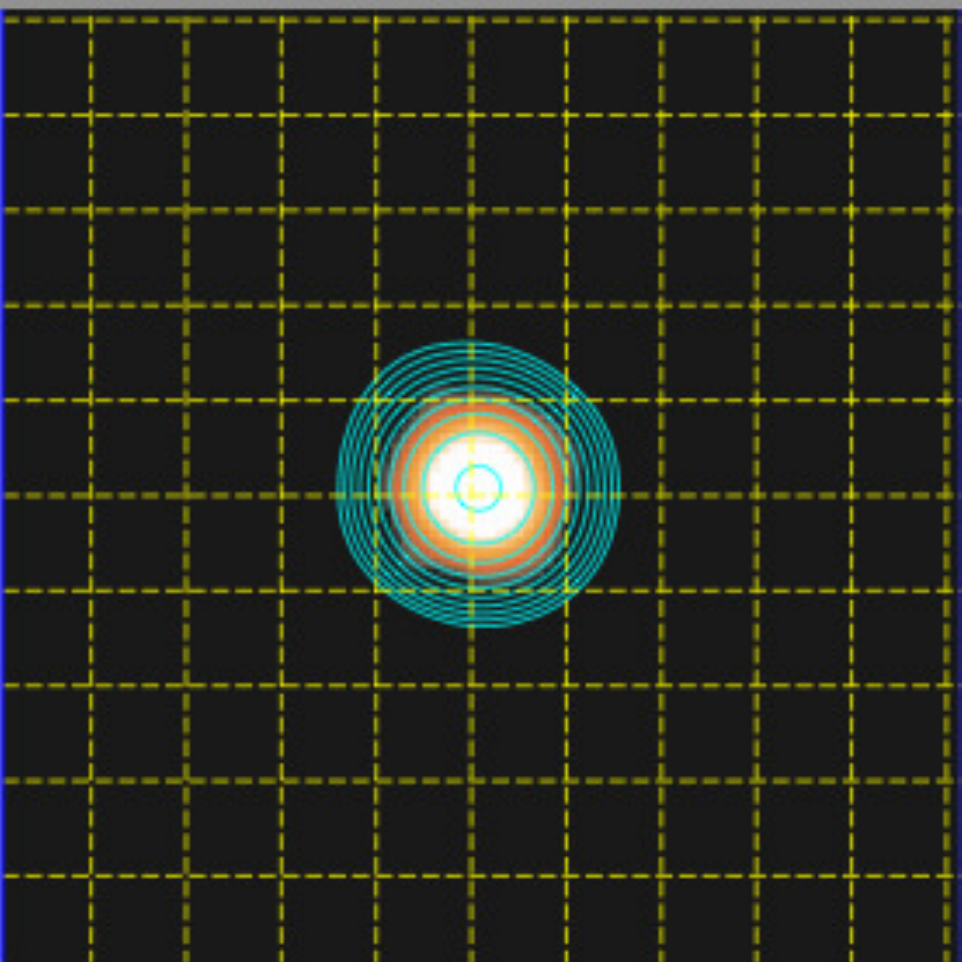}
 \end{minipage}
 \hspace{0.5cm}
 \begin{minipage}{0.45\textwidth}
 \includegraphics[width=\textwidth]{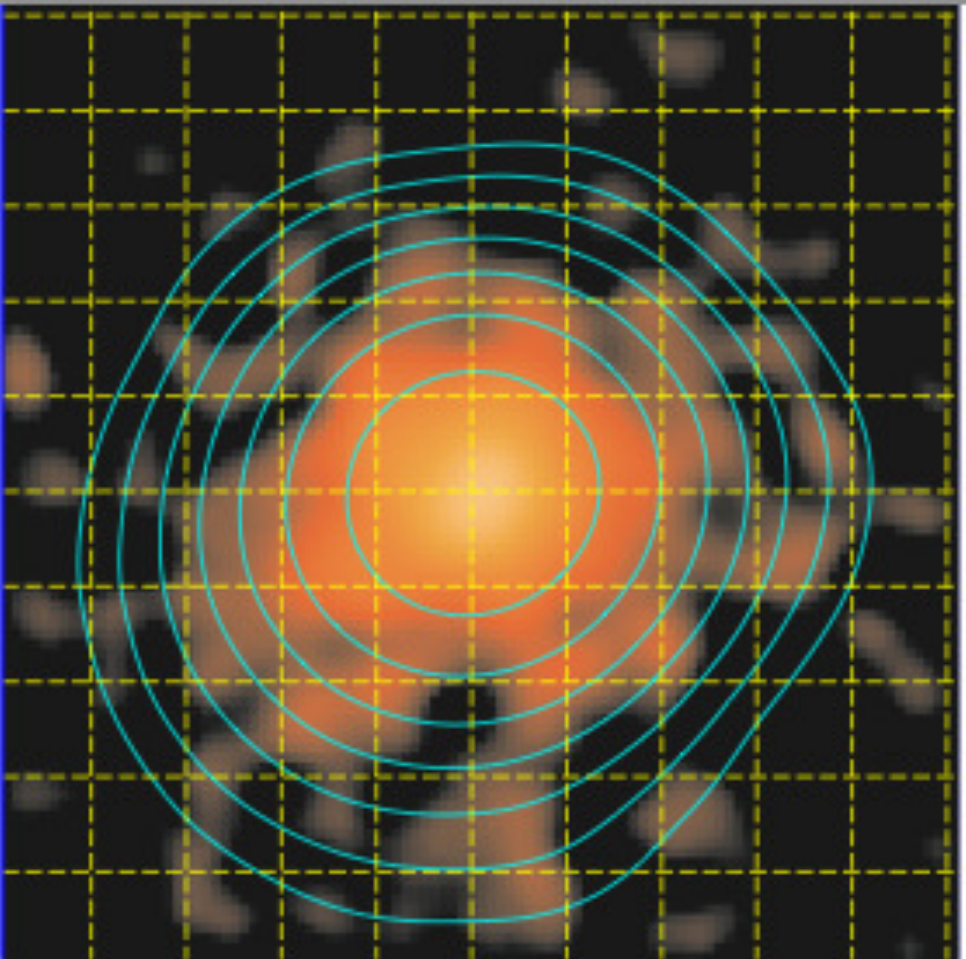}
 \end{minipage}
 \caption{Stacked images from 92 continuum-selected LBGs from \citet{2011ApJ...736..160S}. The regions shown are $20''$ ($\approx 160$ physical kpc at $z = 2.65$) on a side, with a grid spacing of $2''$. \emph{Left}: scaled far-UV continuum image, drawn from three independent fields. Right: the continuum-subtracted, stacked \lya image for the same sample of galaxies. In both panels, the contours are logarithmically spaced in surface brightness with the lowest contour shown at $\approx 2.5 \times 10^{-19} \ergsca$. The \lya emission is seen to extend to $\sim 80$ physical kpc, significantly beyond the continuum emission. Figures from \citet{2011ApJ...736..160S}; ; reproduced by permission of the AAS.}
 \label{fig:Steidel}
 \end{figure*}

Other groups have also attempted to study the CGM through \lya stacking. \citet{2012MNRAS.425..878M} stacked 2128 \lya emitters and 24 protocluster LBGs at $z = 3.1$, observed with a narrow-band filter on Subaru Suprime-Cam. They too found extended \lya haloes down to surface brightness limits of $\sim 10^{-18} \ergsca$, with a trend for larger scale lengths for LAEs in higher density (Mpc) environments. \citet{2013ApJ...776...75F}, however, sound a warning: \emph{surface photometry at very low flux levels is treacherous}, sensitive to large-scale flat-fielding and point-spread-function issues which may not have been taken into account in previous works. They observe smaller ($5-8$ kpc) scale \lya haloes at $z = 3.1$ and no evidence for haloes at $z = 2.1$, down to $6-10 \times 10^{-19} \ergsca$. \citet{2013ApJ...773..153J} find no conclusive evidence of \lya haloes around $z = 5.7$ or 6.5 in a stack of 40 LAEs, reaching $1.2 \times 10^{-19} \ergsca$. Thus, LAEs seem to have significantly fainter \lya haloes than Steidel's LBGs. This is consistent with the order of magnitude smaller star formation rates and lower mass host haloes of LAEs, implying a lower density environment (\emph{a la} Matsuda).

The scattering of \lya photons in the CGM was investigated by \citet{2011MNRAS.416.1723B}, who modelled the scattering of \lya photons emitted by a central source through the ISM and CGM of galaxies drawn from an SPH cosmological simulation. The simulations reveal a \lya halo extending to scales comparable to the virial radius, with an extent that depends on the wind prescription used in the simulation. The more efficient feedback implementations result in reduced column densities at the centre and therefore reduced diffusion in frequency space and narrower spectral profiles. The simulated spectra display a more prominent blue peak than most observed LAEs, which may be due to the chosen galaxies being smaller and having lower star-formation rates than typical LBGs. This  conclusion is consistent with the observations of multiply-peaked \lya emission by \citet{2012ApJ...745...33K}, who note that such systems have systematically slower outflows (as measured by absorption lines). The \lya emitting halo around star-forming galaxies is predicted to extend well beyond the CGM, though at very low surface brightness \citep{2011ApJ...739...62Z,2012MNRAS.424.2193J}.

Because a first principle calculation of the distribution and kinematics of cold gas in galactic outflows are currently not feasible, \citet{2012MNRAS.424.1672D} studied a phenomenological model of a galactic wind, wherein large numbers ($\sim 10^5 - 10^6$) of cold, dense dusty spherical clumps of \hi are outflowing isotropically in pressure equilibrium with a hot component. Following the simple model of \citet{2011ApJ...736..160S}, they consider a wind which accelerates ($a \propto r^{-\alpha}, ~ \alpha > 1$) to $\sim 800$ \kmsec at large distances ($\sim 100$ kpc). The full parameter space of the model is explored using a Monte Carlo Marcov Chain (MCMC) simulation, constraining the parameters using absorption line data. A suite of constrained models is used in a radiative transfer calculation to predict the \lya emission. The emitters are too faint and too centrally concentrated to match observations because the clumps at large distances are too few and too fast to significantly scatter escaping \lya photons. Models in which gravity decelerates the clumps beyond $\sim 10$ kpc and in which the outflow is bipolar are more successful in reproducing both the absorption and emission line data. 

\citet{2012AandA...540A..63N} observed \lya, [O\textsc{ii}] and H$\alpha$ emission from a DLA, inferring a large SFR ($\sim 25 \Msol$ yr$^{-1}$) from a low mass galaxy ($M_\ro{halo} \sim 10^{10} \Msol$). The \lya emission is double-peaked and spatially extended, which would suggest the static Neufeld solution (see Section \ref{SS:lyaRT} and Figure \ref{fig:sphtest}), except that the blue and red peaks arise from spatially distinct regions, separated by a few kpc and on opposite sides of the star-forming region. They successfully reproduce the observed properties of the system with a model in which two ionized jets expel gas through a spherical distribution of cold, neutral infalling clouds. Similarly, \citet{2012MNRAS.427.1973C} and \citet{2013MNRAS.433.3091K} are able to reproduce observed \lya emission spectrum using a model of a dusty, clumpy outflow at $\sim$100 \kmsec. These models use the MoCaLaTA code of \citet{2009ApJ...704.1640L}, and show the need for data other than the \lya spectrum to constrain the RT calculation.

In light of the faintness of the CGM, \citet{2005ApJ...628...61C} investigated the illumination of extragalactic \hi by a nearby quasar. This provides a method of observing otherwise undetectable gas in the CGM and in small, dark galaxies. The observations of \citet{2012MNRAS.425.1992C}, using VLT-FORS to perform deep narrow-band survey for Lyα emission within $\sim$Mpc of a hyperluminous QSO at $z = 2.4$, reveal a population of emitters whose equivalent width distribution, luminosity function and the average luminosity versus projected distance are consistent with detailed radiative-transfer simulations of quasar fluorescence. Their large EW $(> 800$ \AA)  rule out internal star formation as the source of \lya. Some of the emitters resemble extended filaments, compatible with the expectations for circumgalactic cold flows, though this interpretation is not unique. Very recently, \citet{2014Natur.506...63C} reported observations of a filamentary, 460 kpc \lya emission region that includes a radio-quiet quasar but extends well beyond the virial radius. This object is plausibly the cosmic web, lit up in \lya by the quasar, a conclusion supported by cosmological hydrodynamical simulations combined with \lya and ionisation radiative transfer.

\citet{2013ApJ...766...58H} simultaneously analyzed absorption and emission from \hi in the CGM of quasars using close projected quasar pairs. While extended \lya fuzz is detected on 50kpc scales in $\sim 30$\% of the sample (possibly up to $50-70$\% when the effects of the single slit are accounted for), it is significantly less emission than would be naively expected given the covering fraction of optically thick \hi absorbers and the ionizing flux from the quasar. This is argued to be evidence of anisotropic quasar emission.

\paragraph*{\lya Blobs (LABs):}
There is another class of Lyman alpha emitter at high redshift, which are extended ($\sim 10 - 150$ kpc), extremely luminous ($L_{\lya} \sim 10^{43} - 10^{44}$ \ergs) and rare, with a number density of $\sim 10^{-3.8}$ comov.Mpc$^{-3}$ \citep{2009MNRAS.400.1109D}, as compared to $\sim 10^{-2.7}$ comov.Mpc$^{-3}$ for the population of \lya emitters in \citet{2005MNRAS.359..895V}. These are the imaginatively-named Lyman Alpha Blobs (LABs) \citep{1999MNRAS.305..849F,1999AJ....118.2547K,2000ApJ...532..170S,2001ApJ...554.1001F,2004ApJ...602..545P,2004AJ....128..569M,2004ApJ...606...85C,2004MNRAS.351...63B,2005MNRAS.359L...5V,2005ApJ...629..654D,2006ApJ...640L.123M,2006A&A...452L..23N,2009ApJ...702..554P,2013ApJ...762...38P,2013MNRAS.428...28F}. Their physical nature remains mysterious, and it is not clear where they stand in relation to the astrophysical sources of \lya outlined above. Intriguingly, \citet{2009AJ....138..986K} find a dearth of LABs at low redshift ($z = 0.8$). Three mechanisms are suggested as the energy source for LABs. They are not mutually exclusive.
\begin{enumerate}[(a)] \addtolength{\itemsep}{-0.4\baselineskip} 
\item Internal photoionizing sources: LABs are the \lya fuzz predicted by \citet{2001ApJ...556...87H}, or the \lya coronae predicted by \citet{2005ApJ...622....7F}. LABs contain internal sources of ionizing radiation in the form of an AGN or star-forming galaxy, possibly obscured. The cosmological simulations of \citet{2012arXiv1210.3600C} suggest that star-bursts are responsible for the majority of \lya from LABs. This is the scenario suggested, for example, by the multiwavelength studies of \citet{2004ApJ...606...85C}, \citet{2007ApJ...655L...9G,2009ApJ...700....1G}, and \citet{2013ApJ...771...89O}. The polarisation of \lya emission from LAB1 is evidence of scattering, suggesting a central source of \lya photons \citep{2011Natur.476..304H}. However, the 1.1-mm imaging survey of \citet{2013arXiv1301.2596T} places strict a upper limit on ultra-luminous obscured star-formation, indicating that most LABs are not powered by intense, dusty star-formation. The population of dusty \lya emitters discovered by \citet{2012arXiv1205.4030B}, 50\% of which are extended ($\gtrsim 25$ kpc), show warm far-IR colours consistent with being in a short-lived AGN feedback phase. LABs have also been discovered around high-redshift quasars \citep{2009MNRAS.393..309S,2012A&A...542A..91N}.

\item Cooling radiation: LABs consist of gas that is radiating away its gravitational potential energy as it cools into massive galaxies.  \citet{2006ApJ...649...14D}, \citet{2006ApJ...649...37D}, and \citet{2009MNRAS.400.1109D} considered simple analytic models of this scenario, showing that the \lya linewidths and number densities of LABs can be reproduced by cold accretion if $\sim$20\% of the gravitational potential energy of the gas is converted to radiation. More detailed hydrodynamical simulations of the connection between LABs and cold flows have been performed by \citet{2010ApJ...725..633F}, \citet{2010MNRAS.407..613G} and \citet{2012MNRAS.423..344R}. These simulations do not agree on whether cooling radiation alone can power LABs --- the sensitivity of collisional ionisation and the cooling rate to the temperature, density, and metallicity of the gas make such calculations worryingly dependent on resolution, subgrid physics, self-shielding and ionising radiative transfer, heating-cooling balance, and other bugbears of numerical simulation. The simulations of \citet{2010MNRAS.407..613G} support the analytic findings of \citet{2009MNRAS.400.1109D}, that haloes of mass $\sim 10^{12} - 10^{13} \Msol$ at $z \sim 3$ can reproduce the extent, luminosity and irregular morphologies of LABs, with most of the \lya emission coming from 50-100 kpc cold streams. This scenario is argued for observationally by \citet{2006A&A...452L..23N,2007MNRAS.378L..49S,2008MNRAS.389..799S}, usually on the basis of a non-detection of associated AGN or star-formation. (But if the accretion rate is so high, why doesn't it fuel an AGN and/or form stars?)

\item Galactic superwinds: a starburst powers a barrage of supernovae, which sweep cooling, dense, radiating shells of \hi into the IGM \citep{2000ApJ...532L..13T,2004ApJ...613L..97M}. This scenario is favoured observationally by \citet{2003ApJ...591L...9O} and \citet{2008ApJ...675.1076S} on the basis of broad wing emission components on the red side, and a sharp cutoff on the blue side of the \lya line as predicted by wind models (Section \ref{subsubsec:lya_ism}). Observations of  associated, spatially-extended narrow \lya absorption lines in the spectra of active galaxies \citep[e.g.][]{2013MNRAS.428..563H}, which suggest the presence of an \hi bubble outside the \lya nebula, are consistent with a starburst-driven superbubble. However, \citet{2011ApJ...735...87Y} and \citet{2013arXiv1301.0622M} report that the velocity offset of \lya with respect to [O\textsc{iii}] is consistent with zero, suggesting that outflows are not the primary driver of \lya escape.
\end{enumerate}

\begin{figure*}[t]
\centering
	\includegraphics[width=\textwidth]{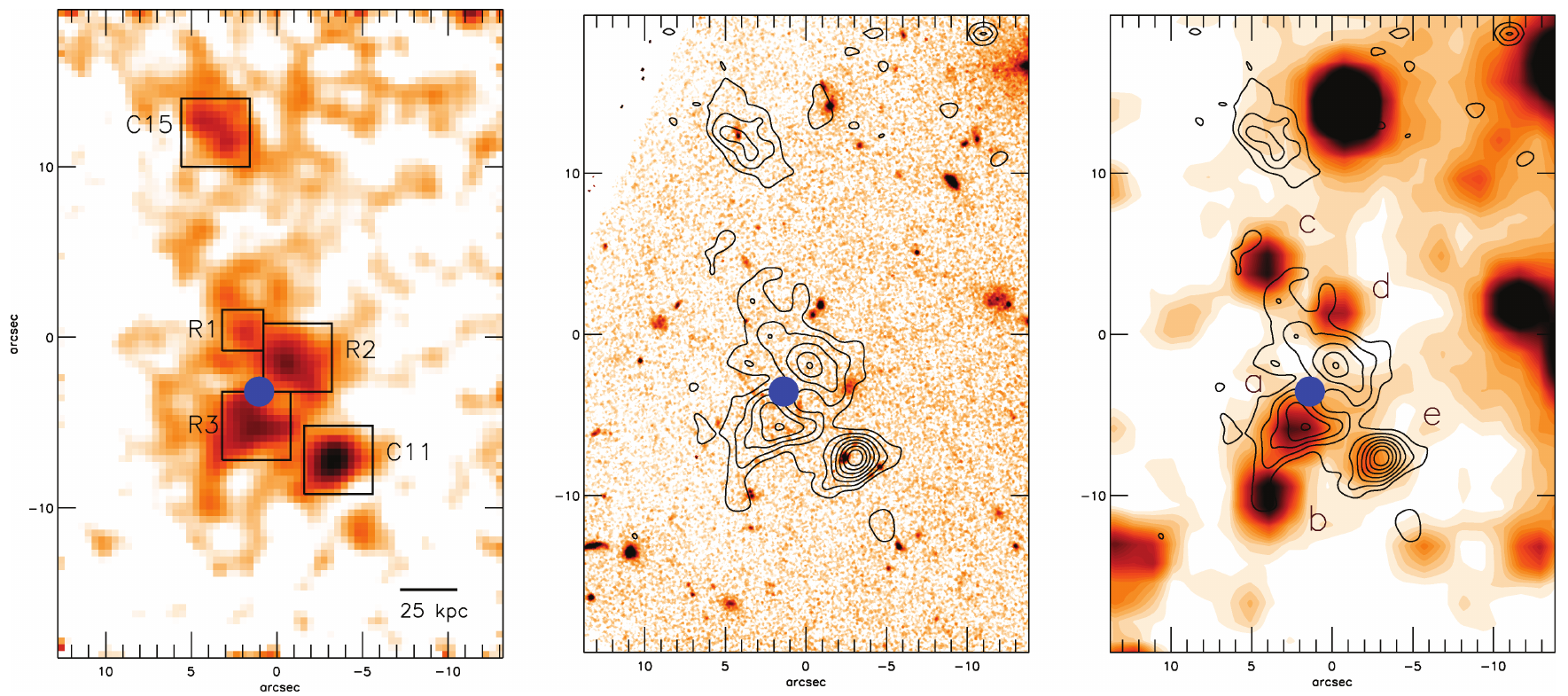}
	\caption{Multi-wavelength observations by \citet{2010MNRAS.402.2245W} of a \lya blob known as LAB1. \emph{Left}: continuum-subtracted \lya emission, obtained from collapsing the SAURON spectra over a narrow wavelength range centred on the emission line. The boxes mark individual sub-clumps of emission. \emph{Middle}: HST/STIS optical image overlaid with \lya contours from the left panel. The faintest contour has a surface brightness of $5.6 \times 10^{-18} \ergsca$ and contours increase in steps of $3.7 \times 10^{-18} \ergsca$. \emph{Right}: same as middle panel, but now a Spitzer/IRAC 3.6 $\mu$m image is displayed. In all plots, the blue dot denotes the position of the radio source \citep{2004ApJ...606...85C}. All images are plotted on the same scale and are orientated such that north is up and east to the left. The \lya blob has a complex morphology, with no simple relationship between \lya, optical and submillimeter emission. Figure from \citet[][Figure 1]{2010MNRAS.402.2245W}; used with permission.} \label{fig:LAB1}
\end{figure*}

Higher resolution observations haven't particularly clarified the situation. \citet{2010MNRAS.402.2245W} observed LAB1 and showed that \lya is emitted from five distinct clumps: two are associated with LBGs, one with a heavily obscured submillimeter galaxy, and two don't appear to be associated with a galaxy. The complex morphology is shown in Figure \ref{fig:LAB1}. Detailed observations of LABd05 by \citet{2012ApJ...752...86P} reveal a blob containing 17 small galaxies, none of which are at the peak of the \lya emission, with a smooth, non-filamentary morphology and a similarly extended UV surface brightness profile. \citet{2013MNRAS.428...28F} presented integral field spectroscopy of an LAB at $z = 2.38$, noting a chaotic velocity structure, two associated compact red massive galaxies, evidence of a superwind, an infalling filament of cold gas that resonantly scatters \lya photons, and bow-shocks and tidally stripped gas in outer subhaloes.

Fainter LABs are similarly complex. \citet{2011MNRAS.418.1115R,2013MNRAS.429..429R,2013arXiv1305.5849R} report long-slit spectroscopic observations of extended, ultra-faint \lya emitters at $z = 2.6 - 3.3$. The first shows a partly obscured ``red core'' of emission, accompanied by a more extended ($\sim$ 37 kpc proper) ``blue fan'' with evidence of substructure. This is interpreted as evidence of an infalling, filamentary cloud of \hi that is illuminated by stellar ionizing radiation. The large inferred ionizing escape fraction, together with \emph{HST} evidence of a stellar tidal tail, hint that the filament may be tidal debris from an interaction that has aided the escape of the ionizing continuum. The second system also suggests the influence of galaxy interactions. The system is extended, asymmetric with spatially irregular stellar components, and shows evidence for a very young starburst and possibly a \lya emitting filamentary structure. The tadpole shape and blue, partly turbulent tails of the filaments seen in the third system is interpreted as evidence of ram-pressure stripping as dwarf galaxies leave behind contrails of gas and stars as they fall into a more massive halo. 

Taken together, these systems suggest that galaxy interactions play an important role in the production and escape of \lya photons in protogalaxies. Such a link has been suggested previously by \citet{2010MNRAS.403.1020C}, who discovered that all LBG pairs in their sample that are separated by $\lesssim 15 h^{-1}$ kpc (projected) exhibit \lya in emission, in stark contrast to 50\% of the LBG population as a whole. Similarly, \citet{2013ApJ...773..153J} find a large fraction ($\sim 50\%$) of $z = 5.7-6.5$ LAEs show signs of merging / interaction, as do the handful of $z = 2.4$ double-peaked \lya emitters of \citet{2013ApJ...775...99C}. This suggests a scenario in which galaxy interactions, as well as providing fuel for star formation \citep{2011MNRAS.418.2196T}, sufficiently disrupt and ionise the ISM to give \lya photons an easy escape.

\citet{2010MNRAS.406..913S} presented a model in which galaxies can not only appear as LAEs when a burst of star-formation occurs \cite[i.e. a \lya-bright phase; see also][]{2007MNRAS.379.1589D,2007ApJ...660..945N}, but also when a young satellite galaxy is accreted onto a more massive object. The simulations of \citet{2012arXiv1210.6440Y} suggest that a merger can cause a burst in both star-formation and cooling radiation, which can result in an LAB. LABs (and their radio-loud counterparts --- see below) are often associated with overdense environments \citep{1999AJ....118.2547K,2000ApJ...532..170S,2004ApJ...602..545P,2004AJ....128..569M,2009MNRAS.400L..66M}, which may suggest a link to galaxy interactions, and/or gas accretion.

Extended \lya emission is also associated with high-redshift radio galaxies. Selecting sources by their radio emission generally finds the most massive high-redshift objects, either galaxies or AGN. Giant \lya haloes (or nebulae) have been discovered around many radio galaxies, and their properties have been studied by \citet{2003ApJ...592..755R,2005MNRAS.359L...5V,2006AN....327..175V,2007MNRAS.375.1299V,2007ApJ...655L...9G,2007MNRAS.378..416V,2007NewAR..51..194V,2008A&ARv..15...67M,2008A&A...488...91C,2009ApJ...694L..31Z}. These objects resemble LABs, except that they are radio loud, have a higher surface brightness (by a factor of $\sim 5$) and contain large, multi-component galaxies \citep{2006AN....327..175V}. \citet{2003ApJ...592..755R} have suggested the following evolutionary sequence: LABs represent the very first stage in the formation of a large galaxy (or a set of smaller galaxies that later merge), and evolve into radio-loud \lya haloes when galaxy merging triggers an AGN.

\subsection{\lya Emission in cosmological context}
\label{lya_sam}

To understand the properties and evolution of the LAE population as a whole, it is essential to investigate \lya emitting galaxies in their cosmological context. Observations of \lya emitters are sensitive to the large scale structure of the universe, and in particular to the ionisation state of the IGM.

\paragraph*{LAEs in Semi-Analytic Models:}

As large samples of LAEs became available in the early 2000s, semi-analytic models of galaxy formation attempted to model the statistical properties of LAEs. These models incorporate the hierarchical evolution of dark matter haloes analytically \citep[][]{1974ApJ...187..425P,2002MNRAS.329...61S} or with N-body simulations \citep[e.g.][]{2005Natur.435..629S}. Galaxies are formed within dark matter haloes according to semi-analytic recipes or SPH simulations. Predictions of observed \lya emission must incorporate the \lya escape fraction (f$_{\rm esc}$).  This has been modelled with simple phenomenological prescriptions \citep[e.g.][]{2007ApJ...670..919K}, and more sophisticated calculations based on numerical \lya radiative transfer \citep[e.g.][]{2012MNRAS.422..310G}.

The simplest approach assumes that a constant fraction of \lya photons escape from each galaxy, with the value of f$_{\rm esc}$ is adjusted to fit \lya LF data \citep[e.g.][]{2005MNRAS.357L..11L,2006MNRAS.365..712L,2008MNRAS.389.1683D,2009MNRAS.400.2000D,2010PASJ...62.1455N}. The typical reported value of f$_{\rm esc}$ = 10\% is strongly model-dependent because the intrinsic \lya LF depends on the underlying dark matter simulation and its cosmology, the baryonic prescriptions used to model galaxy formation, and the IMF. For example, the best-fit f$_{\rm esc}$ value varies from 0.02 to 0.20 when changing from a top-heavy IMF to a more standard Kennicutt IMF \citep{2006MNRAS.365..712L}.

Alternatively, ``duty cycle'' models have been investigated \citep{2009MNRAS.398.2061S,2010PASJ...62.1455N}. In these scenarios, only a fraction of galaxies are \textit{turned on} as \lya emitters at a given time. In the model of \citet{2010PASJ...62.1455N}, based on a cosmological SPH simulation, a duty cycle of $7\%$ ($20\%$) is necessary to reproduce both the UV LFs of LBGs and \lya LFs of LAEs at $z = 3$ (6). 

These simple models quantify how the intrinsic \lya LFs from models have to be modified to reproduce the data, either in terms of \lya luminosity (constant fraction scenario) or LAE density (duty cycle scenario). It remains to be seen whether more physical models support either picture.
 
An obvious method to deal with the dust extinction of the \lya line is to treat it the same way as UV continuum, neglecting resonant scattering. Various models have been tried using simple screen- and slab-like distributions of dust \citep{1999ApJ...518..138H,2007ApJ...667..655M,2007ApJ...670..919K,2011MNRAS.418.2273S}. Additional parameters often included to improve the agreement with the UV and \lya data. In \citet{2007ApJ...670..919K}, a phenomenological implementation is developed for the \lya escape fraction, varying the visibility of an LAE depending on whether the starburst galaxy is in a pre-outflow, outflow, or post-outflow phase. The model reasonably reproduces the UV LFs of LBGs and \lya LFs of LAEs between $z \approx 3 - 6$.

To assess the impact of the Neufeld effect discussed in Section \ref{ism_struct}, clumpy dust distributions have been investigated in the context of cosmological simulations \citep{2008MNRAS.389.1683D,2010ApJ...708.1119K,2011MNRAS.418.2273S,2011MNRAS.410..830D}. The ISM clumpiness is usually described by the $q$ (free) parameter, defined in Equation \ref{eq_q}. This is equivalent to tuning independently the \lya and continuum dust extinction. In addition to UV and \lya LFs, a very clumpy model \citep[e.g.][assume $q=0.15$]{2010ApJ...708.1119K} can reproduce other observed quantities, such as the \lya EW distribution and the anti-correlation between UV magnitudes and \lya EWs.

\citet{2011MNRAS.410..830D} coupled cosmological SPH simulations and ionizing RT calculations to predict the ionisation state of the intergalactic medium on large scales. ￼They use a phenomenological recipe for ISM dust extinction and an analytic model for \lya IGM opacity along the line-of-sight.\footnote{\citet{2011MNRAS.410..830D} computes the IGM attenuation of a Gaussian \lya spectrum, unprocessed by ISM scattering. Internal radiative transfer effects are expected to alter the line emerging from the galaxy. In particular, outflows are expected redshift the \lya line, reducing the effect of IGM attenuation \citep{2010MNRAS.408..352D}.} They conclude that the IGM transmission alone cannot explain the \lya and UV LFs at $z = 5-7$, and that the ISM extinction, described by a clumpy dust distribution, is required to simultaneously match UV and \lya data. 
There is tension between the success of clumpy ISM models, which assume the enhancement of the escape of \lya radiation over that of UV continuum, and the current interpretation of observations, and with detailed numerical RT simulations, as we have seen in Section \ref{ism_struct}.\\

The models discussed above all use phenomenological prescriptions to describe f$_{\rm esc}$; they don't model the resonant scattering of \lya photons in the ISM. Only recently have simulations have incorporated \lya RT calculations into galaxy formation models in a cosmological context.

\citet{2012MNRAS.419..952F} computed the \lya escape fraction of galaxies by post-processing the MareNostrum SPH simulation\footnote{\tt{http://astro.ft.uam.es/marenostrum/}}, approximating the ISM as a dusty slab of gas with the \lya sources being homogeneously distributed. Following \citet{2000ApJ...539..718C}, the dust distribution is described using two components: a homogeneous ISM, and dense birth clouds of young stars. Resonant scattering of \lya photons is taken into account only for the homogeneous phase, using an analytical fit to numerical RT calculations for a slab configuration. The \hi opacity of birth clouds is assumed to be low, and so their dust extinction of \lya photons is taken as for the UV continuum. Their model provides a reasonable match with \lya and UV LF data at $z = 5-6$, although the faint-end is strongly overpredicted. They find that  f$_{\rm esc}$  decreases towards more massive host haloes, echoing the results of \citet{2009ApJ...704.1640L}. In an extension of this model, \citet{2012MNRAS.419..952F} argue that the average lower f$_{\rm esc}$ of galaxies in massive haloes is an important factor in explaining the small fraction of LAEs found in samples of LBGs.

A similar conclusion is drawn in the work of \citet{2012MNRAS.422..310G}, though their physical hypotheses are quite different. Noting that galactic winds are ubiquitous in high-redshift galaxies \citep{2010ApJ...717..289S},  \citet{2012MNRAS.422..310G} consider \lya emission that is powered by star formation and scattered within a neutral outflow represented by a spherical shell. They couple the semi-analytic model of galaxy formation GALICS \citep{2003MNRAS.343...75H} with a library of numerical \lya RT experiments \citep{2011AandA...531A..12S}, based on the expanding shell model of \citet[][]{2006AandA...460..397V}. In their model, higher velocity, denser and more dusty shells are found in more massive, star-forming galaxies. Their \lya LFs, UV LFs of LAEs and LBGs are in good agreement with high-redshift observations. In particular, the model predicts a large abundance of faint line emitters ($10^{41-42}$ erg s$^{-1}$): low-SFR galaxies are mostly dust-free and so have a \lya escape fraction of the order of unity. However, f$_{\rm esc}$ is strongly dispersed (from 0 to 1) for massive/star-forming galaxies as a result of a trade-off between larger shell velocities and \hi/dust opacities.

Shell models are useful for interpreting \lya line profiles (Section \ref{subsubsec:lya_ism}) and statistical properties of LAEs \citep[luminosity functions, mean stellar masses, overlap with LBGs, etc.;][ see also \citealt{2012MNRAS.425...87O}]{2012MNRAS.422..310G}. Nonetheless, they assume a physical picture of galactic outflows that is quite idealised, and more realistic models are needed.

\paragraph*{The LAE-LBG connection:}

Cosmological models of both LAEs and LBGs can illuminate the apparent tension between their physical properties (as discussed in Section \ref{Ss:LAEs}). Although it is reported that LBGs tend to be more evolved  than LAEs, with higher stellar mass and dust content, the simulations of \citet{2012MNRAS.421.2568D} suggest that $z \approx$ 6 LAEs and LBGs have similar physical properties. They find that LAEs are a subset of the LBG population; different observed properties result only from different selection criteria. 

Using their numerical \lya radiative transfer models, \citet{2008AandA...491...89V} argue that the link between LBGs and LAEs is mainly governed by RT effects due to variations in HI and dust opacities. They suggest that galaxy mass is the driver of variation of gas and dust content. This would explain the small fraction of LBGs with \lya in emission \citep{2003ApJ...588...65S,2011ApJ...728L...2S}: LBGs are highly star-forming, massive galaxies, in which the \lya emission can be strongly suppressed \citep[see also][]{2012MNRAS.422..310G}. Environment must also play a role --- using $\sim$57,000 $z \sim 3$ LBGs in the CFHTLS Deep Field, \citet{2013MNRAS.433.2122C} showed from their auto- and cross-correlation functions that LBGs with \lya in emission vs. absorption live in very different environments, e.g. parent haloes of $10^{11} \Msol$ vs. $10^{13} \Msol$.

While the UV LF of LBGs decreases from $z = 3$ to $z = 6$ \citep{2007ApJ...670..928B}, the UV LF of LAEs appears to increase over this redshift range and is a reasonable match for LBGs at $z = 6$ \citep{2008ApJS..176..301O}. This trend can be interpreted as LAEs being a sub-population of LBGs, where the LAE fraction increases towards higher redshift. The dropout technique preferentially detects bright UV-continuum galaxies; hence a significant fraction of low-continuum objects (i.e. LAEs with high equivalent width) are missed by LBG surveys at $z = 6$ \citep[10-46\%,][]{2007ApJ...660...47D}. Therefore, LAEs and LBGs do not necessarily arise from the same population. The apparent overlap of the LAE/LBG UV LF at $z = 6$ may be accidental \citep{2007MNRAS.379.1589D}. 

\paragraph*{Lyman Alpha and the IGM:}
Though this review is focussed on galaxy properties, the interaction of \lya radiation with the intergalactic medium (IGM) can greatly affect the observed properties of \lya emitters. \citet{1965ApJ...142.1633G} first predicted that the presence of \hi in the IGM would absorb radiation blueward of rest-frame \lya as the expansion of the universe redshifts this light into resonance. Following the recombination of the primordial plasma, the universe was neutral until the formation of sources able to reionise the IGM. In spite of intense research, the nature of these sources is still unknown, and the Epoch of Reionisation (EoR) is only partially constrained. Observations of the cosmic microwave background suggest that the EoR began around $z = 11$ \citep{2011ApJS..192...16L}. Based on the analysis of the spectra of high-redshift quasars \citep{2006AJ....132..117F}, the Universe must have been almost completely reionised at $z \gtrsim6$.

The \lya emission line from high-redshift galaxies is a useful tool for probing the variation in the ionisation state of the IGM. One expects that, as the fog of neutral hydrogen clears during reionisation, a population of hitherto obscured \lya emitting galaxies will burst into view. An increase in \lya attenuation could trace a sudden change in the neutral fraction of the IGM ($x_{\hi}$),  providing a diagnostic of the epoch of reionisation \citep{1998ApJ...501...15M,1999ApJ...518..138H,2001ApJ...563L...5R}. Several authors have suggested that a rapid evolution of the \lya luminosity function at high redshift would reflect a change in the IGM neutral fraction \citep{2004ApJ...617L...5M,2004MNRAS.349.1137S,2005ApJ...623..627H,2007ApJ...667..655M}, especially if there were no corresponding evolution of the UV LF of LAEs. While \citet{2010ApJ...725..394H} find that $L^∗_{\lya}$ remains unchanged between z = 5.7 and 6.6, \citet{2010ApJ...723..869O} and \citet{2011ApJ...734..119K} measure a decline of about $30\%$, along with little evolution of the UV LF \citep[see also][]{2006ApJ...648....7K}. This small increase in the \lya attenuation translates into a neutral fraction $x_{\hi} \lesssim 20\%$ according to \citet{2010ApJ...723..869O}, which would imply that the Universe is still highly ionized at $z = $ 6.6. On the other hand, \citet{2007MNRAS.379..253D} have shown that the observed drop of the \lya LF between z = 5.7 and 6.6 is expected from the IGM opacity evolution due to cosmic expansion, even if the Universe remains fully ionized \citep[see also][]{2011ApJ...728...52L}.

The spectral line profile can also constrain reionisation.  \citet{2010ApJ...725..394H} and \citet{2011ApJ...734..119K} find very similar shapes of \lya composite spectra at $z = 5.7$ and $6.5$. Nevertheless, they note that the peak of the rest-frame \lya equivalent width distribution shifts towards lower values from $z = 5.7$ to $6.5$, perhaps due to an increase of the IGM contribution to the damping of the \lya line. The interpretation of this effect is not straightforward given that infalling and outflowing gas in the vicinity of the galaxy can also strongly affect the \lya profile. For instance, \citet{2010MNRAS.408..352D} demonstrate that radiative transfer through galactic winds can increase the visibility of the \lya line even for a highly neutral IGM. 

Few \lya surveys at $z \gtrsim 7$ have been conducted and the number of detections is still small. Nonetheless, small statistics or even non-detections can constrain the change of the LF. Observations tentatively indicate an evolution of the \lya emitting galaxy population $z = 6-7$ \citep{2012ApJ...752..114S,2012AandA...538A..66C,2013arXiv1311.0057C}, which can be accounted for using models positing a rapid increase of $x_{\hi}$ \citep{2007ApJ...667..655M,2010ApJ...708.1119K,2011MNRAS.414.2139D,2013MNRAS.428.1366J}. On the other hand, \citet{2010ApJ...721.1853T}, \citet{2012ApJ...744...89H} and \citet{2012ApJ...745..122K} find no conclusive evidence for a variation of L$^{*}$ up to $z = 7-8$, based on photometric samples. Future spectroscopic follow-up observations will certainly provide more accurate estimates.\\

\begin{figure}[t]
\includegraphics[width=8.5cm,height=7.cm]{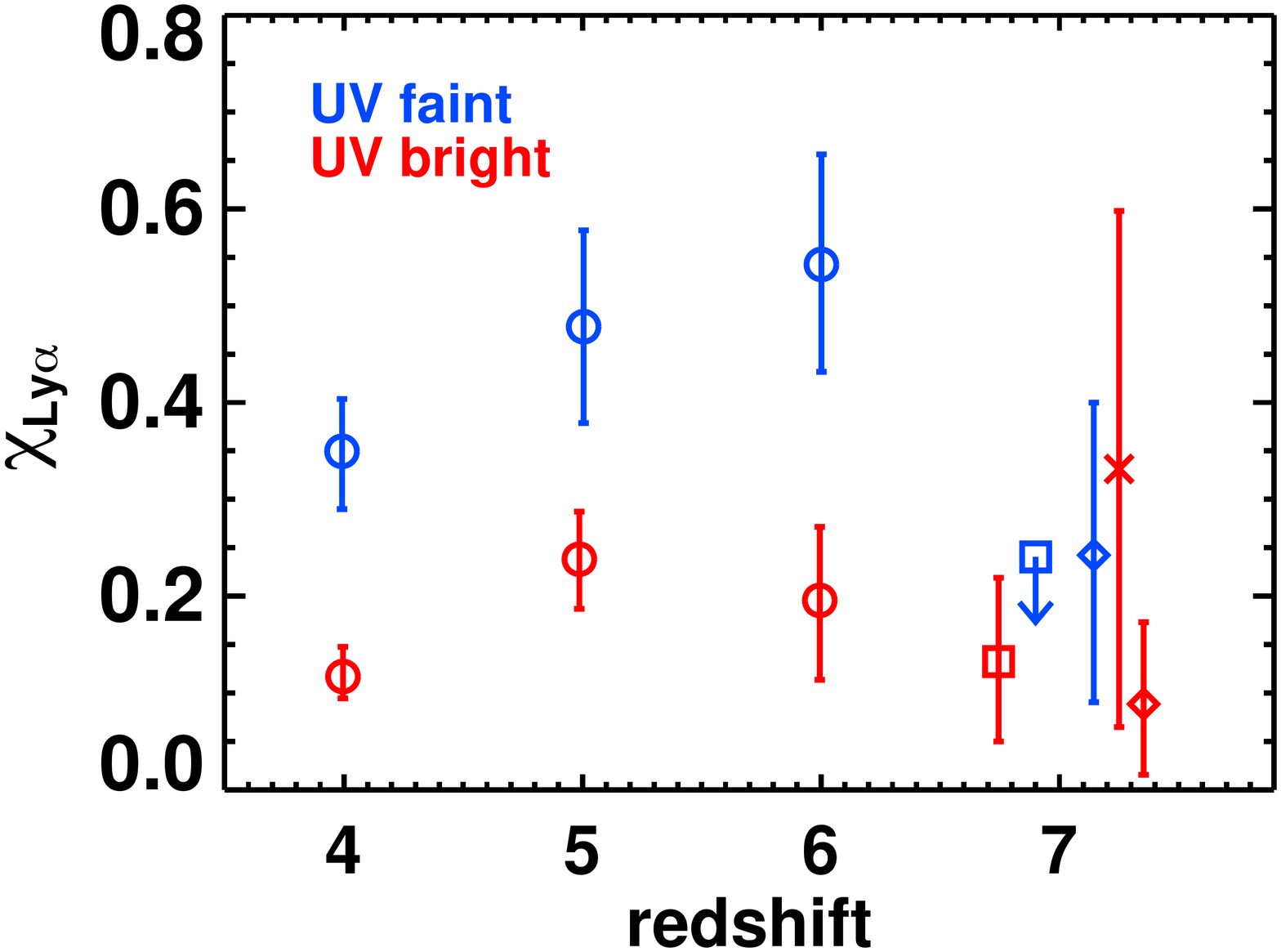}
\caption{Redshift evolution of the fraction \protect \raisebox{2pt}{$\chi^{}_{{\rm Ly}\alpha}$} of \lya emitters (EW$>25$\AA) within the Lyman-Break galaxy population. Blue and red symbols correspond to UV-fainter and UV-brighter galaxies respectively. The $z = 4-6$ data from \citet{2010MNRAS.408.1628S}, represented by circles, show an increase of \protect \raisebox{2pt}{$\chi^{}_{{\rm Ly}\alpha}$} with redshift. The observations at $z \approx$ 7 by \citet{2012ApJ...744..179S}, \citet{2011ApJ...743..132P} and \citet{2012ApJ...744...83O} correspond to the diamonds, the squares and the cross respectively. They show that the fraction of \lya emitters is likely to drop from $z = 6$ to 7, possibly suggesting an increase of the IGM neutral fraction $x_{\hi}$. Also, the decline of \protect \raisebox{2pt}{$\chi^{}_{{\rm Ly}\alpha}$} at $z \approx6$ is stronger for UV faint galaxies than UV bright ones. It may imply that $x_{\hi}$ is evolving more significantly in low-density regions at this epoch.}
\label{schenker12}
\end{figure}

The evolution of the fraction of \lya emitters \raisebox{2pt}{$\chi^{}_{{\rm Ly}\alpha}$} among samples of Lyman-Break galaxies provides another useful probe of the ionisation state of the IGM. As discussed in Section \ref{statprop}, \citet{2010MNRAS.408.1628S} measure an increase of \raisebox{2pt}{$\chi^{}_{{\rm Ly}\alpha}$} from $z = 3$ to $z = 6$. Interestingly, the trend starts to reverse at $z >6$. \citet{2011ApJ...743..132P}, \citet{2012ApJ...744...83O}, and \citet{2012ApJ...744..179S} report that \raisebox{2pt}{$\chi^{}_{{\rm Ly}\alpha}$} drops from $z = 6$ to $z = 7$ , as shown in Figure \ref{schenker12}. Furthermore, the drop of the fraction of LAEs is more significant among the fainter UV-selected galaxies. These galaxies are less clustered than brighter ones and located in low density environments \citep[e.g.][]{2001ApJ...550..177G,2005ApJ...619..697A}, which may indicate that reionisation occurred later in low density regions, privileging an inside-out model. Indeed, galaxies in more clustered regions can blow larger ionized bubbles during the EoR, which should boost the \lya transmission \citep{2007MNRAS.381...75M,2008MNRAS.391...63I,2008MNRAS.386.1990M,2009MNRAS.400.2000D}. A strong signal in the two-point correlation of LAEs in the EoR is predicted by  \citet{2007MNRAS.381...75M}.

Whether or not reionisation is over by $z \sim$ 6 is not yet clear \citep[e.g.][]{2010MNRAS.407.1328M,2011MNRAS.415.3237M,2012MNRAS.421.1969R}, and there is no conclusive evidence that \lya observations at $z =$ 6-7 are the signature of a sharp transition of the IGM neutral fraction. Other reasons can explain the Lyα attenuation around $z = 6 - 7$, such as the intrinsic evolution of the galaxies \citep{2008MNRAS.389.1683D,2007MNRAS.379..253D,2009MNRAS.398.2061S}, a change in kinematics and covering factor affecting the escape of Lyα photons from galaxies \citep{2011MNRAS.414.2139D}, observational uncertainties \citep{2011ApJ...743..132P}, or the increase of the number density of optically thick absorption systems \citep{2012MNRAS.tmp..412B}. \\

\begin{figure*}[t]
\centering
	\includegraphics[width=0.9\textwidth]{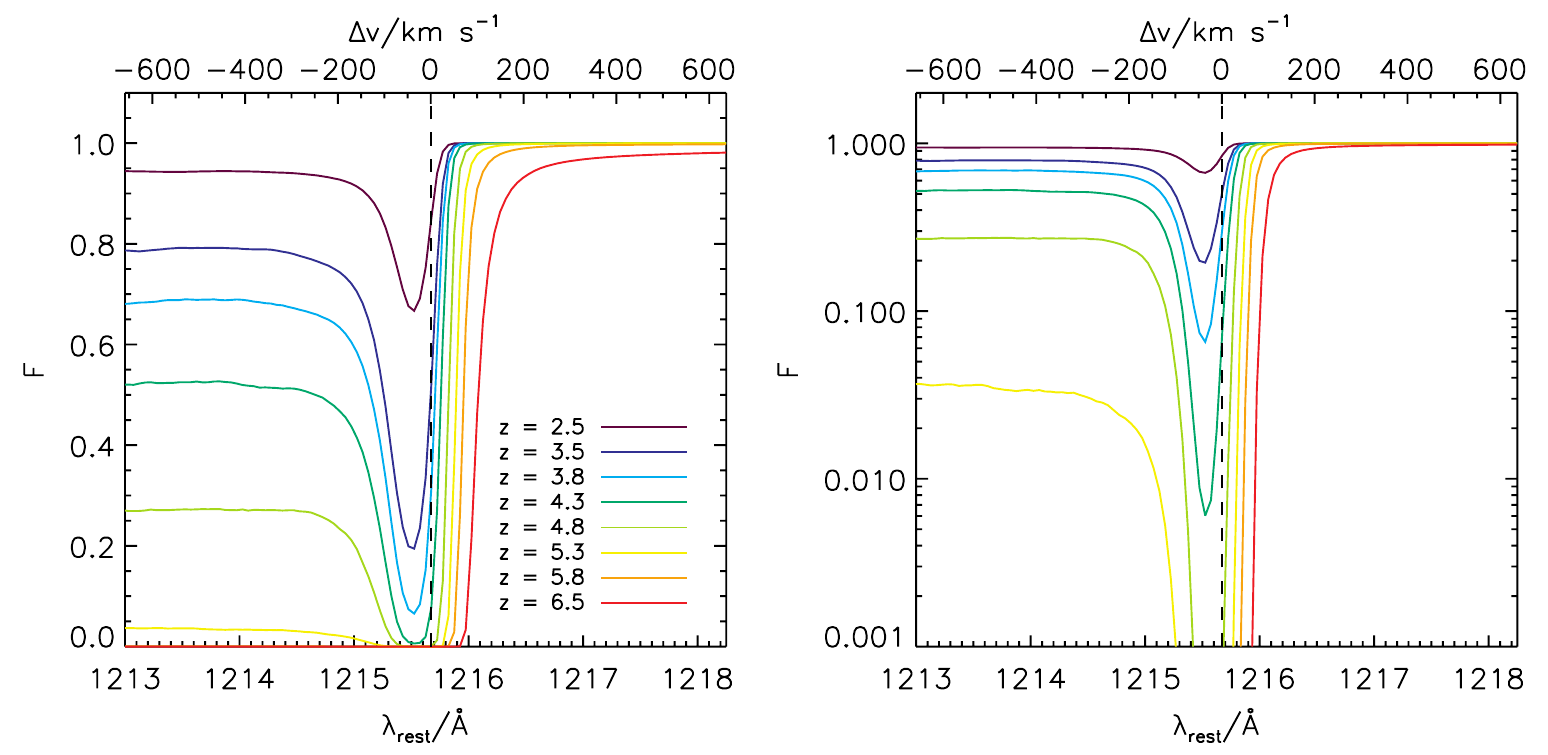}
	\caption{The results of \citet{2011ApJ...728...52L}, showing the normalized, average IGM transmission $F (\lambda)$ at wavelengths around \lya, for different redshifts indicated by the colour. In the left panel, the vertical axis is linear, while in the right it is logarithmic, emphasizing the transmission at high redshifts. Even at fairly low redshifts, the IGM is able to erase a substantial fraction of the blue side of the intrinsic spectrum. At higher redshifts, even the red side of the line can be affected. Figure from \citet{2011ApJ...728...52L}; reproduced by permission of the AAS.} \label{fig:LaursenIGM}
\end{figure*}

\lya radiative transfer through the IGM has been studied by a number of authors. \citet{2011ApJ...728...52L} traced \lya photons from emitters between $z = 2.5$ and 6.5 through a cosmological simulation of the IGM. Figure \ref{fig:LaursenIGM} shows the average transmission of wavelengths around \lya as a function of the redshift of the emitter. We see that, even at fairly low redshifts, the IGM is able to erase a substantial fraction of the blue side of the intrinsic spectrum. At higher redshifts, even the red side of the line can be affected. Note also that this is the average transmission: individual sightlines can vary substantially. For example, at $z = 3.5$ the transmission on the blue side ($\Delta v < -200 \kmsec$) varies between 0.2 and 0.95.

\citet{2011ApJ...728...52L} note that, in some cases, a single, red \lya peak seen in high-redshift galaxies could be the result of a intrinsic double-peaked spectrum with the blue side erased by the IGM. While this effect is not sufficient to explain the predominance of red-peaks in LAEs, it does demonstrate the importance of the IGM in interpreting \lya spectra. Conversely, \citet{2010MNRAS.408..352D} and \citet{2011MNRAS.414.2139D} have emphasised that the intrinsic \lya spectra affects IGM transmission. The presence of a galactic wind in a pre-reionisation galaxy will shift most of the \lya photons to the red-side before they encounter the IGM. Even a mild wind ($\sim 25 \kmsec$) increases ISM transmission relative to predictions that assume that \lya emission peaks at line centre \citep[e.g.][]{2008MNRAS.391...63I}.

\citet{2011ApJ...726...38Z} have emphasised the importance of large-scale structure for LAE visibility --- lower \hi densities and larger velocity gradients aid \lya escape, meaning that, for example, LAEs are more visible for sightlines that look along a filament than across one \citep[see also][]{2011MNRAS.415.3929W,2012arXiv1212.0977G,2013A&A...556A...5B}. Such studies are crucial to controlling the systematics of massive observational programs underway to measure cosmological parameters using LAEs as tracers of the cosmic web, such as the Hobby-Eberly Telescope Dark Energy Experiment  \citep[HETDEX][]{2008ASPC..399..115H}, whose aim is to measure the spectroscopic redshifts of 800,000 LAEs between $1.9 < z < 3.5$.

\lya may provide a way of directly observing the IGM, and thereby mapping the cosmic web. \citet{1987MNRAS.225P...1H} predicted that the UV ionizing background falling on clouds of \hi in the IGM would produce detectable \lya fluorescence. This emission would allow us to study the size and morphology of these clouds, as well as the strength of the UV background. The calculations of \citet{1987MNRAS.225P...1H} were refined by \citet{1996ApJ...468..462G}, who correctly predicted that detecting FLEs would be difficult even with a 10m telescopes. The more sophisticated modelling of \citet{2005ApJ...628...61C} have reduced expectations of observing FLEs further still, by showing that simplifications such as a static, plane slab geometry overpredict the visibility of FLEs. Similarly, the cosmological simulations of \citet{2010ApJ...708.1048K} predict that, in the absence of a strong ionizing continuum source, the highest fluorescent surface brightnesses at z = 3 are $\sim 2 \times 10^{-19} \ergsca$, which would require $\gtrsim 1000$ hours to detect. \citet{2007ApJ...657..135C} report that despite the observational efforts of themselves and others in finding a handful of plausible candidates, there is still some doubt about whether UVB-powered FLEs have actually been detected. As noted above, the observations of \citet{2014Natur.506...63C} of the \lya-emitting environment of a quasar may represent out best hope of seeing the IGM in emission.

A more comprehensive review of the ability of Lyman alpha emitting galaxies to probe cosmic reionisation can be found in \citet{2014arXiv1406.7292D}.

\section{Summary and Future}

We have reviewed what observations and theoretical models of \lya emission, and \lya and {\MgII} absorption have told us about the interstellar, circumgalactic and intergalactic medium of galaxies. Individually, each tracer provides direct insight into the state and physics of the gas in and around galaxies. Much more can be understood by combining these different spectral probes. We saw in Section \ref{Ss:LyaCGM} that modelling of the \lya emission and absorption of DLAs dramatically improves constraints on galaxy properties. The low ion absorption lines of DLAs constrain kinematics, metallicity and, via galaxy formation models, DLA halo mass distribution and even galactic feedback (Section \ref{S:lyaAbs}). The samples of {\MgII} absorbers (and their associated galaxies) in the redshift range that can be probed by \lya from the ground could shed light on the accretion and feedback physics that shapes star-forming galaxies. Observations of the dust, stellar and star-formation properties of \lya galaxies must continue to inform models of \lya emission.

Our understanding of \lya emission remains largely untouched by 21cm observations. The 21cm transition is an ideal complement to \lya: it is emitted by the same atoms, with a different dependence on temperature, is often optically-thin (at least in the warm-neutral medium), and provides a wealth of kinematic and spatial information. 21cm observations would greatly assist theoretical models of \lya emitters, providing a constraint on the extent, structure and kinematics of \hi that is independent of the sources \lya photons. At the moment, resolution and brightness limits have confined 21cm observations to the local universe, while the UV wavelength of \lya keeps ground-based observations to $z \gtrsim 2$. Future radio telescopes, space-based \lya observations, and {\MgII} observations at intermediate redshifts will continue to close the gap.

The increasing number of detections of LAEs at high redshift in recent times has significantly improved our understanding of the physical nature of these galaxies, and put tighter constraints on theoretical models (see Section \ref{S:lyaEmission}). Nonetheless, larger samples (especially with spectroscopic confirmation) and multi-wavelength observations will help greatly to refine existing models. Thanks to MUSE, KCWI, HETDEX and more, samples of LAEs will soon become orders of magnitude larger, deeper and more detailed. Space-based observation of local \lya emitters will also provide important tests of our understanding of the relevant physics. In particular, ISM models and galactic outflows in \lya radiative transfer calculations could be better informed by local observations, allowing models to replace their current simple approximations. From a theoretical viewpoint, high-resolution hydrodynamic simulations with \lya radiative transfer and large cosmological volumes are necessary to provide further insights into the complex mechanisms governing \lya emission in galaxies, and reproduce the wide variety of properties seen in the LAE population.

For DLAs and {\MgII} absorbers, future southern hemisphere surveys, similar to SDSS, will provide a factor of two or so more systems, from which we can learn about the details of gas physics around galaxies. Further advancements in simulations are required to interpret the data. The full range of gas cloud sizes for absorbers --- from a few pc to a few kpc --- has not yet been resolved in cosmological simulations. Higher resolution simulations and the inclusion of important physical processes such as AGN feedback and self-shielding will shed light on gas within and around galaxies. 

\section*{Acknowledgments} %If needed

We would like to thank the anonymous referee, Mark Dijkstra, Chris Churchill, Stephen Curran and James Allison for their insightful comments.

\end{document}